\documentclass[finnish,english,a4paper,12pt]{report}
\usepackage{ucs} 
\usepackage[utf8x]{inputenc} 
\usepackage[T1]{fontenc}
\usepackage{graphicx}
\usepackage{tabularx}
\usepackage[english]{babel}
\usepackage{float}
\usepackage[center]{subfigure}
\usepackage{pslatex}
\usepackage{tabularx}
\usepackage[super,numbers,compress]{natbib}
\usepackage{comment}
\usepackage[version=4]{mhchem}

\usepackage{times,mathptmx}
\usepackage{mathtools}
\usepackage{sectsty}
\usepackage{graphicx} 
\usepackage{lastpage}
\usepackage[format=plain,justification=raggedright,singlelinecheck=false,font={stretch=1.125,small,sf},labelfont=bf,labelsep=space]{caption}
\usepackage{fancyhdr}
\usepackage{fnpos}
\usepackage{array}
\usepackage{droidsans}
\usepackage{charter}
\usepackage[usenames,dvipsnames]{xcolor}
\usepackage{setspace}
\usepackage{amsmath}
\newcommand\mbf{\mathbf}

\newcommand{\ie}{\emph{i.\,e.},\xspace}

\newcommand{\trans}{\emph{trans}\xspace}
\newcommand{\meso}{\emph{meso}\xspace}

\newcommand*\ket[1]{ \left | #1 \right \rangle }

\newcommand*\braOpKet[3]{\left \langle #1 \left | \hat{#2} \right | #3 \right \rangle }

\newcommand*\braOpNoHatKet[3]{\left \langle #1 \left | #2 \right | #3 \right \rangle }

\newcommand*\modulus[1]{ \left | #1 \right | }

\newcommand{\vect}[1]{\mathbf{#1}}

\newcommand{\pd}[3]{\frac{\partial^{#1}{#2}}{\partial{#3}^{#1}}}
\newcommand{\half}{\frac{1}{2}}
\newcommand{\dr}{\mathrm{d}\vect{r}}

\newcommand{\ofR}{(\vect{r})}

\newcommand{\where}{\ensuremath{\; \mathrm{, \, where}}}

\newcommand*{\nAT}[1]{\mbox{\ensuremath{#1 \, \mathrm{nA} \cdot \mathrm{T}^{-1}}}}

\newcommand*{\nATA}[1]{\mbox{\ensuremath{#1 \, \mathrm{nA} \cdot \mathrm{T}^{-1} \cdot} \AA \ensuremath{^{-2} }}}
\newcommand*{\bohr}[1]{\mbox{\ensuremath{#1 \, a_\mathrm{0}}}}

\usepackage{xspace}
\usepackage{listings}

\newcolumntype{C}{>{\centering\arraybackslash}X}
\newcolumntype{R}{>{\raggedleft\arraybackslash}X}
\newcolumntype{L}{>{\raggedright\arraybackslash}X}

\graphicspath{.}

\subsubsectionfont{\normalfont\itshape\bfseries}

\makeatletter

\begin{document}

\title{{\it \Large From the book, Aromaticity: Modern Computational Methods and Applications,
edited by Israel Fern{\'a}ndez, publishing 2021 by Elsevier, Inc. All rights
reserved.}  \\
\vspace{24mm}
{\LARGE  CHAPTER 5} \\
\vspace{24mm}
 {\LARGE \bf Current density, current-density pathways and molecular aromaticity} 
}
\vspace{0.6cm}

\vspace{0.5cm}
\author{Maria Dimitrova and  Dage Sundholm \\
\textit{Department of Chemistry, Faculty of Science,} \\
\textit{P.O. Box 55 (A.I. Virtanens plats 1),} \\
\textit{FIN-00014 University of Helsinki, Finland,} \\
\textit{E-mail: maria.dimitrova@helsinki.fi, dage.sundholm@helsinki.fi}
}

\date{\today}

\maketitle

\linespread{1.5}

\section*{Abstract}

\noindent \normalsize{Current densities are induced in the electronic structure
of molecules when they are exposed to external magnetic fields.  Aromatic
molecular rings sustain net diatropic ring currents, whereas the net ring
current in antiaromatic molecular rings is paratropic and flows in the
opposite, non-classical direction.  We present computational methods and
protocols to calculate, analyse and visualise magnetically induced current
densities in molecules. Calculated ring-current strengths are used for
quantifying the degree of aromaticity. The methods have been demonstrated by
investigating ring-current strengths and the degree of aromaticity of aromatic,
antiaromatic and non-aromatic six-membered hydrocarbon rings. Current-density
pathways and ring-current strengths of aromatic and antiaromatic porphyrinoids
and other polycyclic molecules have been studied. The aromaticity and current
density of M{\"o}bius-twisted molecules has been investigated to find the
dependence on the twist and the spatial deformation of the molecular ring.
Current densities of fullerene, gaudiene and toroidal carbon nanotubes have
also been studied.   
}

\section*{Keywords}

Molecular aromaticity, magnetically induced current-density susceptibility,
ring-current strength, current-density pathway, diatropic ring current,
paratropic ring current, gauge-including atomic orbitals, nuclear magnetic
shielding, magnetic shielding density, magnetic susceptibility

\subsection{Magnetic fields in quantum mechanics} 

Magnetic fields interact with charged particles in motion and induce rotational
motion (precession) of the particle.  Electrostatic fields which are routinely
studied with quantum chemistry methods are scalar fields, whereas magnetic
fields are vector fields, rendering the interpretation of the electronic
response to the magnetic field complicated as it requires specialised
computational approaches. In the presence of an external magnetic field, the
generalised momentum operator $\vect{p}$ comprises the kinetic momentum
operator $\hat{\vect{\pi}}$ 
and the magnetic vector potential $\vect{A}$ of the magnetic field, which can
be expressed as  

\begin{equation}
\hat{\vect{\pi}} = -i \hbar \nabla + e \vect{A}.
\label{eq:kinMomOp}
\end{equation}

\noindent The vector potential defines the magnetic flux $\vect{B}$ of the
external magnetic field, which is commonly referred to as the magnetic-field
strength. A uniform external magnetic field can be obtained as the curl of the
vector potential, 

\begin{equation}
    \vect{B} = \nabla \times \vect{A}.
\end{equation}

\noindent Many vector potentials define the same magnetic flux since adding
the gradient of any differentiable scalar function (a gauge function) to the
vector potential will give the same magnetic field.\cite{Berger:20} Even though
a uniform static magnetic field is independent of the chosen gauge function,
the origin of the magnetic vector potential (gauge origin) must be defined.
However, one has the freedom to choose a convenient gauge origin, $\vect{O}$,
which relates the vector potential $\vect{A}\ofR$ to the magnetic flux
$\vect{B}$ of a constant uniform magnetic field as

\begin{equation}
\vect{A} \ofR = \half \vect{B} \times \vect{r}_{\vect{O}},
\label{eq:gaugeOrigin}
\end{equation}

\noindent where $\vect{r}_{\vect{O}} = \vect{r} - \vect{O}$ is the distance
between the point $\vect{r}$ and the gauge origin. 

\noindent In quantum chemistry, the Coulomb gauge, $\nabla \cdot \vect{A} = 0$,
is usually employed which reduces the number of terms in the Hamiltonian. The
Hamiltonian describing the magnetic interaction consists of three terms, namely
the field-free Hamiltonian $\hat{H}^{(0)}$, a term that depends linearly on the
magnetic field $\hat{H}^{(1)}$, and a term with a quadratic dependence on
$\vect{A}$ denoted by $\hat{H}^{(2)}$, 

\begin{equation}
\hat{H} = \hat{H}^{(0)} + \hat{H}^{(1)}(\vect{A}) + \hat{H}^{(2)}(A^2).
\label{eq:HamMFterms} 
\end{equation}

\noindent
Employing the definition of the magnetic flux in Eq. \ref{eq:gaugeOrigin}, the
following terms are obtained:

\begin{equation}
\hat{H}^{(1)} = \frac{e}{2m_e } \vect{B} \cdot \hat{\vect{L}},
\label{eq:HamMFfirstOrder}
\end{equation}

\begin{equation}
\hat{H}^{(2)} = \frac{e^2}{8m_e } \left( B^2 r^2 - \left(\vect{B} \cdot \vect{r} \right)^2 \right) .
\label{eq:HamMFsecondOrder}
\end{equation}

\noindent The term linear in $\vect{B}$ describes the coupling between the
external magnetic field and the orbital angular momentum of the electron,
whereas the quadratic term can be seen as the coupling between the induced
magnetic moment of the electron and the external field.  Since magnetic fields
on Earth, both in nature and under laboratory conditions, are very weak
compared to Coulomb interactions, the first-order dependence on $\vect{B}$ in
Eq.\ \ref{eq:HamMFfirstOrder} can be studied using perturbation theory. 

\noindent In the Hamiltonian of open-shell systems, a first-order term appears
describing the interaction between the spin of the electron ($\hat{\vect{S}}$)
and the external magnetic field, 

\begin{equation}
    \hat{H}^{(1)}_S = \frac{e}{m_e } \vect{B} \cdot \hat{\vect{S}}.
    \label{eq:HamMFfirstOrderSpin}
\end{equation}

\noindent However, in this chapter we do not account for interactions between
the external magnetic field and the spin of the electron.

\subsection{Current density}

\noindent Current density $\vect{j}(\mathbf{r})$ is a conserved property such
that $\nabla \cdot \vect{j}(\mathbf{r}) = 0$, which is a direct consequence of charge
conservation. In the general case, it was defined by
\citeauthor{Schrodinger:26c} as the continuity equation\cite{Schrodinger:26c}

\begin{equation}
\vect{j}\ofR = -\frac{e}{2m_e} \left (\psi \vect{\hat{p}} \psi^* + \psi^* \vect{\hat{p}} \psi \right).
\label{eq:currentDensity}
\end{equation}

\noindent where $\hat{\mathbf{p}}$ is the momentum operator, $\psi$ is the wave
function and $\psi^*$ is its complex conjugate. Current densities are  a
subobservable that is in principle experimentally accessible and can be
calculated as the expectation value of a quantum-mechanical
operator.\cite{Hirschfelder:78} However, at present there is no experimental
method of directly probing the current density in a single molecule. 

\noindent In the presence of an external magnetic field, the molecular wave
function can be expressed as a sum of the zero-order field-free wave function,
$\Psi_0^{(0)}$, and the first-order magnetically perturbed wave function,
$\Psi_0^{(1)}$,

\begin{equation}
    \Psi_0 = \Psi_0^{(0)} + \Psi_0^{(1)},
    \label{eq:PsiPT}
\end{equation}

\noindent where the perturbed wave function can formally be written as a sum
over the unperturbed wave functions of the excited states $j$,  

\begin{equation}
    \Psi_0^{(1)} = \frac{e}{2m_e \hbar} \sum_{j\neq 0} \ket{\Psi_j^{(0)}} 
                   \frac{ \braOpNoHatKet{\Psi_j^{(0)}}{\hat{\vect{L}} \cdot \vect{B} }{\Psi_0^{(0)}}  }{E_j^{(0)}-E_0^{(0)}},
    \label{eq:PsiPerturbed}
\end{equation}

\noindent where $E_0^{(0)}$ and $E_j^{(0)}$ in the denominator of
Eq.~\ref{eq:PsiPerturbed} are the energies of the ground and the
$j^{\mathrm{th}}$ excited state. The terms in the numerator are the transition
moments of the angular momentum operator between the ground state and the
excited states. The angular momentum exerts the rotational motion
around the gauge origin $\vect{O}$ so that $\hat{\vect{L}} =
\vect{r}_{\vect{O}} \times \hat{\vect{p}}$.

\noindent 
Inserting the wave function with the first-order perturbation correction from
Eq.~\ref{eq:PsiPT} into the expression in Eq. \ref{eq:currentDensity} 
leads to two terms that depend on the chosen gauge origin. The contributions to
the current density that involve the ground-state wave function $\Psi_0^{(0)}$
are called diamagnetic, $\vect{J}_d^B$, whereas the terms representing the sum
over states are called paramagnetic, $\vect{J}_p^B$. This division is entirely
arbitrary. The terms will change depending on the choice of the gauge origin.
For example, the paramagnetic current density can be made to vanish
completely.\cite{Heller:77,Pelloni:09} Only the sum of the diamagnetic and
paramagnetic has a physical meaning. The total current density is expressed as
the sum of the diamagnetic and paramagnetic contributions as

\begin{equation}
\vect{J}^\mathbf{B} = \vect{J}^\mathbf{B}_d + \vect{J}^\mathbf{B}_p \qquad  
\begin{dcases}
    \vect{J}^\vect{B}_d & = -\frac{e^2}{2m_e}\vect{B} \times \vect{r}_{\vect{O}} \left( \Psi_0^{(0)} \right)^2 ,\\
    \vect{J}^\vect{B}_p & = -i\hbar \frac{e^2}{m_e} \sum_{j \neq 0} \left( \Psi_j^{(1)} \nabla \Psi_0^{(0)} + \Psi_0^{(0)} \nabla \Psi_j^{(1)} \right)       . 
\end{dcases}
    \label{eq:currentDensityDiaPara}
\end{equation}

\noindent The diamagnetic term can be interpreted as the Larmor precession of an
electron in a magnetic field as in classical electrodynamics, while the
paramagnetic term does not a have a classical counterpart.  Since the
diamagnetic term depends only on the ground-state wave function, it means that
all atoms and molecules exhibit a diamagnetic response.

\noindent Using the expression for the perturbed wave function in Eq.\
\ref{eq:PsiPerturbed} for the paramagnetic term $\vect{J}^B_p$ and employing
the expressions for the magnetic field and the angular momentum in the chosen
gauge origin in Eq.\  \ref{eq:gaugeOrigin} leads to terms with two sets of
integrals namely, transition moments with rotational symmetry involving the
angular momentum operator, $\braOpKet{\Psi_j^{(0)}}{\vect{L}}{\Psi_0^{(0)}}$,
and the terms proportional to $\braOpKet{\Psi_j^{(0)}}{\vect{p}}{\Psi_0^{(0)}}$
which are, in turn, the transition moments involving the linear momentum
operator.  A thorough derivation of the formalism was recently presented by
Berger, Monaco and Zanasi.\cite{Berger:20}

\subsection{Current-density susceptibility}

The current density $\vect{J}^B\ofR$ induced by a weak magnetic field is a
vector field can be expanded in a Taylor series with respect to the
magnetic-field strength $\vect{B}$, 

\begin{equation}
\vect{J}^\vect{B}\ofR = \vect{j}_0 \ofR + \sum_{\beta \in \{x,y,z\}} \,  \pd{}{\vect{J}^\vect{B} \ofR}{B_\beta} \Biggr|_{\substack{B_\beta = 0}} B_\beta + \mathcal{O} \left( B_\beta ^2 \right ). 
    \label{eq:currentDensityExpansion}
\end{equation}

\noindent The zero-order term $\vect{j}_0 \ofR$ vanishes for closed-shell
molecules.  The first-order terms are the first derivatives of the current
density with respect to the $x$, $y$ and $z$ components of the magnetic field.
These derivatives are the current-density susceptibility of the
molecule,\cite{Lazzeretti:94}

\begin{equation}
\mathcal{J}^{B_\beta}_\alpha\ofR = \pd{}{J_\alpha^\vect{B} \ofR}{B_\beta}\Biggr|_{\substack{B_\beta = 0}} 
 \where \,\, \alpha,\beta \in {\{ x,y,z \}}.
\label{eq:susceptibilityElem}
\end{equation}

\noindent The current-density susceptibility is a rank-2 tensor describing the
current-density flux in the $\alpha$ direction when the magnetic field is
applied in the $\beta$ direction.  Contracting the tensor with a particular
magnetic field direction gives the first-order current-density vector field for
the chosen orientation of the molecule with respect to the magnetic field.

\subsection{Current-density vector field}

The magnetically induced current density $\vect{J}^\mathbf{B}\ofR$ is a unique
fingerprint of the magnetic response of the electronic structure of a molecule.
Since the first-order perturbed wave function $\Psi_0^{(1)}$ can be written as a
sum-over-state expression containing transition moments of operators with a
given symmetry, it is obvious that orbital symmetry plays an
important role for the topology of the current-density vector field.
\cite{Berger:20,Fowler:04,Gomes:83,Keith:93,Lazzeretti:16,Lazzeretti:00,Lazzeretti:18}

\noindent Saddle-stagnation points define closed surfaces (separatices) that
divide the current density into multiple
domains.\cite{Gomes:83,Hirschfelder:77} A closed separatix encases a vortex
with a singular axis, around which the electrons move in a closed loop.
\citeauthor{Gomes:83} showed that there is always a vortex which completely
encloses the entire molecule.\cite{Gomes:83} Inside that global domain, many
other vortices can exist, however, due to charge conservation, the separatrices
may not cross. 

\noindent The tropicity concept, originally introduced in fluid mechanics, can
be employed to analyse the direction of the rotational motion in the
current-density vortices with respect to the magnetic field.\cite{Ricca:01} In
our convention, a vortex is labelled \emph{diatropic} when the direction of the
current density is clockwise when looking towards the negative direction of the
magnetic field vector, which is in the classical direction according to Lenz's
law.\cite{Pelloni:09,Sondheimer:72}  When the electrons follow a pathway in the
opposite direction, the current-density flux is termed \emph{paratropic}. 
The tropicity is a property of the total current density carrying a physical meaning. 
It is not related to the similarly sounding diamagnetic and paramagnetic terms, 
which arise as a mathematical formality and depend on the gauge origin.

\noindent Diatropic current densities induce a secondary magnetic field which
opposes the external magnetic field, while paratropic current densities give
rise to a secondary magnetic field that strengthens the external field.  In
spite of the similarly of the names, diatropic and paratropic current densities
cannot be mapped to the gauge-origin dependent diamagnetic and paramagnetic
contributions to the current density. The diatropic current density generally
originates from the diamagnetic current density expressed in
Eq.~\ref{eq:currentDensityDiaPara}, whereas the sum-over-states expression in
Eq.~\ref{eq:PsiPerturbed} contains both linear- and angular-momentum transition
moments which are responsible for diatropic and paratropic current densities,
respectively.\cite{Berger:20,Fowler:04} 

\noindent Furthermore, diamagnetism and paramagnetism denote the response of a
material to an external magnetic field, which has no physical connection to the
names of the terms derived in the perturbation-theory
formalism.\cite{Pelloni:09} Materials which expel the magnetic field are termed
diamagnetic, whereas paramagnetic materials strengthen the field. Historically,
molecules with a diamagnetic response to a magnetic field were called diatropic
molecules, while molecules with paramagnetic response were termed
paratropic.\cite{Sondheimer:72}  However, nowadays these labels are only used
for current-density pathways, since there are usually many diatropic and
paratropic vortices in the same molecule.

\subsection{Nuclear magnetic shielding}

Atomic nuclei with non-zero spin give rise to a magnetic vector potential
$\vect{A}_{\mathrm{nuc}}$,

\begin{equation}
    \vect{A}_{{\mathrm{nuc}}} = \frac{\mu_0}{4 \pi } ~ \frac{\vect{m}_{{\mathrm{nuc}}} \times \vect{r}}{r^3} .
    \label{eq:nucMF}
\end{equation}

\noindent In the presence of an external magnetic field, the total magnetic
vector potential experienced by electron $i$ becomes 

\begin{equation}
\label{eq:vectorpotential}
    \vect{A}_i = \half \vect{B} \times \vect{r}_{i \vect{O}} + \frac{\mu_0}{4 \pi} \sum_K \frac{\vect{m}_K \times \vect{r}_{iK}} {r^3_{iK}} ,
\end{equation}

\noindent where $\vect{r}_{i \vect{O}}$ is the distance between point
$\vect{r}_i$ and the gauge origin, 
the index $K$ spans all nuclei, $\vect{m}_K$ is the magnetic moment of nucleus
$K$, and $\vect{r}_{iK}$ is the distance between electron $i$ and nucleus $K$.  Adding
the total magnetic vector potential in Eq.\ \ref{eq:vectorpotential} to the
molecular Hamiltonian leads to an expression for the coupling between the
nuclear magnetic moment and the magnetically induced current
density,\cite{Lazzeretti:94,Lazzeretti:00,Lazzeretti:18,Jameson:79,Jameson:80}

\begin{equation}
    E^{\vect{m}_K B} = - \int \vect{A}^K_{{\mathrm{nuc}}}\ofR \cdot \vect{J}^\vect{B}\ofR \, \mathrm{d}\vect{r}.
\label{eq:Enuc}
\end{equation}

\noindent
The magnetically induced current density gives rise to a secondary magnetic
field according to Biot-Savart's law,

\begin{equation}
\vect{B}_{\mathrm{loc}} = \frac{\mu_0}{4\pi} \int \frac{\vect{r} \times \vect{J}^\vect{B}\ofR }{r^3} \mathrm{d}\vect{r}.
\label{eq:BLocal}
\end{equation}

\noindent
Hence, at every point in space, the magnetic field is the sum  $\vect{B} =
\vect{B}_\mathrm{ext} + \vect{B}_{\mathrm{loc}}$, which can also be expressed as

\begin{equation}
\vect{B}_{\mathrm{loc}} = \left( 1 - \sigma \right) \vect{B} ,
\label{eq:localB}
\end{equation}

\noindent where $\sigma$ is the trace of the nuclear magnetic shielding tensor.
Nuclear magnetic shieldings can be probed in nuclear magnetic resonance (NMR)
spectroscopy, where the magnetic shielding constants of the nuclei are compared
to a reference. The nuclear magnetic shieldings $\sigma$ and the NMR chemical
shifts $\delta$ relative to the reference are small and reported in parts per
million (ppm).

\subsection{Nuclear magnetic shieldings and current densities}

\noindent Nuclear magnetic shieldings can be calculated as the
second derivative of the electronic energy with respect to the nuclear magnetic
moment $\vect{m}_K$ and the strength of the external magnetic field
$\vect{B}$,\cite{Wolinski:90}

\begin{equation}
\sigma_{\alpha \beta}^K = 
    \frac{\partial^2 E^{\vect{m}_K \vect{B}} }{\partial m_\alpha^K \partial B_\beta}
    \Biggr|_{\substack{m_\alpha^K, B_\beta = 0}} 
    \where \quad \alpha,\beta \in {\{ x,y,z \}}.
\label{eq:shieldingConstHellmann}
\end{equation}

\begin{figure}[H]
    \subfigure[$^{13}$C NMR shielding density]{
    \includegraphics[width=0.45\linewidth]{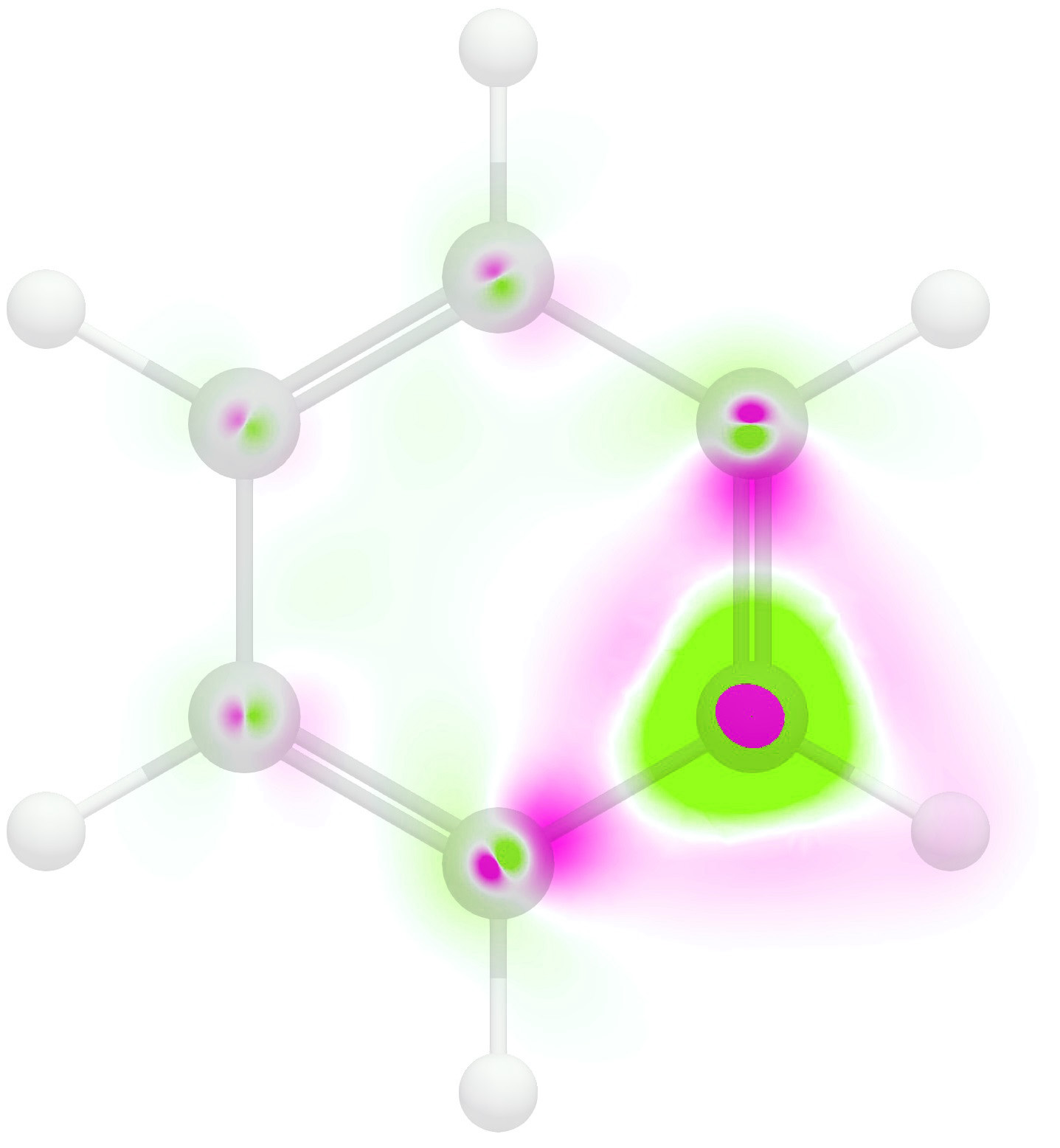}
    \label{fig:sigma-C}}
\hspace{10mm}
    \subfigure[$^{1}$H NMR shielding density]{
    \includegraphics[width=0.45\linewidth]{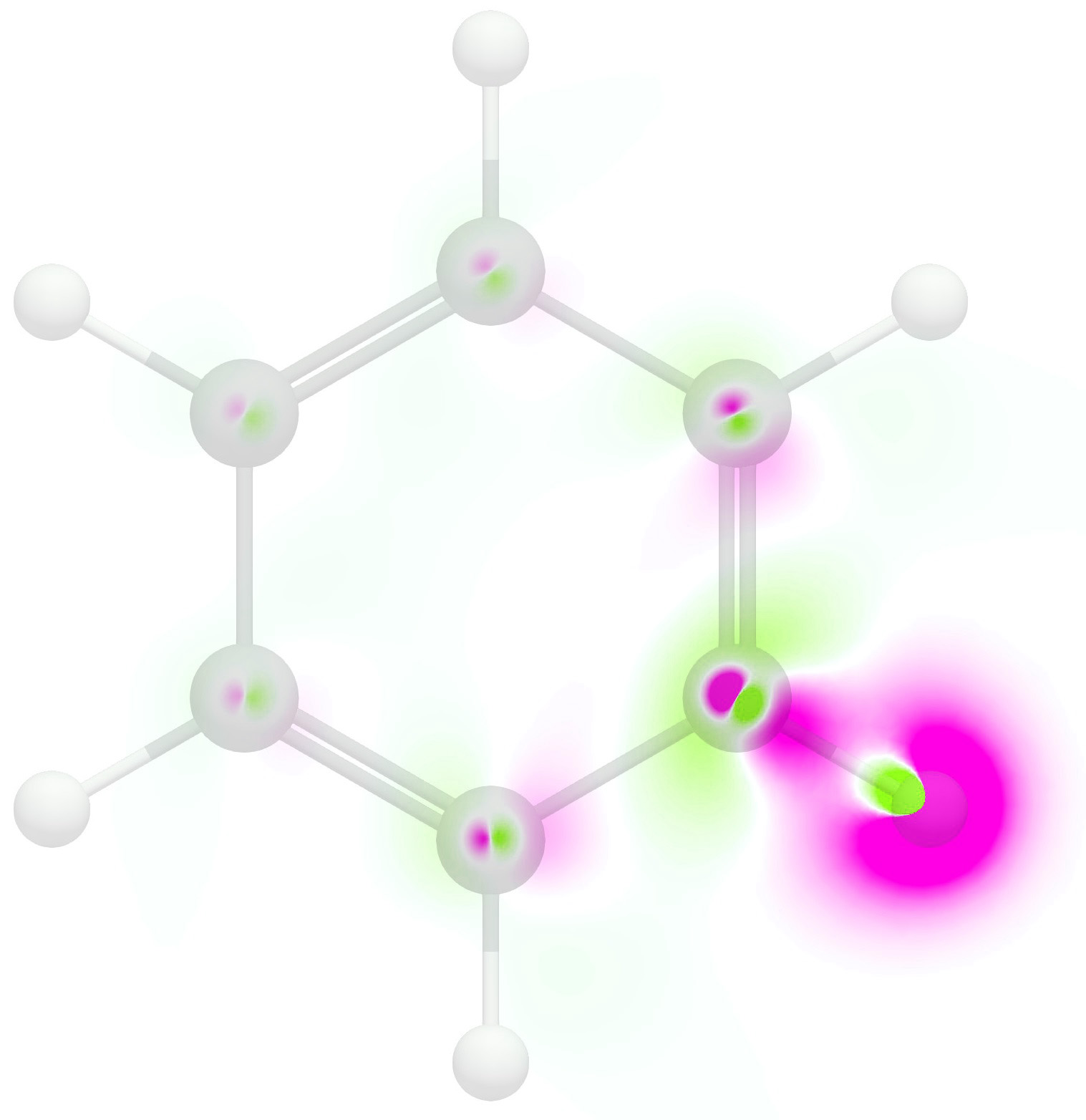}
    \label{fig:sigma-H}}

\caption{The spatial contributions to the $^{13}$C NMR shielding density (left)
and the $^1$H NMR shielding density (right) in the molecular plane of benzene.
Shielding and deshielding contributions are shown in pink and green,
respectively. \label{fig:sigma}}

\label{fig:shieldingdensityofbenzene}
\end{figure}

\noindent Jameson and Buckingham showed that NMR shieldings can
alternatively be obtained by differentiating the Biot-Savart expression in Eq.\
\ref{eq:Enuc} with respect to the magnetic field and the nuclear magnetic
moment.\cite{Lazzeretti:94,Lazzeretti:00,Lazzeretti:18,Jameson:79,Jameson:80,McWeeny:86}

\begin{equation}
\sigma_{\alpha \beta}^K = -\frac{\mu_0}{4 \pi} \sum_{\gamma \delta} \varepsilon_{\alpha \delta \gamma} \int \frac{r_\delta - R_{K\delta}}{\modulus{\vect{r} - \vect{R}_K} ^3} \, \mathcal{J}^{B_\beta}_\gamma\ofR~\mathrm{d}\vect{r},
\label{eq:shieldingConstBiotSavart}
\end{equation}

\noindent where $\alpha, \beta, \gamma, \delta$ are Cartesian directions, 
$\varepsilon_{\alpha \delta \gamma}$ is the Levi-Civita
symbol, $\vect{R}_K$ the distance from nucleus $K$. 
The advantage of the latter approach is that
NMR shieldings can be analysed by studying the spatial contributions to them.
Orbital contributions to the NMR shieldings can also be unambiguously
calculated by determining orbital contributions to the current density and
integrating their contributions to the Biot-Savart expression of the NMR
shieldings.\cite{Steiner:04,Acke:18}  The integrand, which is called the
magnetic shielding density,\cite{Jameson:80} can be visualised, yielding a
rigorous physical basis for interpreting NMR shieldings.  The magnetic
shielding density of a carbon atom in benzene is shown in
Fig.~\ref{fig:sigma}(a) where one sees that the $1s$ core electrons of the
carbon shield the magnetic field at the nucleus. The nucleus is surrounded by a
more distant deshielding region which in turn is embedded in a diffuse
shielding region due to contribution from the electrons near the other atoms as
well as from the electrons of the $\sigma$ bonds and the conjugated $\pi$
sextet.  The shielding density of the hydrogen atom in benzene is illustrated
in Fig.~\ref{fig:sigma}(b). The inner part of the $1s$ orbital of the hydrogen
is deshielding the nucleus, whereas shielding regions are found outside the
hydrogen and along the \ce{C-H} bond. The alternating shielding and deshielding
contributions from the other carbon atoms are due to their atomic current
densities.

\noindent The contributions to the shielding density from each orbital can be
obtained by partitioning the magnetic shielding tensor into localised
orbitals.  However, it leads to mixing of occupied orbitals,\cite{Bohmann:97}
which results in unphysical orbital contributions when the orbital mixing is
not prevented by symmetry.  For planar molecules one can divide
magnetic shieldings into $\sigma$ and $\pi$ contributions, since unitary
transformations do not mix $\sigma$ and $\pi$ orbitals.\cite{Acke:18,Orozco:20}

\subsection{Magnetic susceptibilities and closed-shell paramagnetic molecules}

The magnetic susceptibility $\chi_{\alpha \beta}$ with $\alpha,\beta=\{x,y,z\}$,
also known as magnetisability, is the second derivative of the electronic
energy with respect to the external magnetic field,\cite{Ruud:93} 

\begin{equation} \chi_{\alpha \beta}  = -\frac{\partial^2 E^{\vect{B}
\vect{B}}} {\partial B_\alpha \partial B_\beta}\Biggr|_{\substack{B_\alpha=0\\
B_\beta=0}} ~~\mathrm{with}~~E^{\vect{B} \vect{B}} = - \int
\vect{A}^{\vect{B}}\ofR \cdot \vect{J}^\vect{B}\ofR \, \mathrm{d}\vect{r},
\end{equation}

\noindent
where $\vect{A}^{\vect{B}}\ofR$ is the vector potential of the external magnetic field. 
Magnetisability is a second-rank tensor that similarly to the nuclear
magnetic shielding, $\sigma$, can formally be divided into diamagnetic and
paramagnetic contributions,\cite{Corcoran:80} 

\begin{equation}
\label{eq:magnetisability}
    \chi_{\alpha \beta} = \chi^\mathrm{d}_{\alpha \beta} + \chi^\mathrm{p}_{\alpha \beta} = \quad 
    \begin{dcases}
	-\frac{1}{4} \sum_i  \braOpNoHatKet{ \Psi_0^{(0)} }
		{ \vect{r}_{\vect{O}i}^2\delta_{\alpha \beta} - \vect{r}_{\vect{O}i,\alpha} \vect{r}_{\vect{O}i,\beta} }
		{ \Psi_0^{(0)} } \\
	+ \half \sum_{\substack{i\\ j \neq 0}} 
	\frac{\braOpNoHatKet{ \Psi_0^{(0)} }{ (\hat{\vect{r}}_{\vect{O}i} \times \hat{\vect{p}}_i)_\alpha  }{ \Psi_j^{(0)} }   
	    \braOpNoHatKet{ \Psi_j^{(0)} }{ (\hat{\vect{r}}_{\vect{O}i} \times \hat{\vect{p}}_i)_\beta  }{ \Psi_0^{(0)} }   
	    } 
	{E_j^{(0)} - E_0^{(0)} }. 
    \end{dcases}
\end{equation}

\noindent where $\vect{r}_{\vect{O}i}$ is the distance vector from a common
gauge origin, $\hat{\vect{p}}_i$ is the momentum operator of electron $i$, and
the energies $E_0^{(0)}$ and $E_j^{(0)}$ are the unperturbed energies of the
ground state and the j$^{\mathrm{th}}$ excited state, respectively. 

\noindent The diamagnetic term depends only on the ground-state wave function
and it is always negative, whereas the paramagnetic contribution that is
expressed as a sum-over-states expression in Eq.\ \ref{eq:magnetisability} may be
positive or negative. Most molecules are diamagnetic with a negative
magnetisability, because the diamagnetic term dominates. Open-shell molecules
might be paramagnetic due to contributions from the spin of the electrons,
whereas a few closed-shell molecules are paramagnetic due to a very large
positive contribution from the paramagnetic term in Eq.\
\ref{eq:magnetisability}.\cite{Stevens:65,Pelloni:09b} Magnetisability can also
be expressed using the current-density susceptibility, ${\cal
J}_\gamma^{B_\beta}(\vect{r})$, 

\begin{equation}
\label{eq:Biot-Savart-magnetisability}
\chi_{\alpha \beta} = \frac{1}{2} \sum_{\delta \gamma}  \varepsilon_{\alpha \delta \gamma} \int r_\delta {\cal J}_\gamma^{B_\beta} \ofR  \dr, 
\end{equation}

\noindent where $\varepsilon_{\alpha \delta \gamma}$ is the Levi-Civita symbol.\cite{Lazzeretti:00} 
Contributions from the diatropic current density are
negative, whereas paratropic current-density vortices have positive
contributions to the magnetisability. The size of the positive contribution
depends on the strength of the paratropic ring current and its radius.
Antiaromatic porphyrins can be closed-shell paramagnetic molecules provided
that the strength of the paratropic ring current be strong
enough.\cite{Valiev:20} 

\noindent Experimental magnetisabilities are the trace of the magnetisability
tensor. For ring-shaped molecules, the external magnetic field induces a ring
current in the molecular ring when the magnetic field has a vector component
perpendicular to the ring. Planar aromatic and antiaromatic molecules i
are diamagnetic for the two other Cartesian components of the magnetic field.
Thus, the paratropic ring current of antiaromatic molecules must be very strong
in order to yield a paramagnetic contribution that cancels the diamagnetic ones
arising from the other two magnetic field directions. Calculations on
isophlorin and other antiaromatic porphyrinoids showed that the studied
molecules become closed-shell paramagnetic when the strength of the paratropic
ring current exceeds \nAT{-20}.\cite{Valiev:17} Calculations on expanded
porphyrins such as expanded isophlorins with eight furan moieties connected by
methine bridges and the corresponding structure with \ce{NH} instead of \ce{O}
yielded strong paratropic ring currents and a large positive magnetisability
due to the large radius of the paratropic ring current.\cite{Valiev:20} A large
positive magnetisability has been measured for a closed-shell tetracationic
hexaporphyrin nanowheel, which sustains a strong paratropic ring current of
\nAT{-65} with a radius of 13\AA.\cite{Valiev:20,Peeks:17} Construction of
paramagnetic porphyrinoid molecules with more than one paratropic ring-current
pathway requires that each of the vortices sustain a paratropic ring current
with a strength exceeding \nAT{-20},\cite{Valiev:20} since the diamagnetic
contribution to the magnetisability scales about linearly with the size of the
molecule according to Pascal's rule.\cite{Pascal:10}

\subsection{Treatment of the gauge origin in quantum chemistry}

An important conclusion drawn by \citeauthor{Keith:93} is that the gauge origin
need not be the same for all positions in space.\cite{Keith:93} Gauge
independence can be achieved, for example, by using different gauge origins for
each point in space, as in the continuous translation of the origin of the
current density (CTOCD) methods
\cite{Lazzeretti:00,Coriani:94,Zanasi:95,Lazzeretti:95,Zanasi:96,Ligabue:99,Lazzeretti:12,Lazzeretti:12e,Steiner:02,Soncini:08b}
or by employing gauge-including atomic orbitals (GIAOs), also known as London
orbitals.\cite{Wolinski:90,London:37,Hameka:58,Ditchfield:74,Helgaker:91} GIAOs
are physically meaningful since they involve a complex pre-factor in analogy
with the one obtained by gauge transformations of wave functions.  An
ordinary Gaussian-type basis function $\chi_K\ofR$ centred on nucleus $K$ is
re-defined as $\omega_K\ofR$ with an explicit dependence on the magnetic vector
potential of the external magnetic field 
$\vect{A}_K^\vect{B}\ofR = \half \vect{B} \times ( \vect{R}_{K} - \vect{O} )$ 
with its gauge origin at the nucleus as

\begin{equation}
\omega_K ( \vect{r}, \vect{A}_K^\vect{B} ) = \mathrm{exp} (-i \vect{r} \cdot \vect{A}_K^\vect{B}\ofR) \chi_K\ofR .
\label{eq:GIAO}	
\end{equation}

\noindent The wave function of a quantum system exposed to an external magnetic
field is not gauge invariant because gauge transformations introduce a complex
pre-factor to the wave function. The electronic interaction integrals as well
as the electron density and the current density are gauge-origin
independent,\cite{Helgaker:91} which implies that molecular properties
calculated as an expectation value of the electron density or obtained by
integrating Biot-Savart expressions involving the current density are also
independent of the chosen gauge function. The magnetically induced current
density is gauge-origin independent when using GIAOs.  Even though the current
density has no reference to the gauge origin, the use of finite basis sets
breaks the complete gauge invariance and leads to a leakage of the current
density, \ie the charge conservation of the current density is not entirely
fulfilled.\cite{Pedersen:99}

\subsection{The gauge-including magnetically induced current method} 

The expression for the nuclear magnetic shielding constant in Eq.\
\ref{eq:shieldingConstBiotSavart} can be combined with the expression obtained
by using analytical derivative theory. Rearrangement of the terms leads to an
alternative expression for the current-density
susceptibility.\cite{Juselius:04,Taubert:11b,Fliegl:11b,Sundholm:16a} Using the
density matrix in the atomic-orbital (basis-function) basis, $D_{\mu\nu}$, and
the magnetically perturbed density matrices, $\pd{}{D_{\mu \nu}}{B_\beta}$,
yields the expression for calculating the current-density susceptibility which
lies is the core of the gauge-including magnetically induced current
method (GIMIC),

\begin{equation}
\begin{split}
\mathcal{J}_\alpha^{B_\beta} \ofR & = 
 \sum_{\mu \nu} \,
    D_{\mu \nu} \pd{}{\omega_\mu^* \ofR}{B_\beta} \pd{}{\tilde{h}\ofR}{{m}_\alpha^K} \, \omega_\nu \ofR + 
 \sum_{\mu \nu} \,
    D_{\mu \nu} \omega_\mu^* \ofR \pd{}{\tilde{h}\ofR}{{m}_\alpha^K} \pd{}{\omega_\nu \ofR}{B_\beta} \\
& + 
 \sum_{\mu \nu} \, 
    \pd{}{D_{\mu \nu}}{B_\beta} \omega_\mu^* \ofR \pd{}{\tilde{h}\ofR}{{m}_\alpha^K} \, \omega_\nu \ofR 
 - \sum_\delta \varepsilon_{\alpha \beta \delta } \sum_{\mu \nu} \,
    D_{\mu \nu} \omega_\mu^* \ofR  \frac{\partial^2 \tilde{h}\ofR}{\partial {m}_\alpha^K \partial B_\delta} \omega_\nu \ofR ,
\end{split}
\label{eq:GIMIC}
\end{equation}

\noindent where $\omega\ofR$ are the GIAOs as defined in Eq.~\ref{eq:GIAO} and
$\varepsilon_{\alpha \beta \delta}$ is the Levi-Civita pseudotensor. 
The magnetic
interaction is represented by the operators 

\begin{equation}
\frac{\partial \tilde h\ofR}{\partial \vect{m}^K} =
    (\vect{r}-\vect{R}_K) \times \hat{\vect{p}} \qquad \mathrm{and}
\label{h_m}
\end{equation}

\begin{equation}
\frac{\partial^2\tilde h\ofR} {\partial \vect{m}^K \partial \vect{B}}
    = \half
    \Big[  (\vect{r}-\vect{O}) \cdot (\vect{r} -\vect{R}_K)\vect{1}
    	- (\vect{r} - \vect{O}) (\vect{r}-\vect{R}_K) \Big]
\label{h-mb}
\end{equation}
 
\noindent where $\vect{O}$ is the gauge origin and $\vect{R}_K$ are the nuclear
coordinates. The singular expression $\modulus{\vect{r}-\vect{R}_K}^{-3}$
appearing in all terms cancels and has been omitted in Eqs~\ref{h_m} and 
\ref{h-mb}. Even though the expression in
Eq.~ \ref{eq:GIMIC} seems to depend on both the gauge origin and the
coordinates of the nuclear positions, they cancel and the dependences vanish. 

\noindent For open-shell systems, it is possible to write two separate
equations like Eq.\ \ref{eq:GIMIC} for calculating the orbital contributions to
the current-density susceptibility from the spin-up ($\alpha$) or spin-down
($\beta$) electrons.\cite{Taubert:11} The expression in Eq.~\ref{eq:GIMIC} is
independent of the employed computational level because the many-body
information from the electronic-structure calculation is compressed into the
density matrix and the magnetically perturbed density matrices. Keith has
derived an alternative expression for calculating current densities at the
Hartree-Fock level using GIAOs.\cite{Keith:96} 

\noindent The GIMIC method has been implemented and is available for download
from GitHub.\cite{gimic} It is interfaced to commonly used quantum-chemistry
program packages including Turbomole,\cite{Furche:14}
Gaussian,\cite{g16-short} Dalton\cite{dalton2016} and Cfour\cite{cfour-short}.
GIAOs improve the basis-set convergence, which avoids the need to use very
large basis sets. Leakage of the current density and charge conservation
problems are very small even when using rather small basis sets of
\mbox{double-$\zeta$} polarisation quality, rendering current-density
studies on very large molecules consisting of more than 2000
atoms feasible.\cite{Reiter:19}

\subsection{Investigating current-density pathways in molecules}

\begin{figure}[H]
    \centering
    \includegraphics[width=0.75\linewidth]{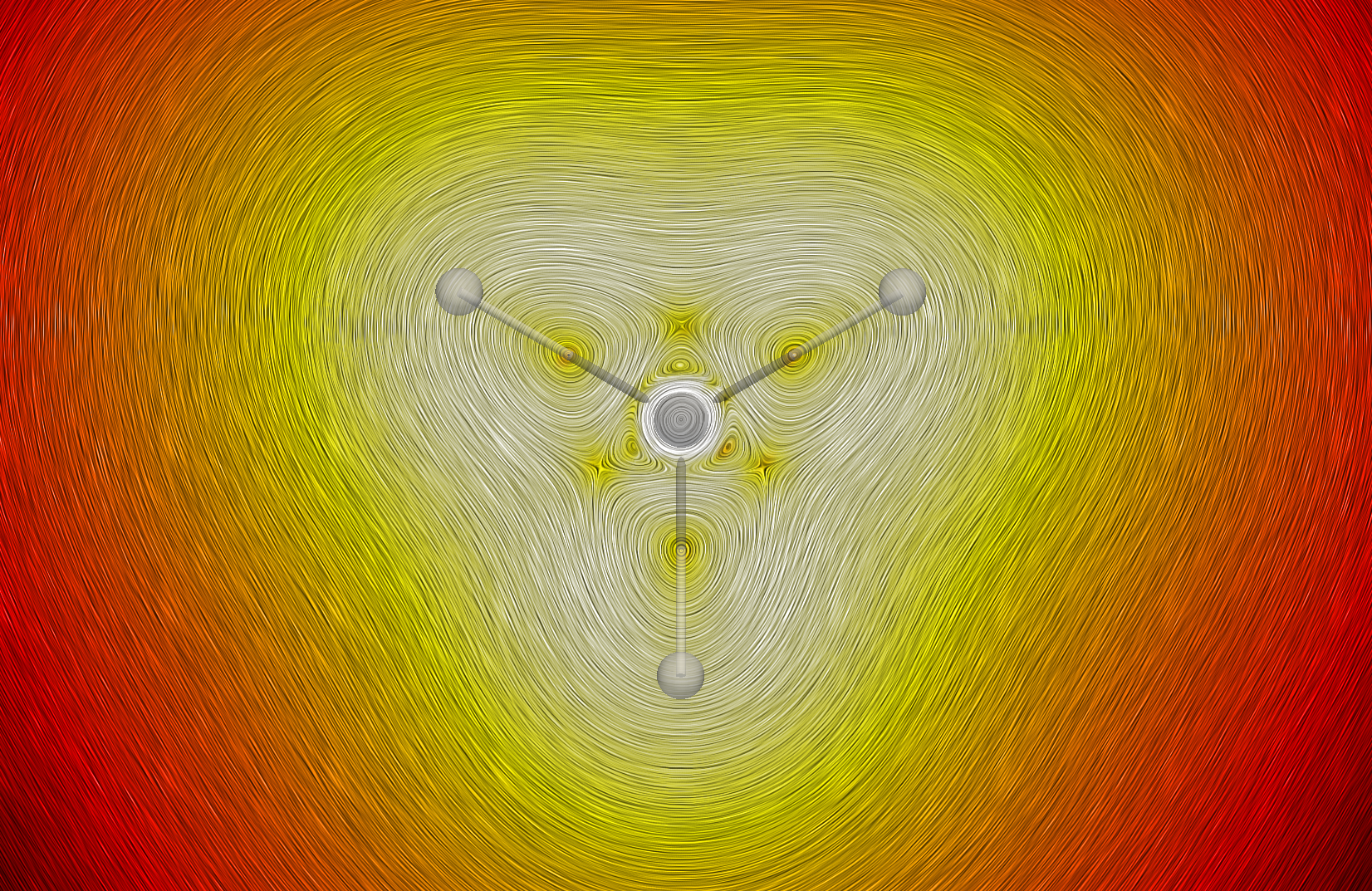}

\caption{The current-density susceptibility of ethane in a plane perpendicular to the \ce{C-C}
bond that crosses the nucleus of a carbon atom. The colour scale is logarithmic
and it gives the strength of the modulus of the current-density susceptibility in the range of
    $[3.57 \times 10^{-6}; 0.71 ]$ \nATA{}.
\label{fig:ethane-front}}

\end{figure}

Magnetic fields induce current densities in all atoms and molecules. For
example, when the magnetic field is pointing along the \ce{C-C} bond in ethane,
atomic and bond vortices can be identified, as well as the global diatropic
current-density flux along the perimeter of the molecule, as shown in Fig.\
\ref{fig:ethane-front}.  Such global diatropic current-density pathways exist
in all molecules.\cite{Gomes:83}

\noindent Cyclic molecules are of interest in studies of magnetic properties,
since aromatic and antiaromatic species can be potentially useful in solar
cells, opto-electronics and as building blocks in conductive materials. The
degree of aromaticity can be estimated from the strength of  the ring-current
(susceptibility) according to the ring-current
model.\cite{London:37,Pauling:36,Lonsdale:37,Pople:58a,McWeeny:58} The aromatic
ring-current model is nowadays an accepted concept even though illustrations of
ring currents shown in textbooks are often oversimplified.\cite{Fliegl:09}

\begin{figure}[H]
\centering
    \subfigure[An inspection plane passing the \ce{C-C} bond.]{
	\includegraphics[width=0.48\linewidth]{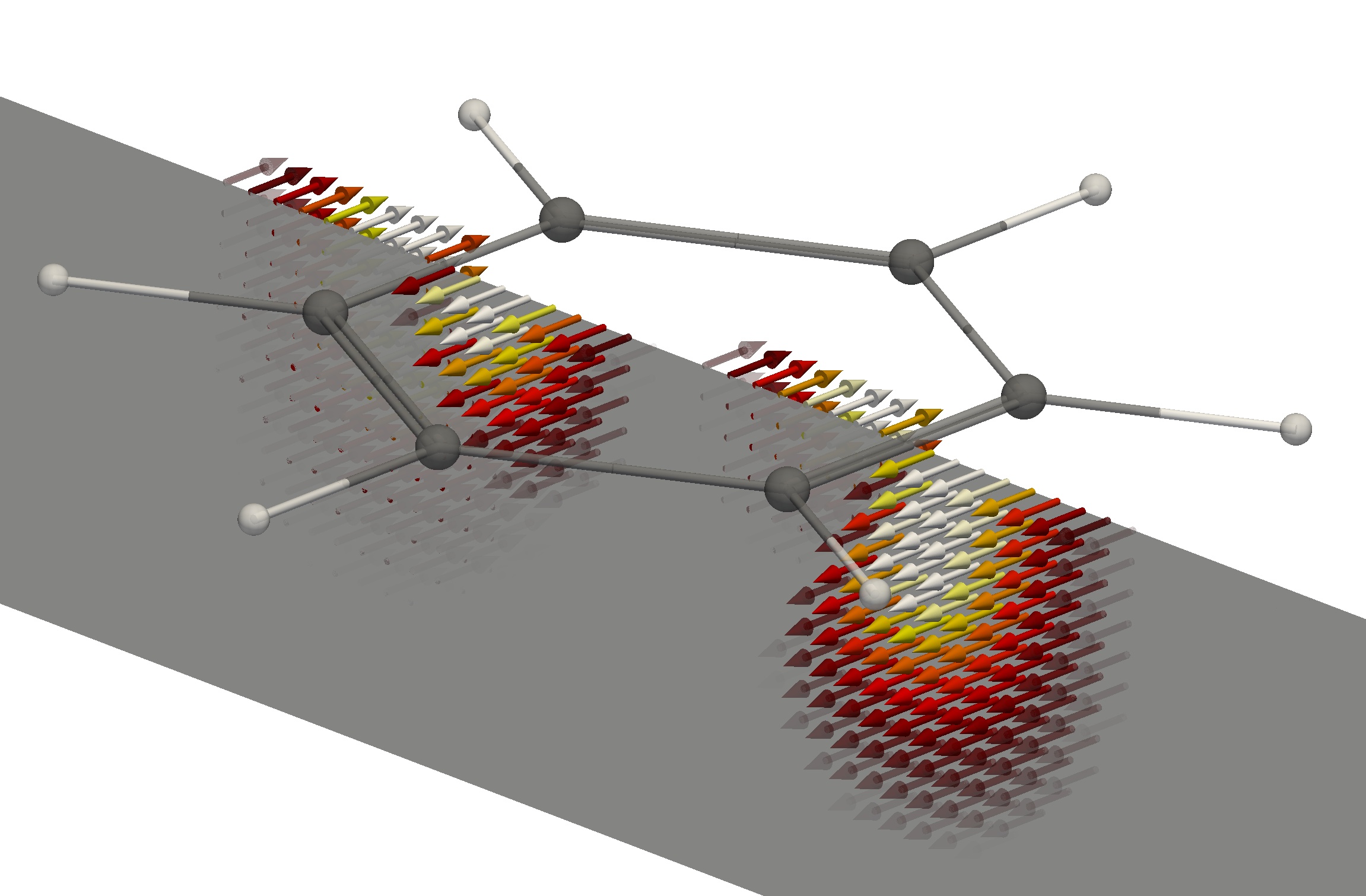} \label{fig:benzene-intplane-bond}}
    \vspace{3mm}
    \subfigure[The plane passes through a carbon atom.]{
	\includegraphics[width=0.48\linewidth]{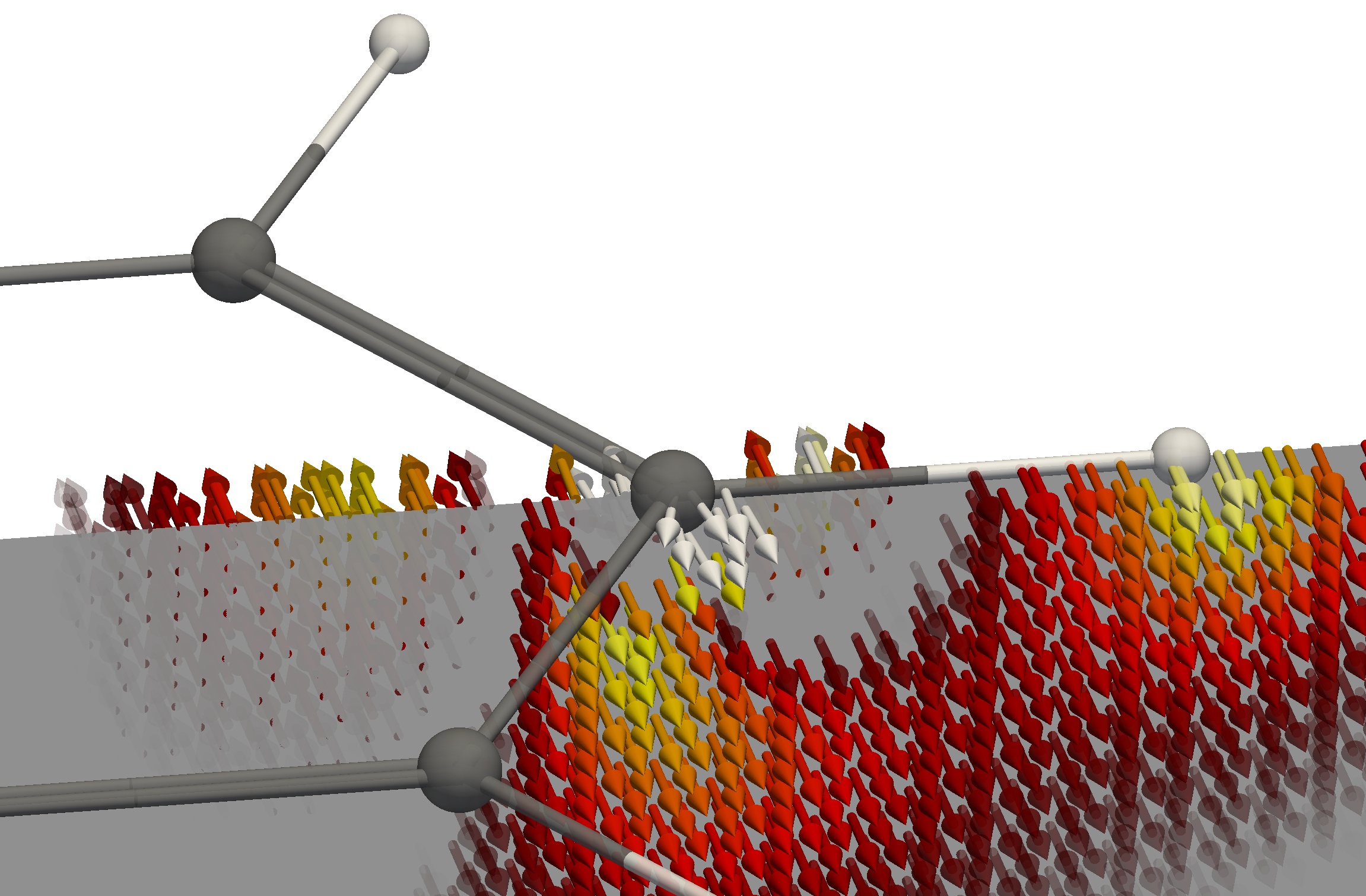} \label{fig:benzene-intplane-atom}}

\caption{Cross-sections of the magnetically induced current-density field in
benzene. The molecule lies in the $(x,y)$ plane and the magnetic field points
in the $z$ direction. Arrows show the direction of the magnetically
induced current density. The colour scale indicates the strength of the current
density. \label{fig:benzene-intplanes-vectors}}

\end{figure}

\begin{figure}[H]
    \centering
    \includegraphics[width=0.45\linewidth,angle=90]{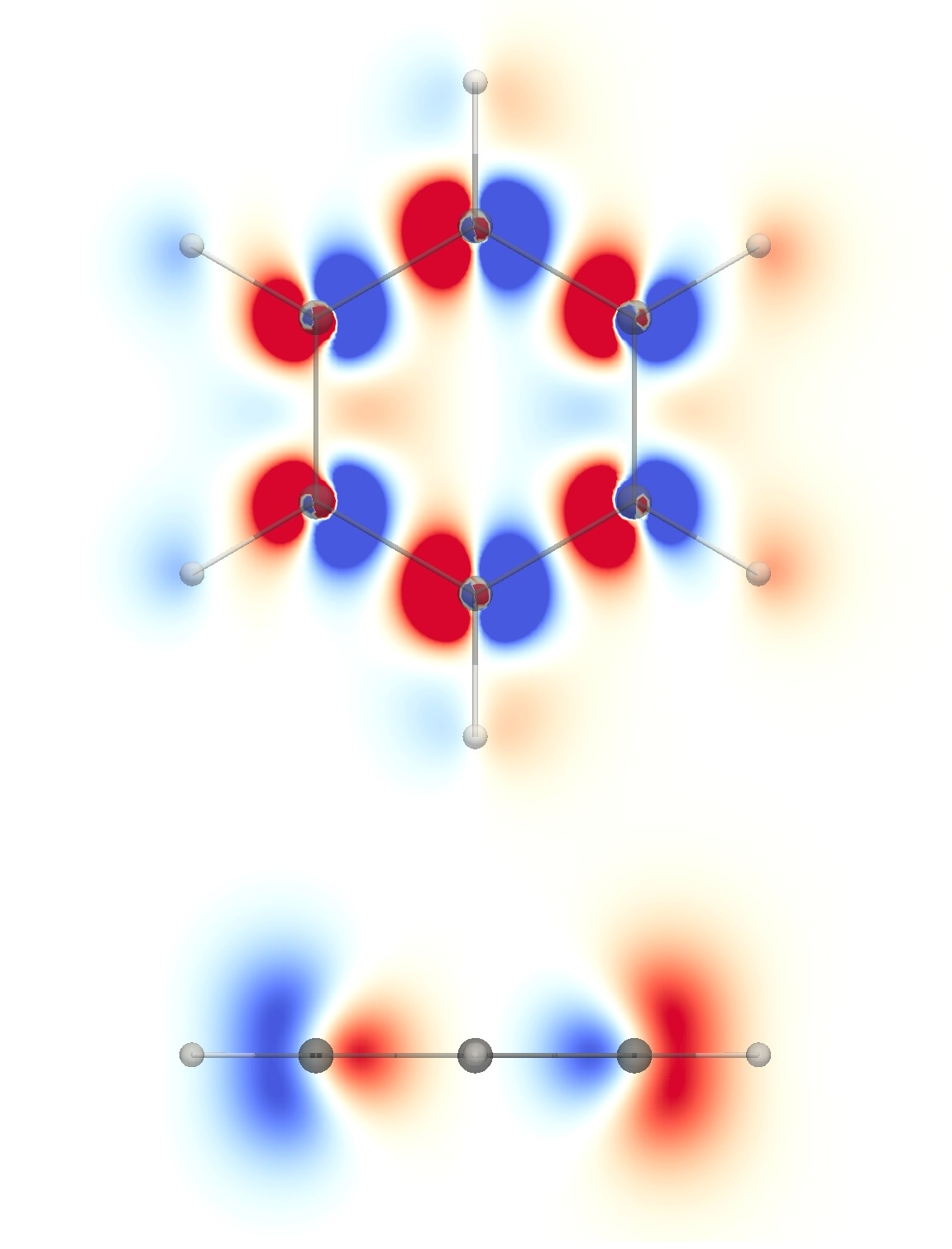}

\caption{The current density in the benzene molecule.
    The magnetic field is parallel to the
    molecular plane of the benzene molecule to the left and perpendicular to
    the molecular plane of the benzene molecule on the right. The
    strength the current density is in the range of
    $[3.57 \times 10^{-6}; 0.43 ]$ \nATA{} where blue areas show the
    current-density flux towards the reader, and the current-density flux away
    from the reader is shown in red. \label{fig:benzene-dimer}  }

\end{figure}

\noindent The current-density susceptibility of benzene in a cross-section
perpendicular to the molecular plane is illustrated in Fig.\
\ref{fig:benzene-intplanes-vectors}. Arrows show the direction of the
current-density flux. The colour scheme reflects the strength of the
current-density susceptibility.  Note that we will use the term \textit{current density} 
in the discussion below when referring to the \textit{current-density susceptibility}.
For weak magnetic fields, the current density is equal to the current-density
susceptibility multiplied by the magnetic field strength. In the exterior of
the molecule, there is a clockwise global diatropic current-density pathway,
while  inside the benzene ring, there is a paratropic ring current following
the counter-clockwise direction. The cross-section of the ring current of
benzene induced by a magnetic field perpendicular to the benzene ring is shown
in the right part of the picture in Fig.\ \ref{fig:benzene-dimer} as well as in
Fig.~\ref{fig:benzene-intplanes-vectors}.  The cross-section of the diatropic
ring current is crescent-shaped on the outside of the carbon atoms and the
paratropic ring current is seen inside the ring. The ring current is strong
also in the molecular plane. 

\noindent The ring current gives rise to the diffuse shielding and deshielding
regions shown in Fig.~\ref{fig:shieldingdensityofbenzene}. The tropicity does
not determine whether the current density shields or deshields the nucleus.
Rather, this depends on the relative direction of the current density with
respect to the nucleus under consideration. Similarly, the atomic current-density vortices
of the other carbon atoms cause both shielding and deshielding based on the
relative direction of the current density. The relative circulation direction
of the global paratropic ring current inside the ring with respect to the
carbon nucleus is the same as the one for the diatropic ring current, leading
to a shielding contribution even though the inner contribution to the ring
current is paratropic. 

\noindent The pathways of the current-density flux depend on the direction of
the magnetic field due to the presence of the angular momentum operator in the
expression for the current density in Eq.~\ref{eq:currentDensityDiaPara}.
Thus, due to the molecular symmetry, there is no paratropic ring current in the
benzene ring when the magnetic field is parallel to the molecular ring as
illustrated for benzene in Fig.~\ref{fig:benzene-dimer}.
Atomic current-density vortices with different spatial distribution and
strength exist regardless of the direction of the magnetic field. The same also
holds for $\sigma$ bonds due to their local cylindrical symmetry along the bond
axis.  

\begin{figure}[H]
    \centering
    \subfigure[The magnetic field is perpendicular to the ring ($B_\perp$)]{\includegraphics[width=0.75\linewidth]{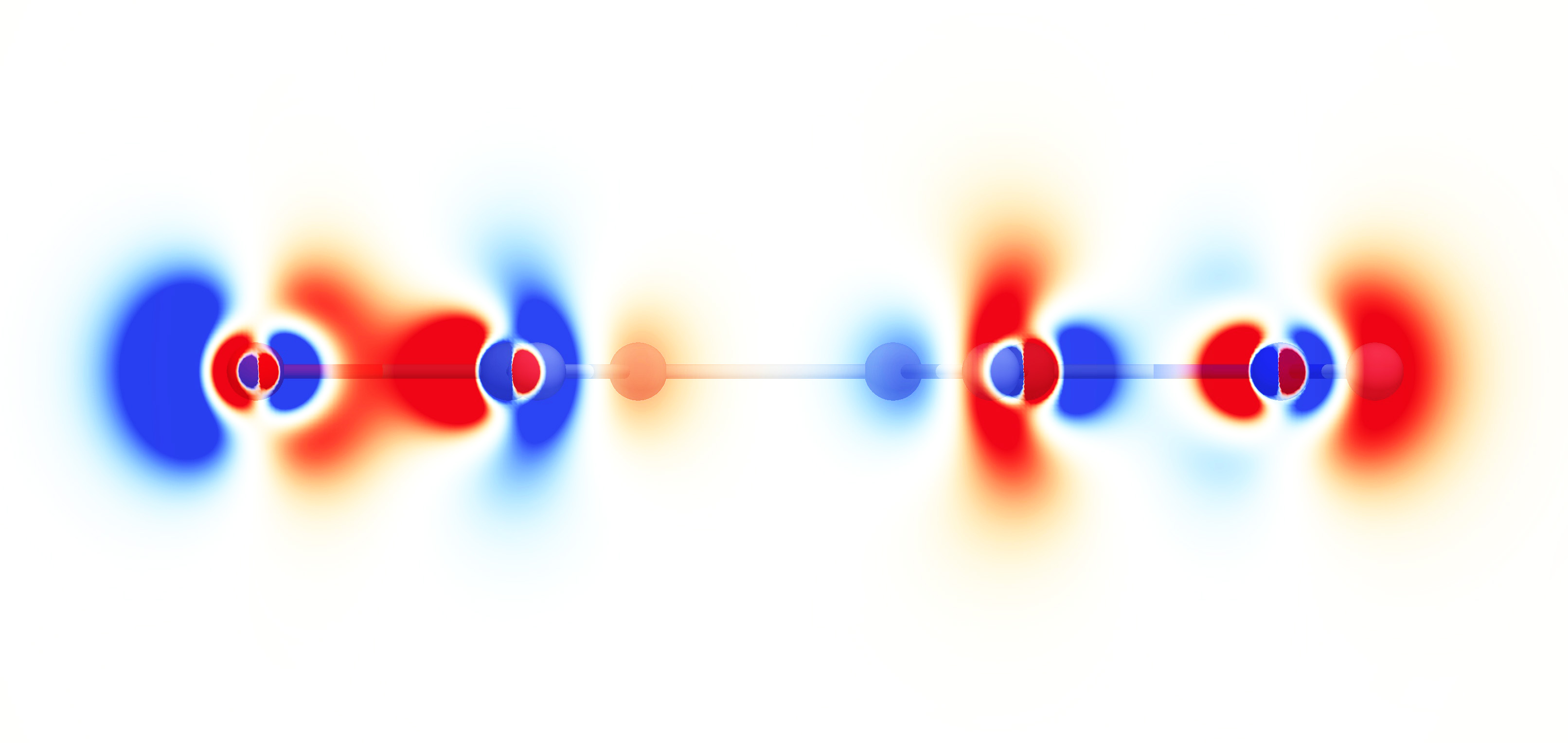}\label{p-aminophenol-B_perp}}
    \subfigure[The magnetic field is parallel to the ring ($B_\parallel$)]{\includegraphics[width=0.75\linewidth]{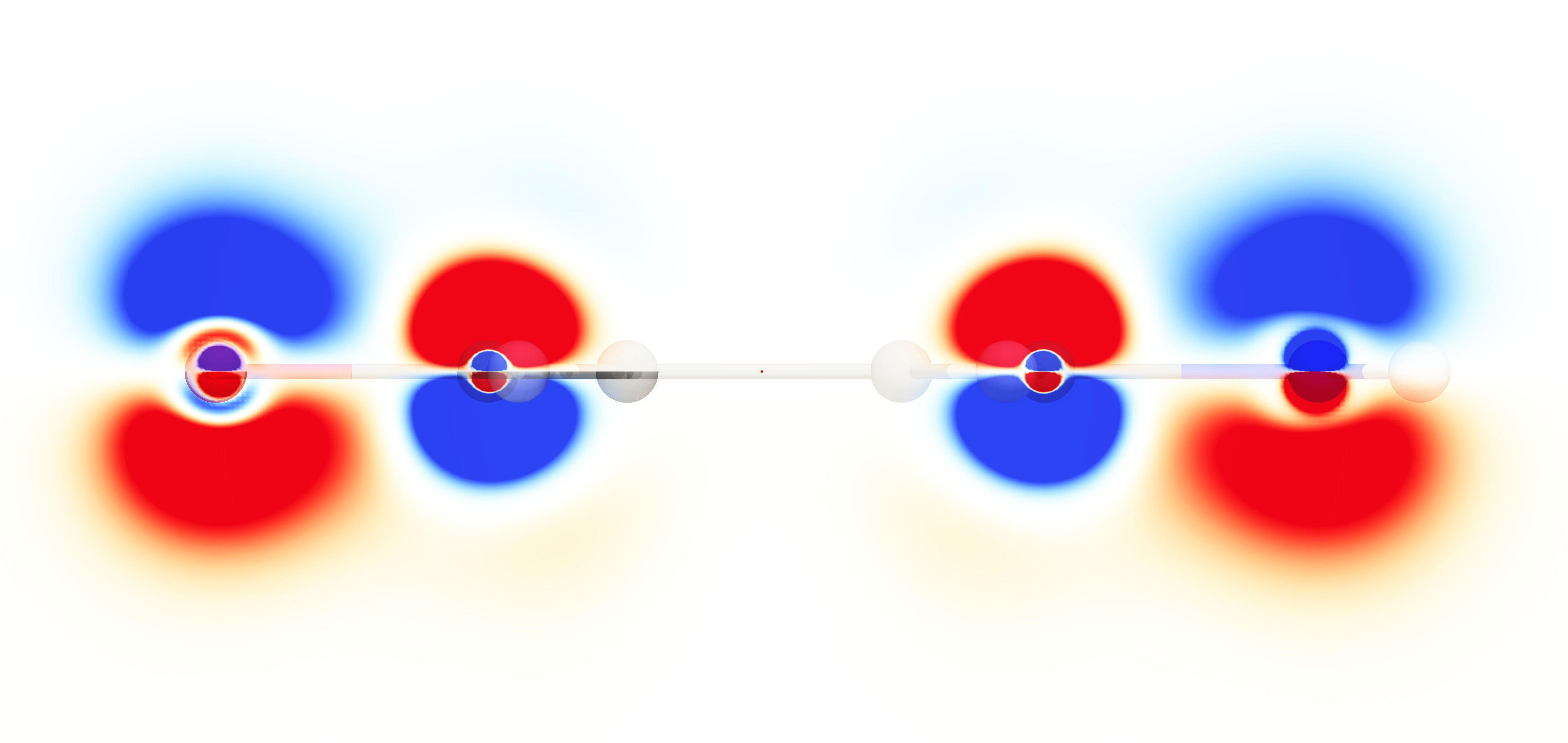}\label{p-aminophenol-B_para}}

\caption{The current density of \textit{p}-aminophenol in an external magnetic
field (a) perpendicular to the benzene ring; (b) parallel to the benzene ring.
The \ce{OH} group is on the left-hand side of the picture and the \ce{NH2}
group is to the right. The current-density flux pointing towards the viewer is
illustrated in red, while the blue areas show the current density flowing away
from the viewer.  \label{p-aminophenol}}

\end{figure}

\noindent Current-density vortices have the shape of a distorted torus, whose
axis points more or less in the same direction as the magnetic field. In Fig.\
\ref{p-aminophenol}, a plane is positioned such that it crosses the oxygen and
the nitrogen atoms of \textit{p}-aminophenol with the magnetic field either
perpendicular to the benzene ring or parallel to the line connecting the
heteroatoms. The diatropic ring current at the carbon atoms with the
substituents and the paratropic ring current inside the benzene ring are shown
in Fig.\ \ref{p-aminophenol}(a). The current-density flux is also seen in the
molecular plane. The \ce{OH} and \ce{NH2} substituents sustain a strong current
density. There are several atomic vortices with different tropicity embedded in
each other like onion shells. The atomic vortices are spatially larger when the
magnetic field is parallel to the molecular plane, since in the perpendicular
direction, there are also strong global current-density pathways on the outside
of the molecule and inside the molecular ring.  The direction of the current
density in the valence orbitals of the atoms alternates when the magnetic field
is parallel to the benzene ring.

\begin{figure}[H]
    \centering
    \subfigure[2H-1,2-azaborolium cation]{\includegraphics[width=0.30\textwidth]{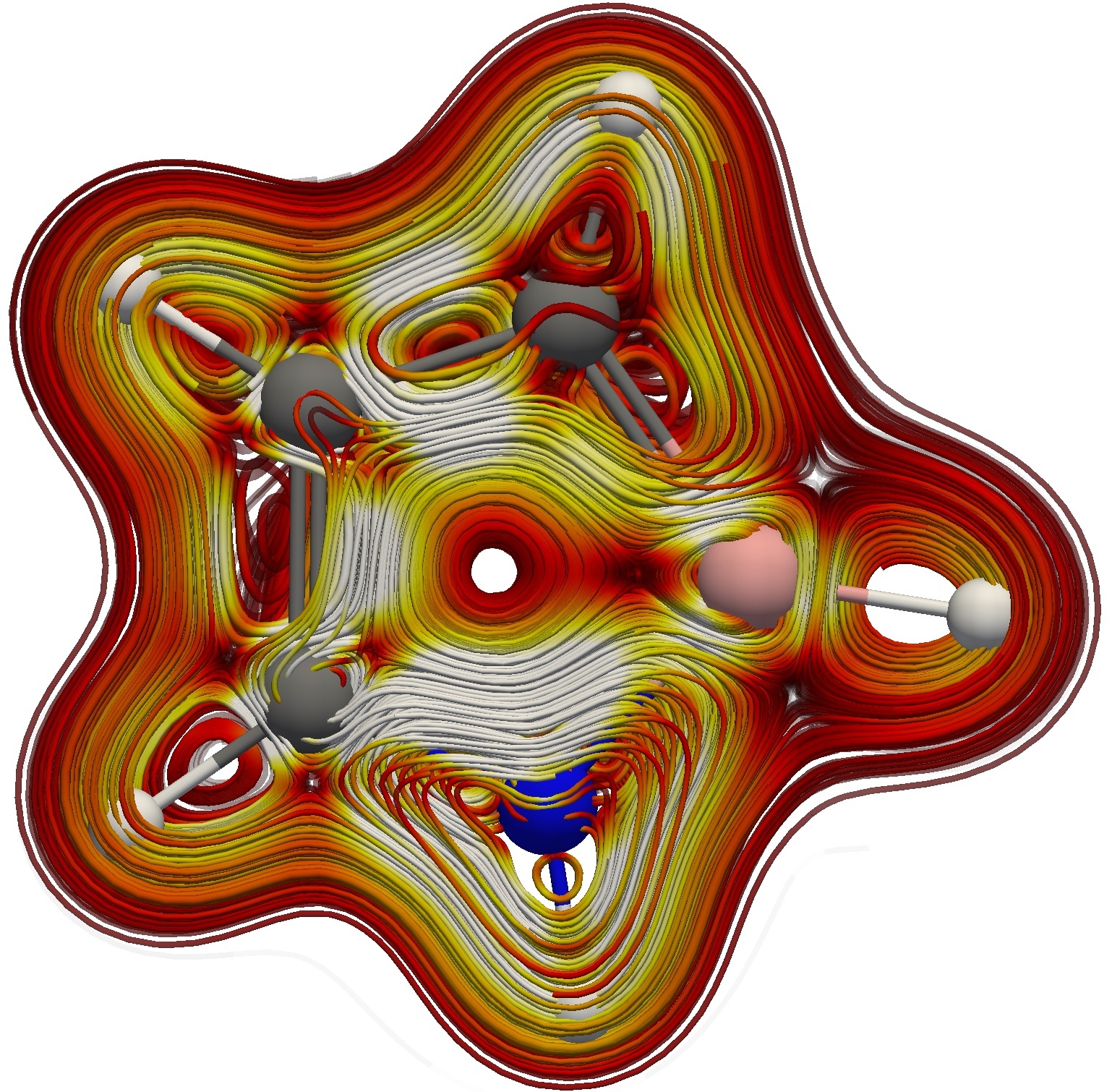}\label{fig:azaborolium}}
\hspace{4mm}
    \subfigure[1,3-imidazole]{\includegraphics[width=0.30\textwidth]{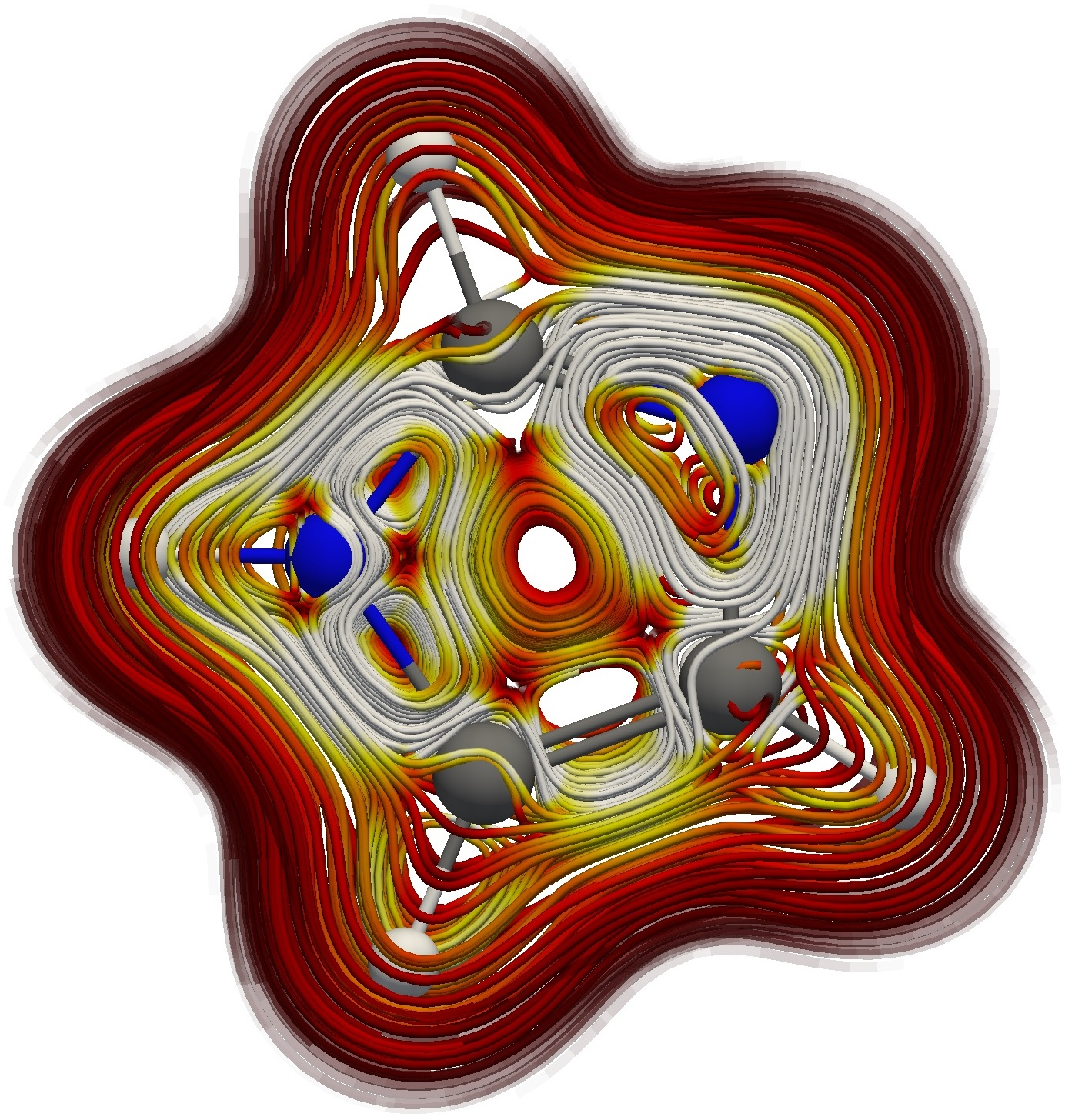}\label{fig:imidazole}}
\hspace{4mm}
    \subfigure[2,5-dihydro-1H-1,2,5-azadiborole]{\includegraphics[width=0.28\textwidth]{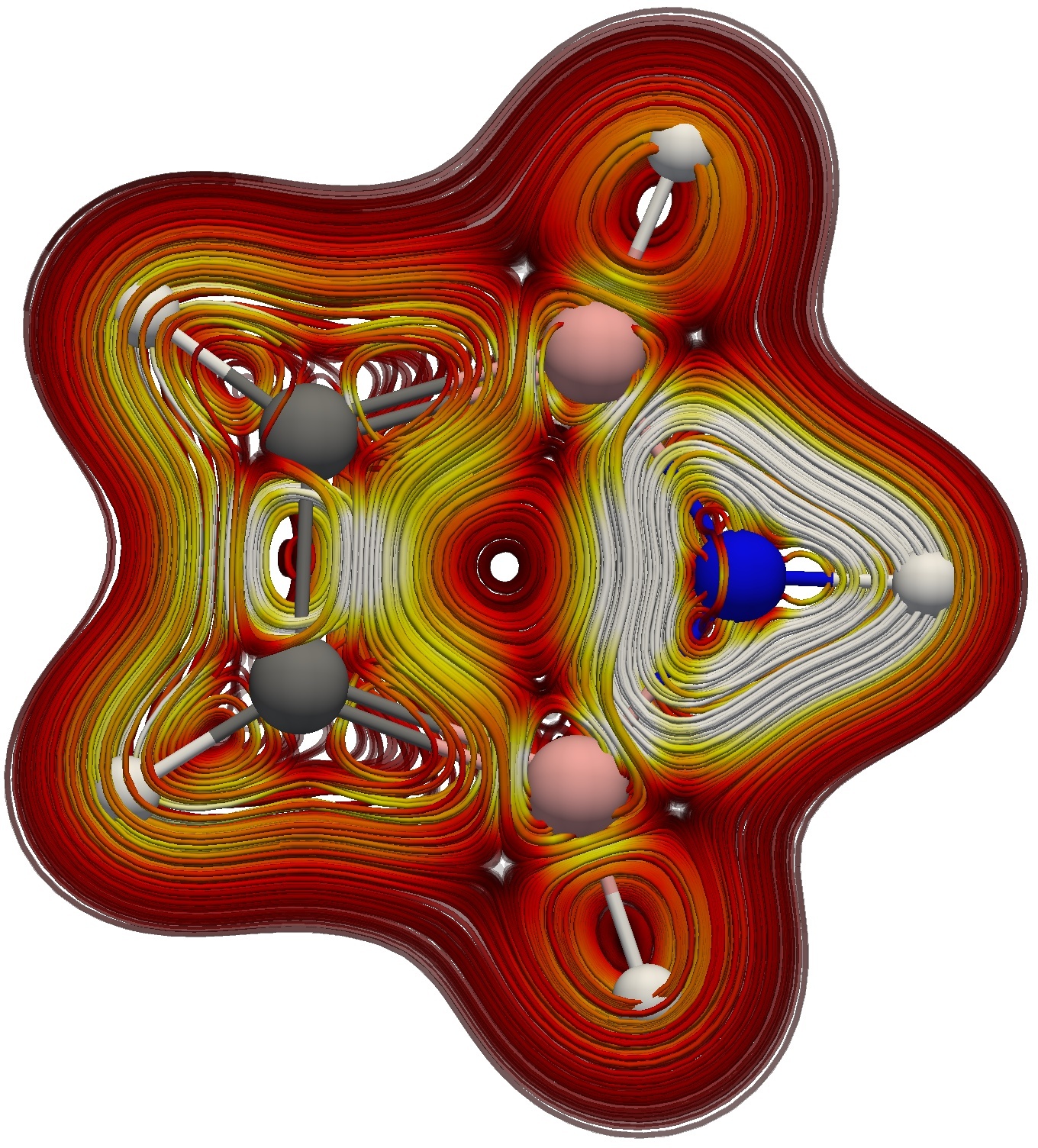}\label{fig:azadiborole}}

\caption{The current densities of the 2H-1,2-azaborolium cation, imidazole and
2,5-dihydro-1H-1,2,5-azadiborole are  illustrated with streamlines. The external
magnetic field is perpendicular to the molecular plane. The colour scheme
corresponds to the strength of the current density with white being the
strongest. \label{fig:heterocycles}}

\end{figure}

\noindent Streamline plots are a useful visualisation tool as an analogue to
classical trajectories. The Runge-Kutta method as implemented in
Paraview\cite{paraview} can be used in studies of the current-density flux
calculated on a 3D grid. Streamline plots are employed to illustrate the effect
of heteroatoms in the 2H-1,2-azaborolium cation, 1,3-imidazole and
2,5-dihydro-1H-1,2,5-azadiborole in Fig.\ \ref{fig:heterocycles}.  There are
strong vortices around each nitrogen atom and the \ce{N - C} bonds, which resemble
the local current-density pathways found in porphyrinoids.\cite{Bartkowski:19}

\subsection{Integrating the strength of the current density}

\begin{figure}[H]
\centering
\subfigure[]{\includegraphics[width=0.25\linewidth]{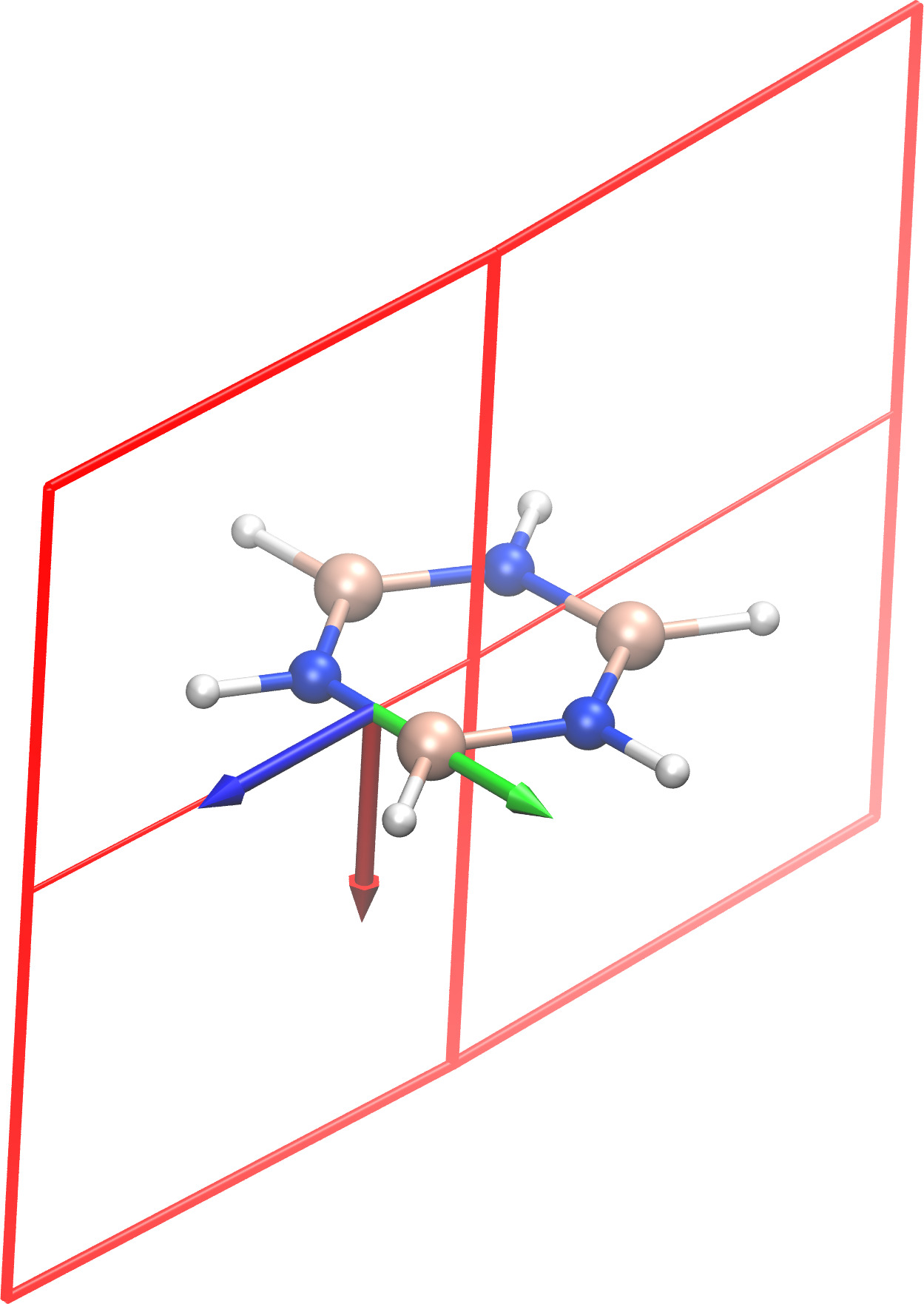}\label{fig:borazine-intplane}}
\hspace{8mm}
\subfigure[]{\includegraphics[width=0.60\linewidth]{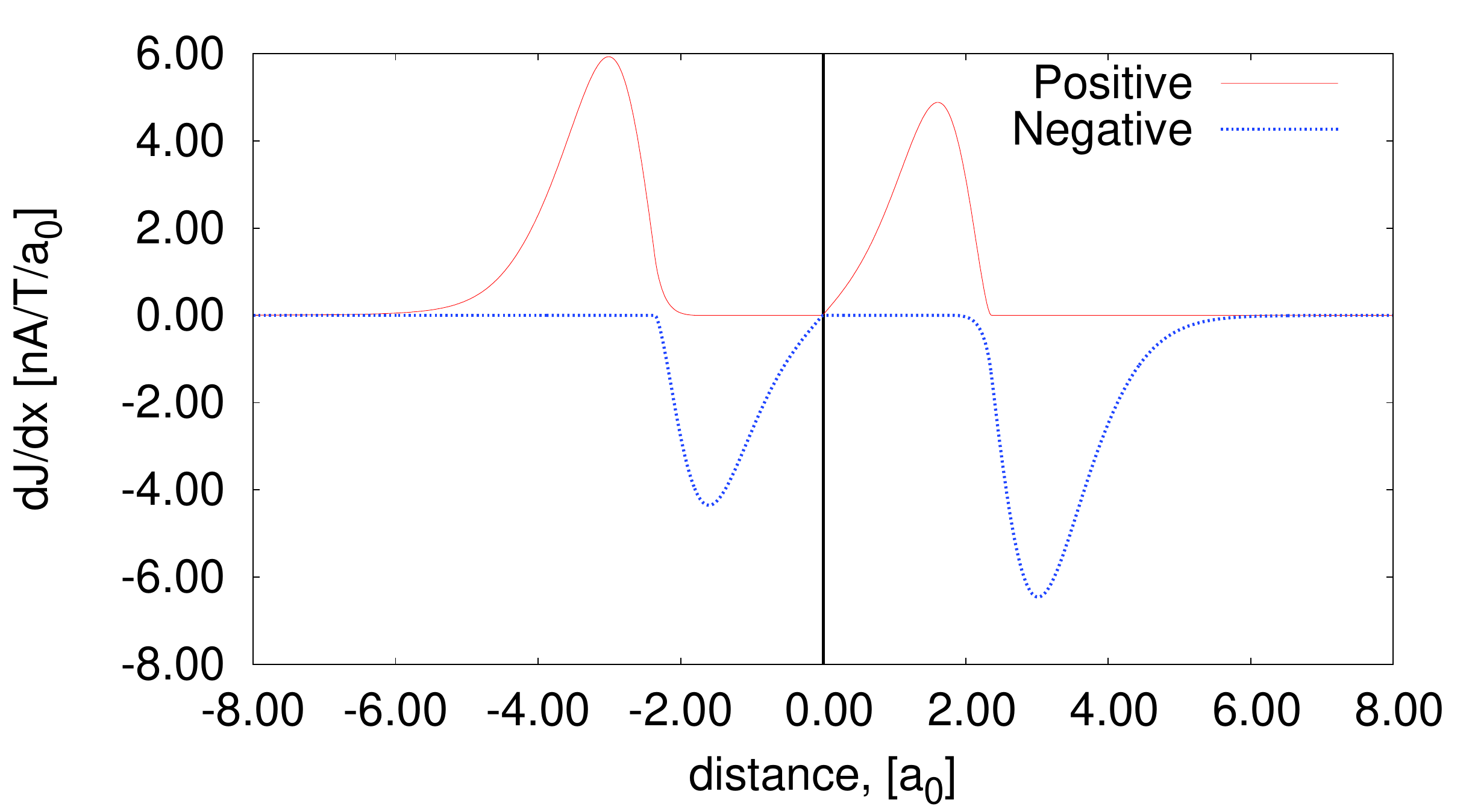}\label{fig:borazine-profile}
\label{fig:borazine-intplane-profile}}

\caption{(a) An integration plane is positioned in borazine such that it is
perpendicular to the molecular plane and crosses a \ce{B-N} bond.  The basis
vectors of the integration plane are shown as red, green and blue arrows.  (b)
The differential profile of the current density of borazine obtained by
numerically integrating the current-density flux in thin vertical slices of the
plane crossing the \ce{B-N} bond and through the whole molecule. The net ring-current strength of borazine is \nAT{3.23}. The positive
contributions refer to the current density, which seems to be flowing away from
the viewer. The vertical bar at $x=0$ is the vortex origin. The values $x<0$
correspond to the left side of the integration plane shown in (a), while the
right side of the plane with $x > 0$ is the returning current on the other side
of the vortex.  \label{fig:borazine-integration} }

\end{figure}

Current density can be investigated by identifying the current-density pathways
and determining their strengths by integrating the current-density flux through
a plane as shown in Fig.~\ref{fig:borazine-integration}(a).  The origin of the
integration plane is in the centre of the vortex and the plane extends far
outside the molecular ring where the current density vanishes. In the vertical
direction, the integration plane needs to be extended far enough where the
current density vanishes. The same ring-current strengths are obtained
regardless of whether the integration plane goes through a chemical bond or
through an atom, because the current density formally fulfils the charge
conservation condition.  Small deviations can be obtained due to the use of
finite basis sets. However, larger uncertainties are introduced for molecular
rings with heteroatoms or for non-planar rings, since the axis of the
current-density vortex is often shifted from the geometrical centre of the ring
and bent with respect to the normal vector of the ring.

\noindent We employ a numerical integration scheme for calculating the strength
of the current-density flux, $I$, which is obtained as\cite{Juselius:04} 

\begin{equation}
    I = \int_S \sum_\beta \frac{B_\beta}{|\mbf{B}|} \mathcal{\hat{J}}^{B_\beta} \ofR \cdot \hat{\vect{n}}\, \mathrm{d}s
\end{equation}

\noindent where the tensor elements of the current-density susceptibility
$\mathcal{\hat{J}}^{B_\beta}\ofR$ are contracted with the three components of
the external magnetic field, which is normalised to one. 

\noindent The strength of the current-density flux through the integration
plane is then obtained as the scalar product of the current-density
susceptibility for a given direction of the magnetic field,
$\mathcal{\hat{J}}^{B_\beta}\ofR$, and the normal of the integration plane,
$\hat{\vect{n}}$. $S$.  The external
magnetic field is typically applied parallel to one of the axes of the
integration plane. The strongest ring currents are obtained when the magnetic
field is perpendicular to the molecular ring, \ie along the red arrow in
Fig.\ \ref{fig:borazine-integration}(a). The SI unit for current density is
A$\cdot$T$^{-1} \cdot$ m$^{-2}$ and we usually report strengths of ring-current
susceptibilities in \nAT{}. We employ atomic units of distance, \ie bohr,
$\bohr{1} = 0.529~177~210~903(80)\,$ \AA.

\noindent The direction of the current-density flux at the integration plane is
used for assigning the tropicity of the current-density pathway. However, this
assignment has to be considered with care because this simple algorithm cannot
distinguish between diatropic current densities and returning paratropic
current densities, and likewise, between paratropic current-density fluxes and
returning diatropic current-density fluxes.  Current densities with a given 
tropicity have different directions when projected on the plane on the opposite sides of the
vortex origin. 

\noindent The profile of the strength of the current density at an integration
plane through the geometric centre of borazine is shown in
Fig.~\ref{fig:borazine-integration}(b). The point $x = 0$ marks the centre of
the ring.  The current density is integrated far out on both ends of the
molecule.  For monocyclic molecules, the positive contribution in one half of
the molecule can be assigned as the diatropic ring current while the paratropic
ring current appears as negative. For polycyclic molecules, the assignment of
diatropic and paratropic ring-current contributions is difficult, especially
when diatropic and returning paratropic ring currents mix as in naphthalene
annelated with two pentalene moieties,\cite{Sundholm:16} where the strongly
antiaromatic pentalene units force the diatropic ring current of naphthalene to
turn back, following the paratropic ring current inside its rings. 

\noindent Thus, it is very important to have a clear visual understanding of
the current-density flux before assigning the tropicity. Tropicity is a global
property that can only be determined from the circulation direction with
respect to the magnetic field vector by following the trajectories of the
current-density vector field. For molecules consisting of fused rings, the assignment of the
tropicity is complicated, because from the perspective of a given molecular
ring, a paratropic vortex in an adjacent ring seems to be diatropic.  When the
tropicity is correctly assigned, diatropic current densities can be assumed to
be positive and the paratropic ones are negative.  The strength of the
current-density flux is then the  sum of the diatropic and paratropic
contributions.  A more detailed picture of the current-density flux can be
obtained by separately investigating its diatropic and paratropic
contributions rather than studying only the net strength.

\subsection{The aromatic ring-current criterion}

Electron delocalisation in the conjugated chemical bonds of aromatic molecules
leads to an energetic stabilisation of the molecule correlating with the
strength of the magnetically induced ring
current.\cite{Kumar:17,Monaco:19,Patra:19} Ring currents can be experimentally
detected indirectly by measuring the anisotropy of the magnetic
susceptibility\cite{London:37,Pauling:36,Lonsdale:37,Pochan:69} and $^1$H~NMR
chemical shifts, which are, therefore, an important indicator of aromaticity.
The $^1$H~NMR signals of the protons located on the exterior part of
an aromatic ring are deshielded, leading to a downfield shift.  The
ring-current definition of aromatic molecules is that they sustain a net
diatropic ring current when exposed to an external magnetic field, whereas
antiaromatic molecules sustain a net paratropic ring current.\cite{Gomes:01}
H{\"u}ckel's original $\pi$-electron counting rule states that molecular rings
with $(4n + 2)~\pi$ electrons are aromatic and that antiaromatic molecules have
$4n~\pi$ electrons.\cite{Huckel:31a,Huckel:32,Breslow:65,Breslow:73}  

\noindent The existence of a strong current-density flux does not make a
molecule aromatic or antiaromatic. Non-aromatic molecules sustain a weak net
ring-current strength that may consist of strong diatropic and paratropic
contributions canceling each other. The concept of aromaticity has been extended to
triplet-state aromaticity,\cite{Baird:72} the aromaticity of excited
states\cite{Rosenberg:14,Vijay:20} and of M{\"o}bius-twisted
molecules.\cite{Heilbronner:64,Herges:06,Rappaport:08} 

\noindent There are also other criteria to determine molecular aromaticity
that are not always concordant.\cite{Lazzeretti:04}  For example, the
magnetic and structural criteria for aromaticity correlate for aromatic
M{\"o}bius-twisted hydrocarbon rings, whereas for the corresponding
antiaromatic planar structures, the two aromatic criteria have opposite
correlation.\cite{Herges:06} Antiaromaticity does not always imply an
increasing bond-length
alternation.\cite{Herges:06,Fowler:01,Fowler:02a,Soncini:02} The electron
delocalisation of antiaromatic molecules resulting in paratropic ring currents
leads to a less pronounced bond-length alternation than in non-aromatic
molecules. Although the energetic criterion states that aromatic molecules are
characterised by lower reactivity and greater chemical stability, this does
not necessary imply that the lowest-energy conformer of a molecule is the most
aromatic one according to the ring-current criterion.\cite{Dimitrova:18} 

\begin{table}[H]

\caption{Calculated diatropic and paratropic contributions to the net
ring-current strength susceptibility (in \nAT{}) of benzene, the lowest singlet
state of the benzene dication, and cyclohexadiene representing aromatic,
antiaromatic, and non-aromatic molecules.  The calculations were performed at
the B3LYP/def2-TZVP level. \label{tab:currents}} 

\begin{tabularx}{\textwidth}{LCCC}
\hline
\hline
Molecule &  Diatropic & Paratropic & Net current strength \\
\hline
Benzene$~^a$ & 17.00 & -4.95 & 12.05 \\
Benzene$~^b$ & 17.96 & -6.20 & 11.76 \\
Benzene$^{2+}$$~^a$ & 2.57 & -68.88 & -66.32 \\
Benzene$^{2+}$$~^b$ & 6.42 &  -73.75 & -67.33 \\
Cyclohexadiene$~^a$ & 9.76 & -10.25 & -0.49 \\
Cyclohexadiene$~^b$ & 7.92 & ~~-8.66 & -0.74 \\
\hline
\hline
\end{tabularx}
$^a$ Through the middle of the \ce{C-C} bond.\\ 
$^b$ Through the \ce{CH} moiety. \\
\end{table}

\noindent Benzene is the archetypal aromatic molecule with 6 $\pi$ electrons
fulfilling H{\"u}ckel's rule for aromaticity. Benzene sustains a diatropic ring
current of \nAT{17.00} in the $\pi$ orbitals on both sides of the molecular
ring as well as outside it. It sustains a paratropic ring current of
\nAT{-4.95} inside the ring. However, note that these values, as well as the
other diatropic and paratropic contributions listed in Table~\ref{tab:currents}
for the plane crossing the \ce{C-C} bond also contain positive and negative
contributions from the bond vortex.  The net ring-current strength of
\nAT{12.05} can be used as reference value for the degree of aromaticity of
molecules. 

\begin{figure}[H]
    \centering
    \subfigure[Through a \ce{C-C} bond]{\includegraphics[width=0.49\linewidth]{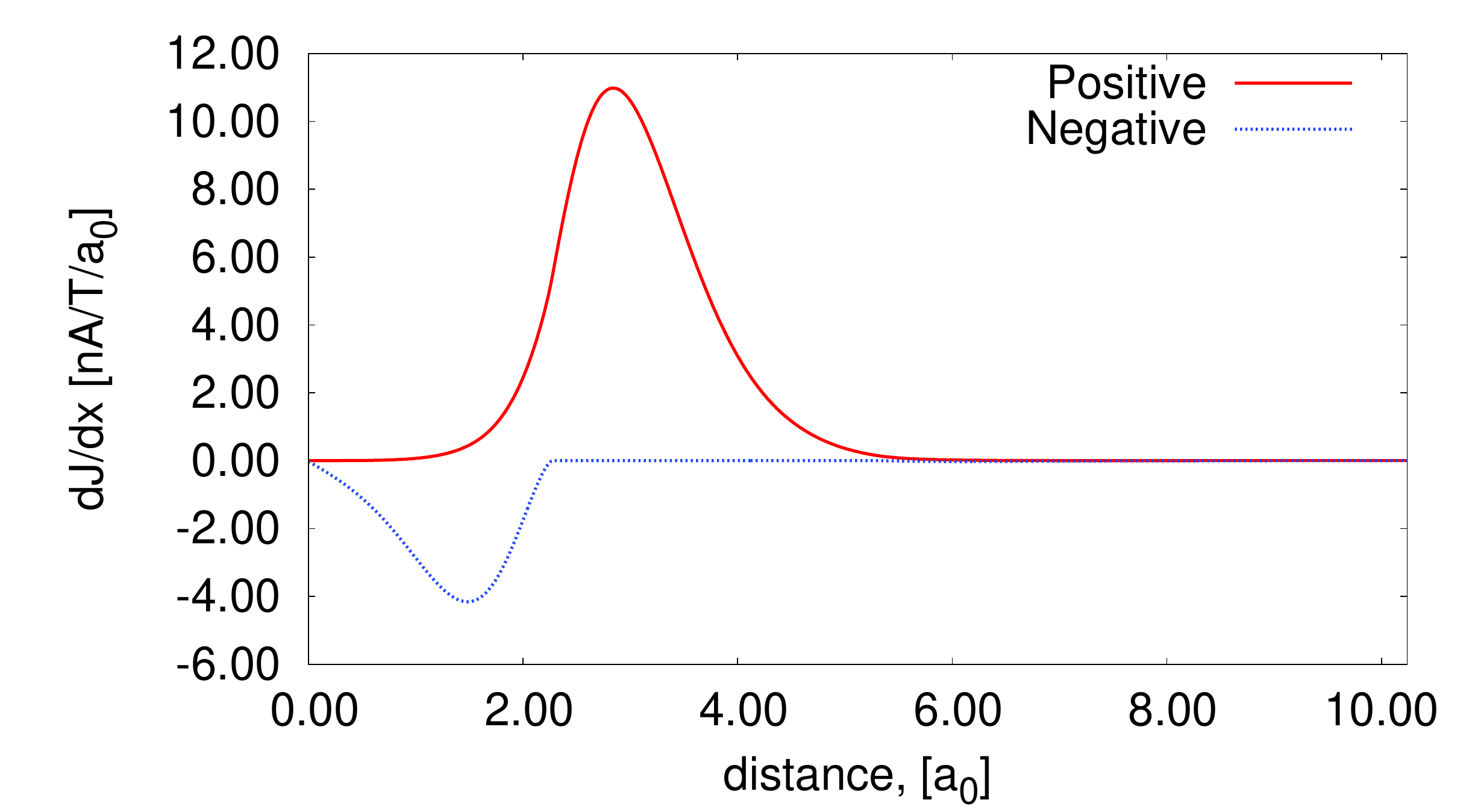}\label{fig:benzene-profile-bond}}
    \subfigure[Through the atoms]{\includegraphics[width=0.49\linewidth]{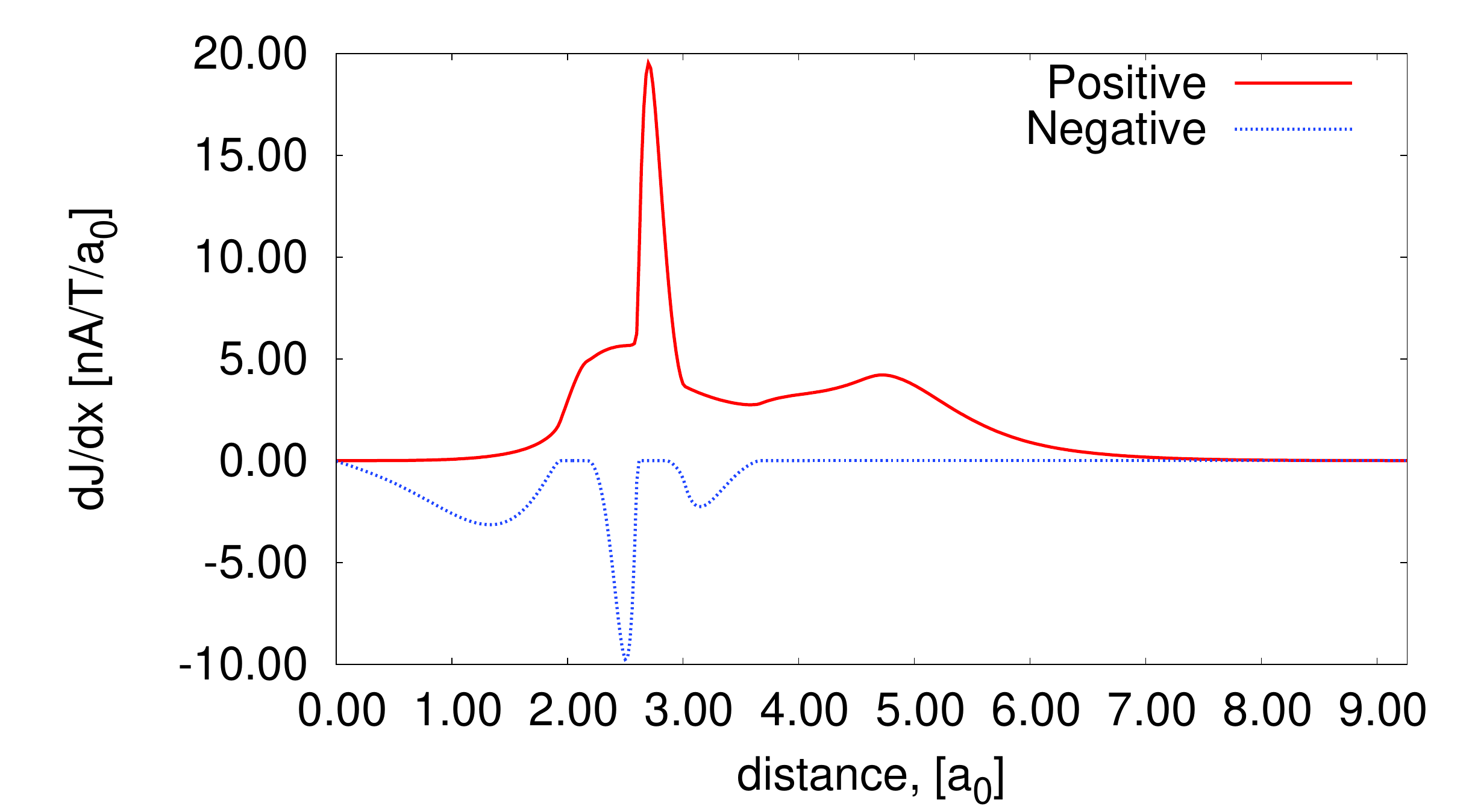}\label{fig:benzene-profile-atom}}

    \caption{The profile of the current-density susceptibility in benzene
    obtained by placing the integration plane (a) through a \ce{C-C} bond; (b)
    through the \ce{C} and \ce{H} atoms.  \label{fig:benzene-profiles}}

\end{figure}

\noindent The profile of the current density passing a plane in the middle of
the \ce{C-C} bond and a plane through the carbon and hydrogen atoms are shown
in Fig.\ \ref{fig:benzene-profiles}. Contributions to the ring current can be
obtained by integrating the diatropic and paratropic domains separately. The actual
paratropic contribution to the ring current of benzene is \nAT{-3.38}, which
excludes contributions from other vortices. Integration of the sharp peak
of the paratropic contribution yields the strength of the atomic
current-density vortex of the carbon atom of \nAT{2.0}. The diatropic
contribution calculated at the centre of the bond consists of the diatropic
contribution to the ring current and the contribution from the diatropic bond
vortex of \nAT{1.6}. The diatropic ring-current contribution is then
\nAT{16.4}. Accurate contributions to the ring-current strengths can be
obtained by integrating the often well-defined paratropic domains inside and
near atoms, whereas in the middle of the bond, the paratropic contribution to
the ring current and the contribution from returning diatropic current density
of the bond vortex overlap.

\begin{figure}[H]
    \centering
    \subfigure[Through a \ce{C-C} bond]{\includegraphics[width=0.45\linewidth]{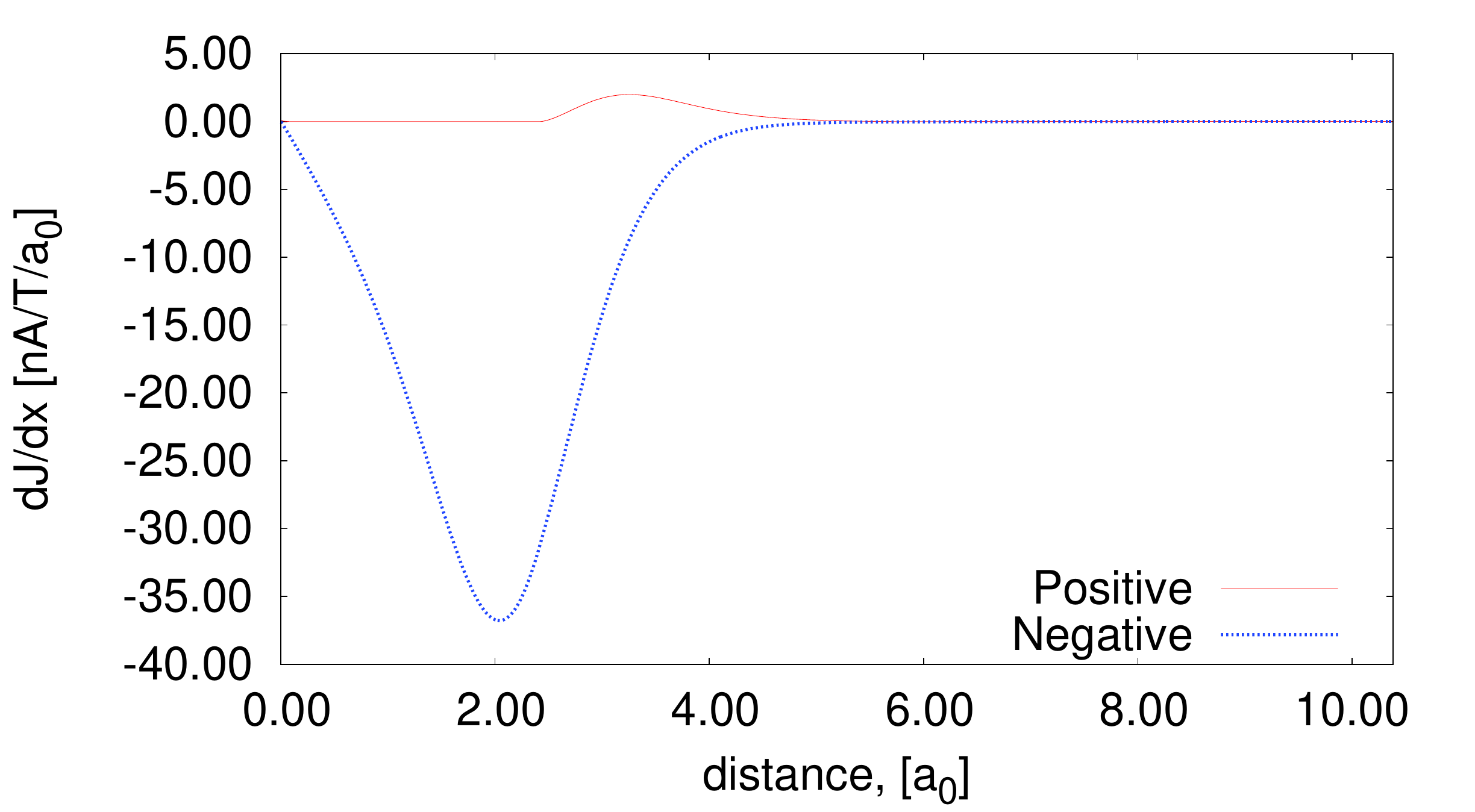}\label{fig:benzene2+-profile-bond}}
    \subfigure[Through the atoms]{\includegraphics[width=0.45\linewidth]{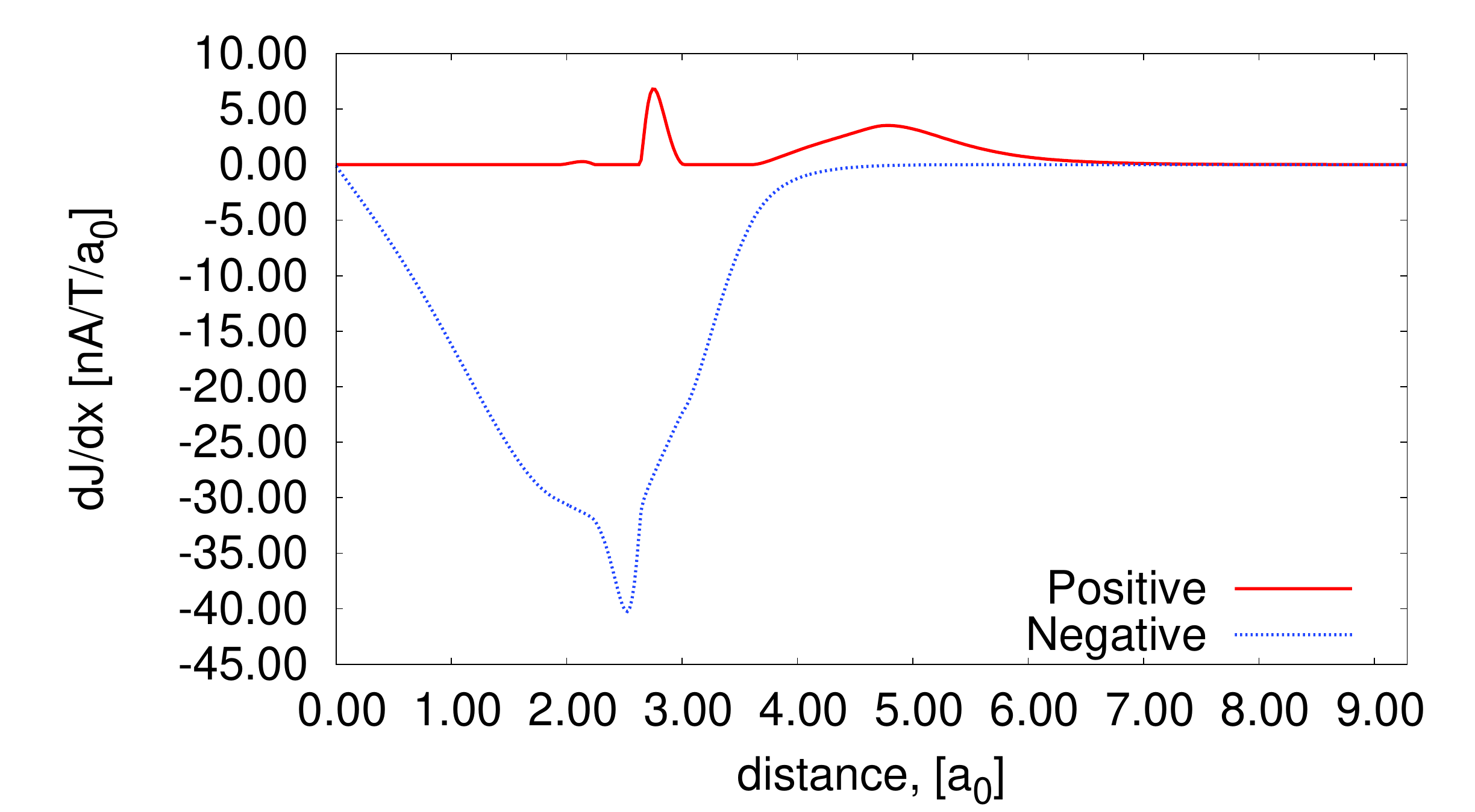}\label{fig:benzene2+-profile-atom}}

    \caption{The profile of the current-density susceptibility in the benzene dication
    obtained by placing the integration plane (a) through a \ce{C-C} bond; (b) through
    the \ce{C} and \ce{H} atoms.  \label{fig:benzene2+-profiles}}

\end{figure}

\noindent The lowest singlet state of the benzene dication with $4~\pi$
electrons fulfils H{\"u}ckel's rule for antiaromaticity. It sustains a strong
paratropic ring current of \nAT{-66.32} which consists of a diatropic
contribution of \nAT{2.57 } and a paratropic contribution of \nAT{-68.88 }. The
profiles of the integrated current density in planes through the middle of the
chemical bond and through the \ce{CH} moiety are shown in Fig.\
\ref{fig:benzene2+-profiles}. The current density is strongly paratropic inside
the ring and it has a weak diatropic ring current outside the ring, which is
typical for antiaromatic rings.\cite{Fliegl:09} The current-density flux is
always diatropic far away from the molecule,\cite{Gomes:83} even for strongly
antiaromatic molecules. The ground-state of the benzene dication is a triplet state, whose molecular structure belongs to the $D_\mathrm{6h}$ point group like benzene. 

\noindent Berger and Viel used group theoretical arguments to show that
Jahn-Teller distorted molecules belonging to the $C_\mathrm{n}$,
$C_\mathrm{nv}$, $C_\mathrm{nh}$, $D_\mathrm{n}$, $D_\mathrm{nh}$ for
$\mathrm{n}>2$, and $D_\mathrm{nd}$ and $S_\mathrm{2n}$ for $n>1$ point groups
are antiaromatic and applied their conclusions on cyclobutadiene and
cyclooctatetraene.\cite{Berger:20a} They proposed a general symmetry principle
for antiaromaticity that reads: ``First-order and primoid second-order
Jahn-Teller-distorted molecules out of non-isometric point groups are prone to
induced paramagnetism in magnetic fields parallel to the main axis of
symmetry''.\cite{Berger:20a} The antiaromaticity of the benzene dication
belonging to the $D_\mathrm{2h}$ point group also follows the proposed symmetry
principle.

\begin{figure}[H]
    \centering
    \subfigure[Through a \ce{C-C} bond]{\includegraphics[width=0.45\linewidth]{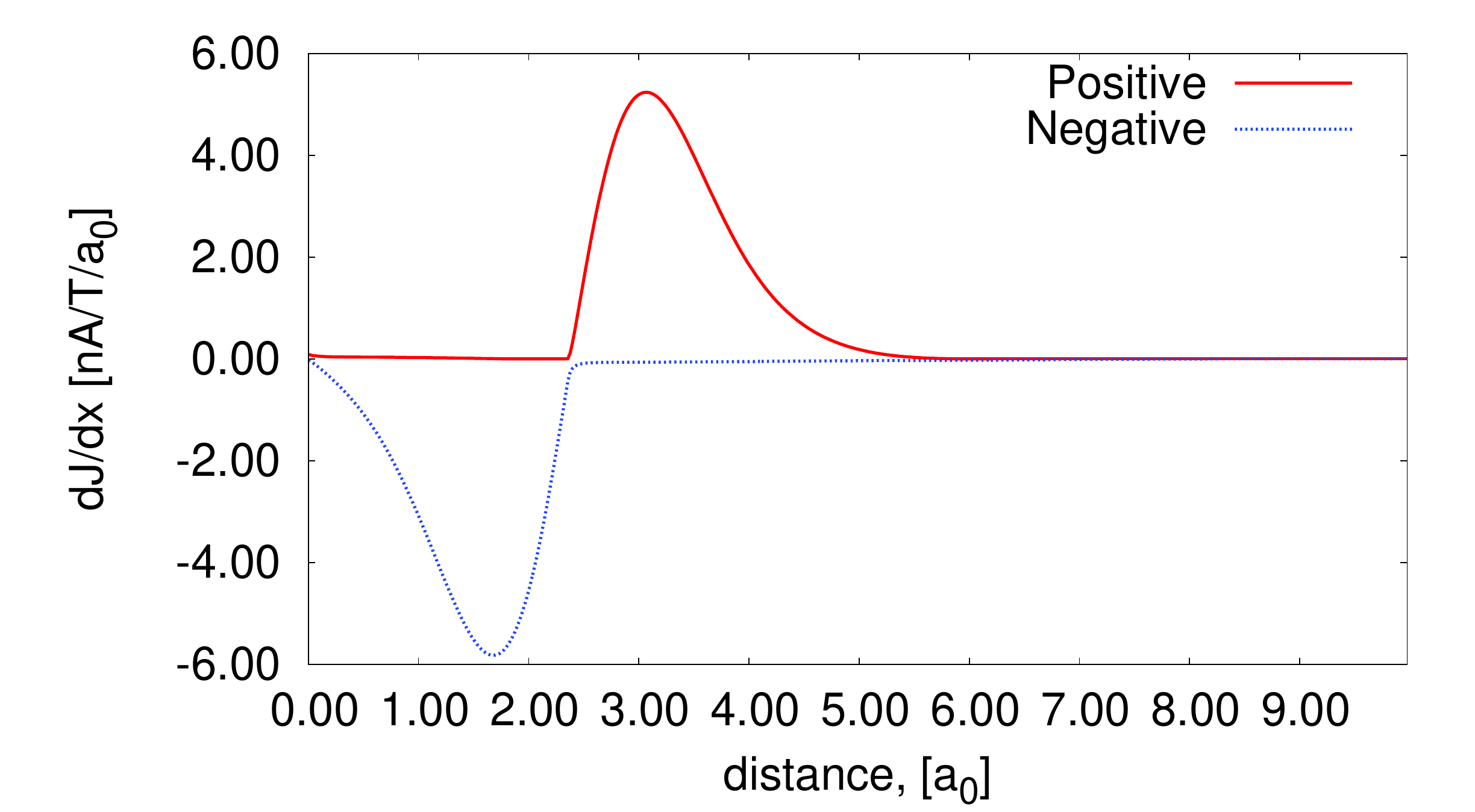}\label{fig:cyclohex-profile-bond}}
    \subfigure[Through the atoms]{\includegraphics[width=0.45\linewidth]{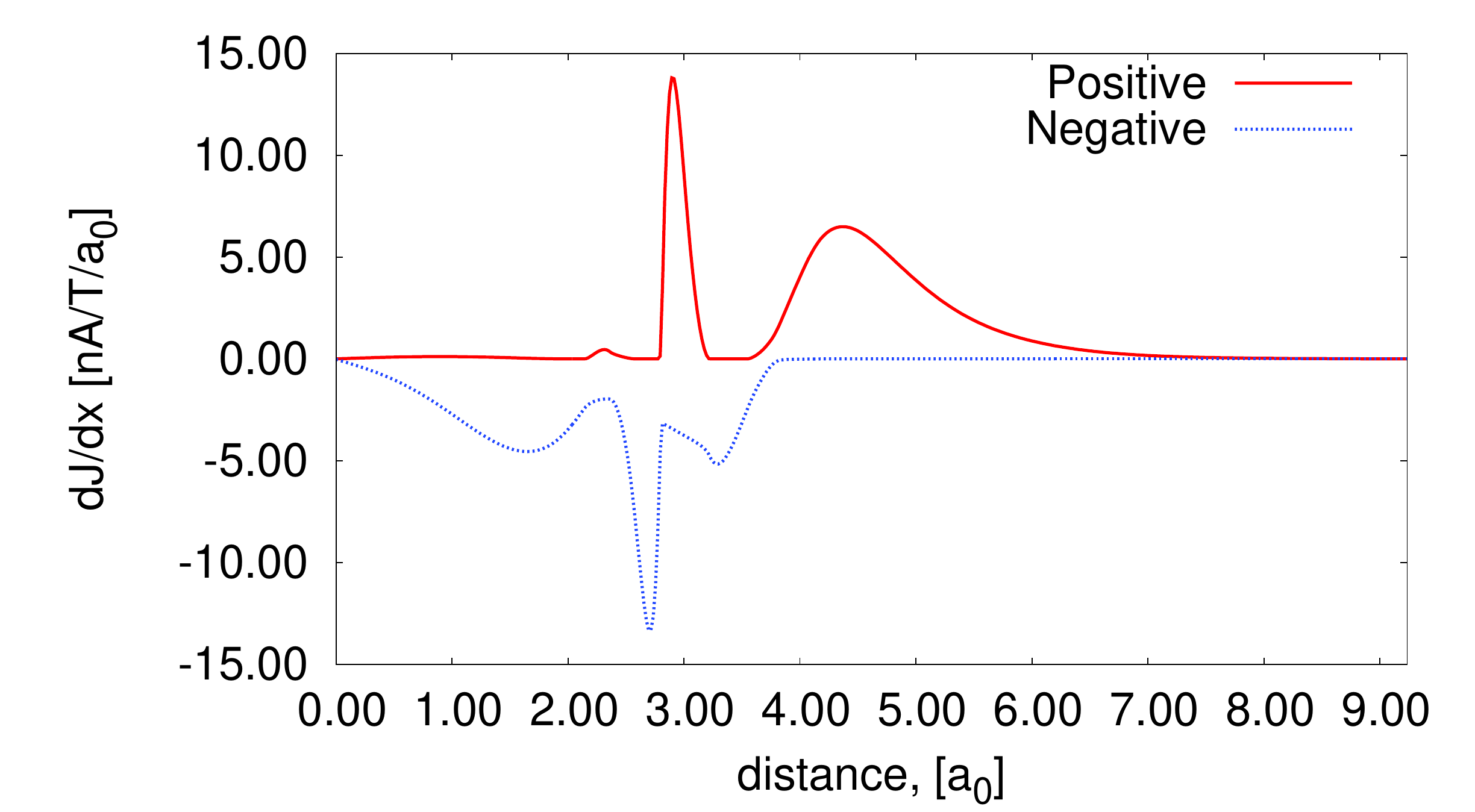}\label{fig:cyclohex-profile-atom}}

\caption{The profile of the current-density susceptibility in
    1,4-cyclohexadiene obtained by placing the integration plane (a) through a
    \ce{C-C} bond; (b) through the \ce{CH} moiety.
    \label{fig:cyclohex-profiles}}

\end{figure}

\noindent 1,4-cyclohexadiene is a non-aromatic molecule.  A diatropic
current-density flux of \nAT{9.76} and a paratropic current-density flux of
\nAT{-10.25} pass through a plane in the middle of the \ce{C=C} double bond.
The ring-current profiles in Fig.\ \ref{fig:cyclohex-profiles} show that the
current density is diatropic outside the bond and paratropic inside it. The
strengths of the diatropic and paratropic current densities are of the same
size and cancel. In the integration profile through the \ce{CH} moiety, one
sees the paratropic current-density flux mainly inside the ring and the
diatropic current-density flux passing on the outside of the carbon atom near
the hydrogen. The sharp peaks are due to the atomic current-density vortex of
the carbon atom. The diatropic and paratropic contributions to the current
density are somewhat weaker when placing the integration plane through the
\ce{CH} moiety.  This reveals that 1,4-cyclohexadiene has a strong
current-density vortex at the \ce{C=C} bond.  The small difference in the net
ring-current strength of 1,4-cyclohexadiene is due to charge conservation
problems. When using the larger \mbox{quadruple-$\zeta$} basis sets augmented
with polarisation and diffuse functions
(def2-QZVPD),\cite{Weigend:05,Rappoport:10} the net ring-current strength of
1,4-cyclohexadiene is \nAT{-0.57}  and \nAT{-0.60 } through the bond and the
\ce{CH} moiety, respectively.

\subsection{Aromatic pathways in polycyclic molecules} 

Determining the aromaticity of polycyclic aromatic hydrocarbons (PAH) and
heterocyclic PAHs is complicated. The routes of the current-density flux can be
determined by employing a variety of integration procedures that yield very
detailed and reliable information about the current-density pathways in these
molecules. Such calculations showed that a strong diatropic edge current
dominates the current-density flux in ovalene,\cite{Kaipio:12} whose molecular
structure consists of a naphthalene moiety surrounded by eight benzene rings.
The benzene rings at both ends of ovalene and those in the middle of the long
sides of ovalene are Clar rings, sustaining a local ring
current,\cite{Clar:72,Balaban:05} whereas the other benzene rings act as
bridges that do not sustain any local ring current. Similar calculations on
hexabenzocoronene showed that it sustains a global ring current around the edge
of the molecule and the Clar rings sustain local ring
currents.\cite{Juselius:04,Benkyi:20,Soncini:03}

\begin{figure}[H] 
    \centering
\subfigure[]{\includegraphics[width=0.45\linewidth]{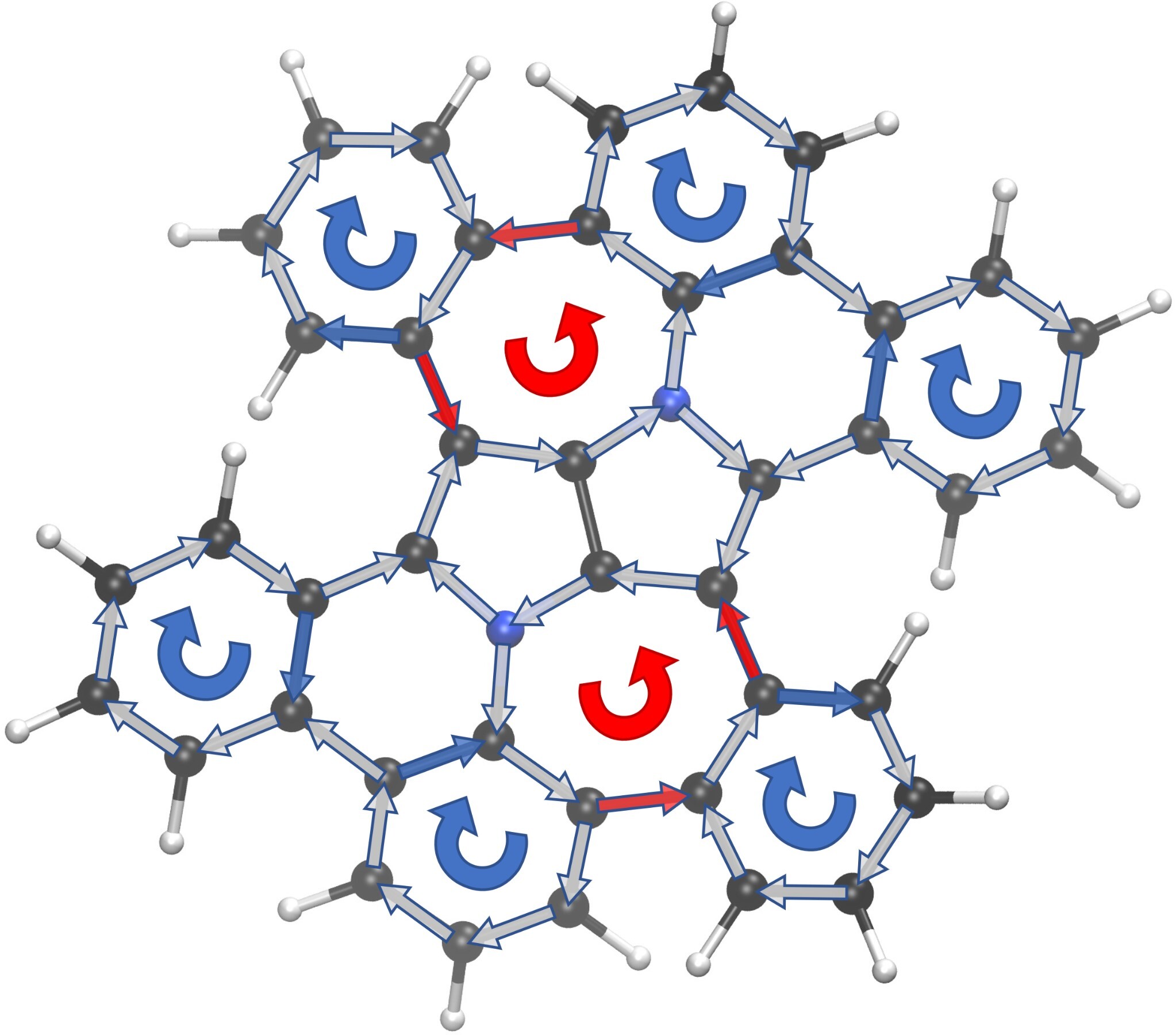}
}
\subfigure[]{\includegraphics[width=0.45\linewidth]{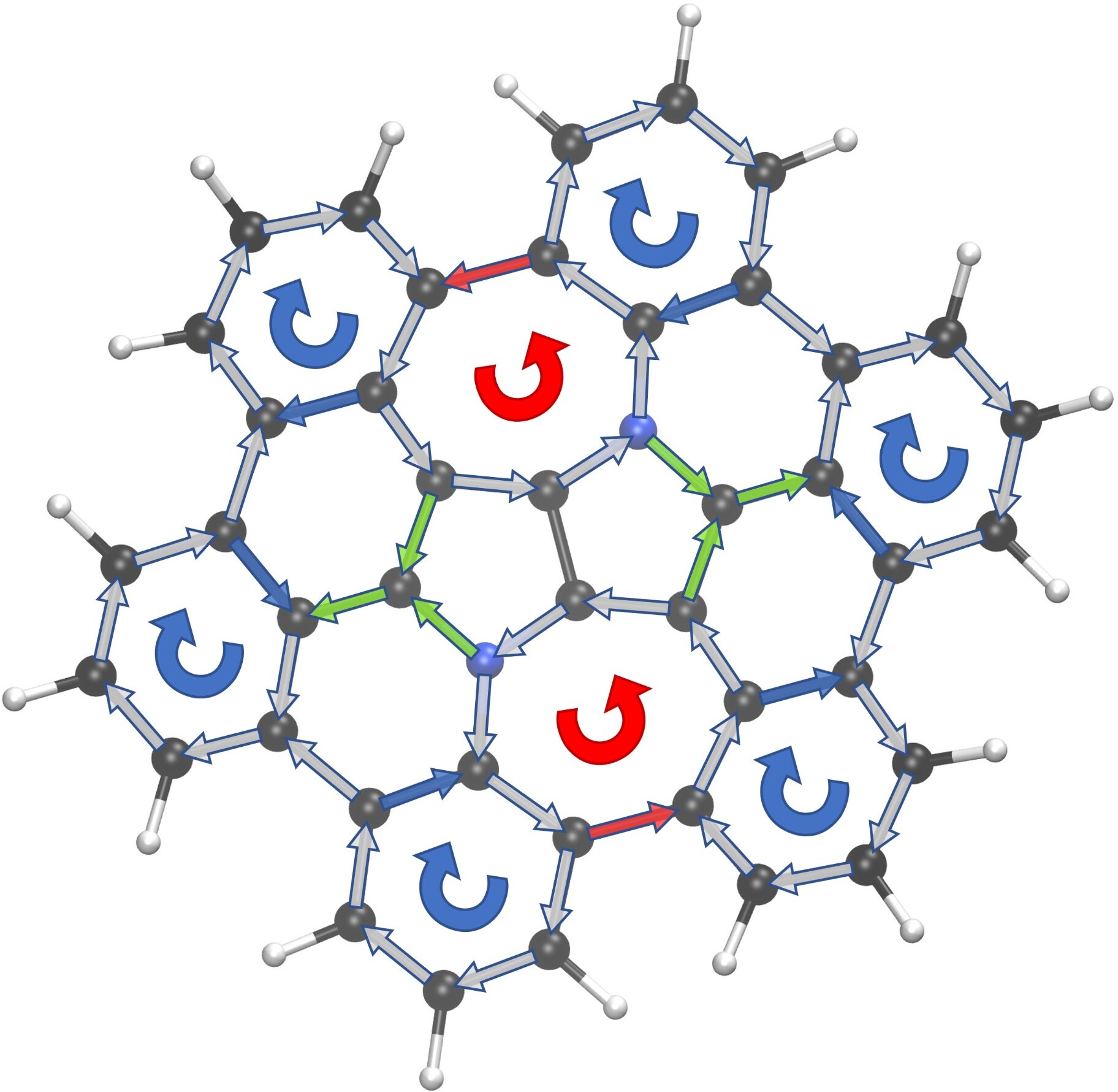}
}

\caption{The current-density pathways in (a) a heterocyclic hydrocarbon and (b)
a fully conjugated nitrogen-embedded buckybowl.\cite{Benkyi:20,Mishra:18} The
gray arrows show the global aromatic pathway. The red and blue arrows along the
bonds indicate the local current-density fluxes. The green arrows show how the
current-density flux splits and joins. The blue bent arrows denote local
diatropic ring currents and the red ones show rings sustaining local paratropic
ring currents.  \label{fig:nitrogen-buckybowl}}

\end{figure}

\noindent The current-density pathways
in the heterocyclic PAHs shown in Fig.\ \ref{fig:nitrogen-buckybowl} are even
more complicated. The seven-membered rings sustain local paratropic ring
currents. The outer benzene rings sustain weak local diatropic ring currents
and the rest of the rings act as bridges where the ring-current pathways split and
join. A global diatropic ring current flows around the entire molecule, except
at the paratropic seven-membered rings, where it takes the inner route,
following the direction of the local paratropic ring current. 

\begin{figure}[H]
    \centering
    \subfigure[Naphthalene]{\includegraphics[height=5cm]{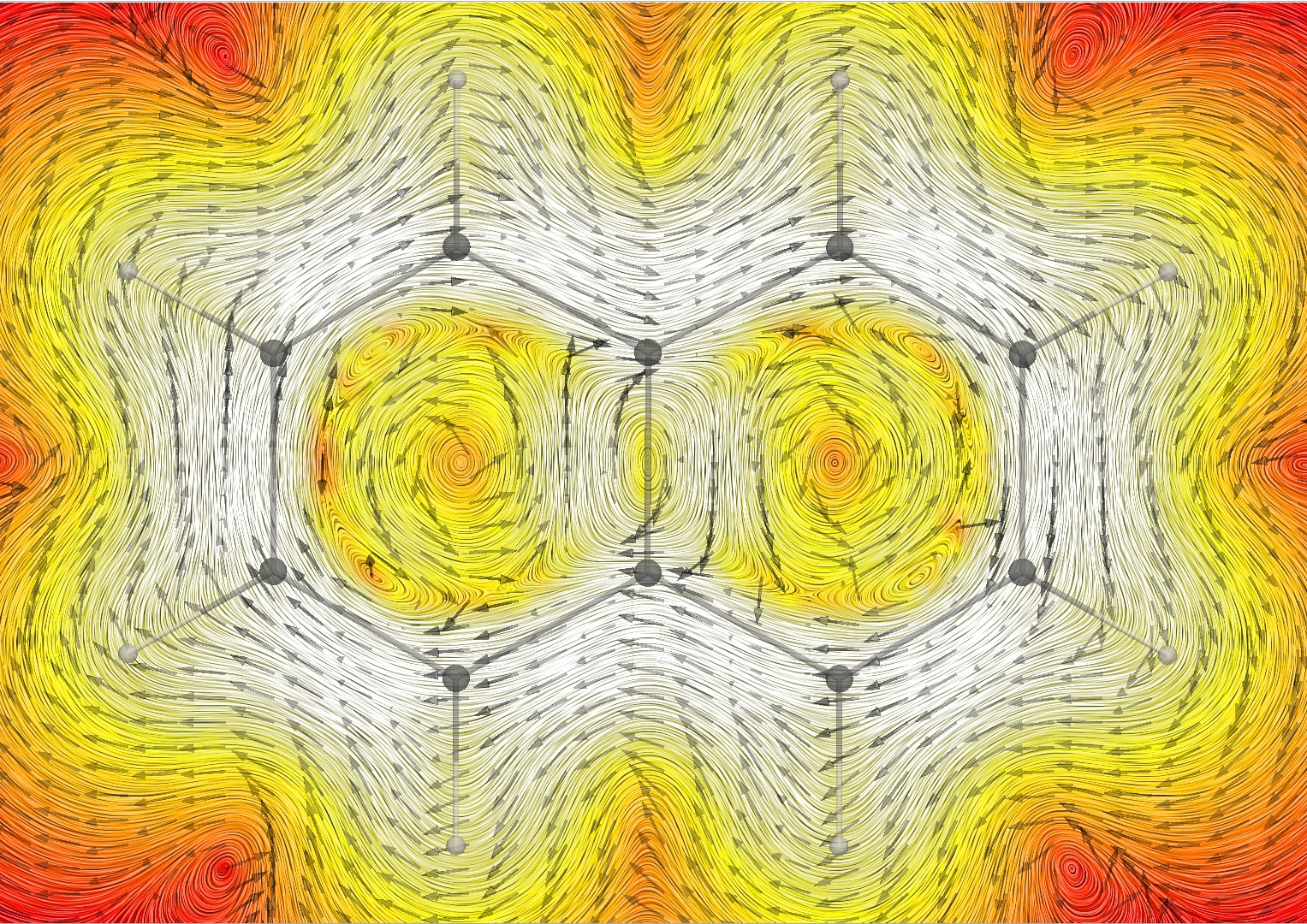}\label{fig:naphthalene}}
    \subfigure[Pentalene]{\includegraphics[height=5cm]{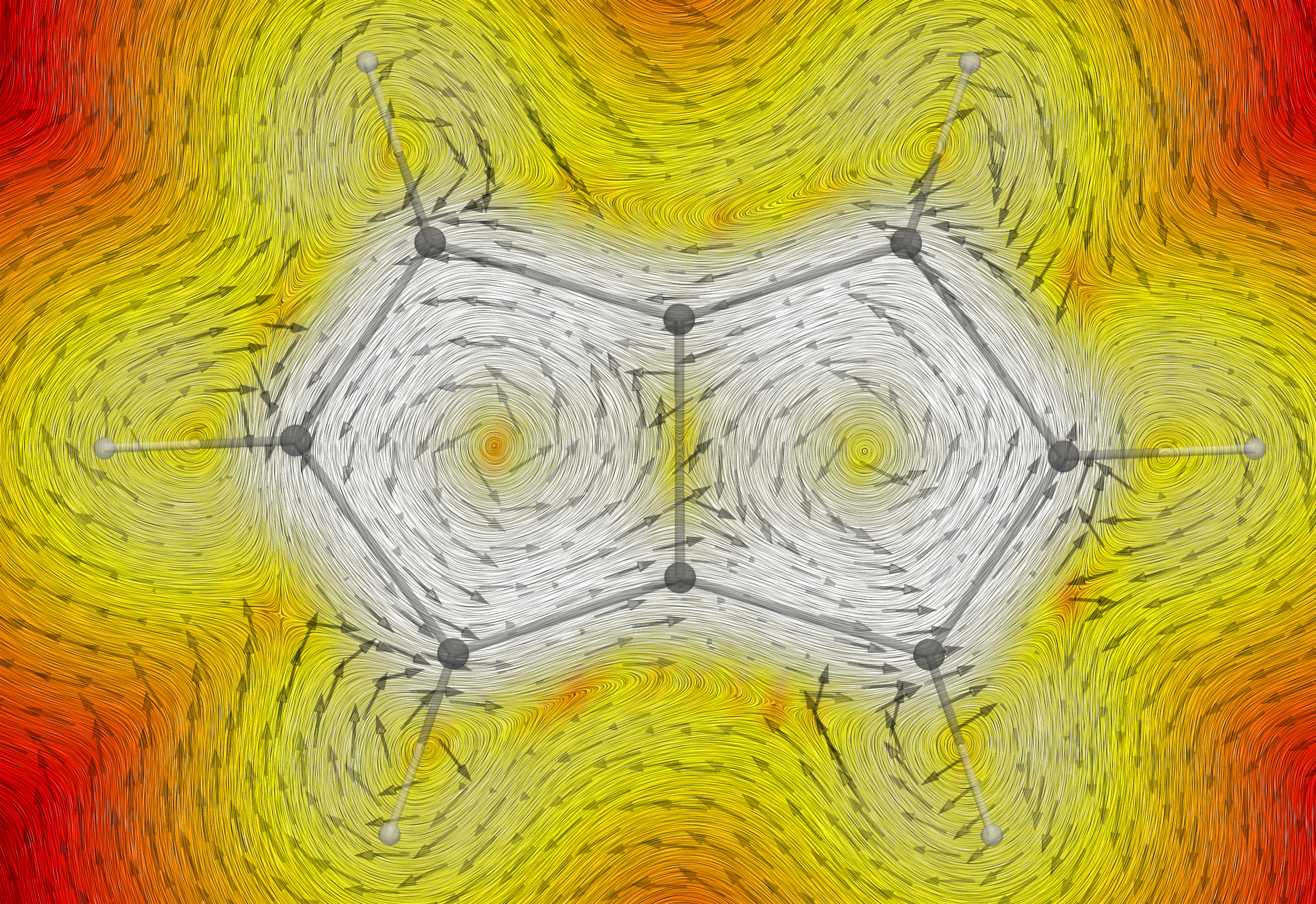}\label{fig:pentalene}}

\caption{The current density in naphthalene and pentalene in a plane positioned
    \bohr{1} above the molecular plane. The colour scale is logarithmic and
    gives the modulus of the strength of the current density in the range of
    $[3.57 \times 10^{-6}; 0.54 ]$ \nATA{}.  \label{fig:naph-pent} }

\end{figure}

\noindent Naphthalene with $10~\pi$ electrons is aromatic, consisting of two
annelated benzene rings. It sustains a ring current of \nAT{12.60} around the
entire molecule. Inside each benzene ring, there is a local paratropic ring
current which is distorted and tilted compared to the one of benzene.
Naphthalene has a bond vortex in the $\pi$ orbitals of the shared \ce{C-C} bond
as shown in Fig.\ \ref{fig:naph-pent}(a). Its strength is about \nAT{4.5}. A
better accuracy cannot be obtained without identifying the tropicity of the
current-density vortices at the \ce{C-C} bond.  

\noindent Pentalene, consisting of two annelated five-membered rings, is
antiaromatic with $8~\pi$ electrons. The current-density plot in Fig.\
\ref{fig:naph-pent}(b) shows that the individual five-membered rings sustain
strong paratropic ring currents mainly inside the ring. Pentalene exhibits also
a global paratropic ring current around the entire molecule.  The
current-density flux is weakly diatropic along the outer edge of the molecule.
The sum of the strengths of the local and the global paratropic ring currents
is about \nAT{-23}, whereas the diatropic ring current is about \nAT{3}. A more
accurate division into diatropic and paratropic contributions is difficult
without identifying the tropicity of the current density by following the
vector flux.

\subsection{Aromatic and antiaromatic porphyrinoids}

\begin{figure}[H] 
    \centering
    \subfigure[]{\includegraphics[width=0.40\linewidth]{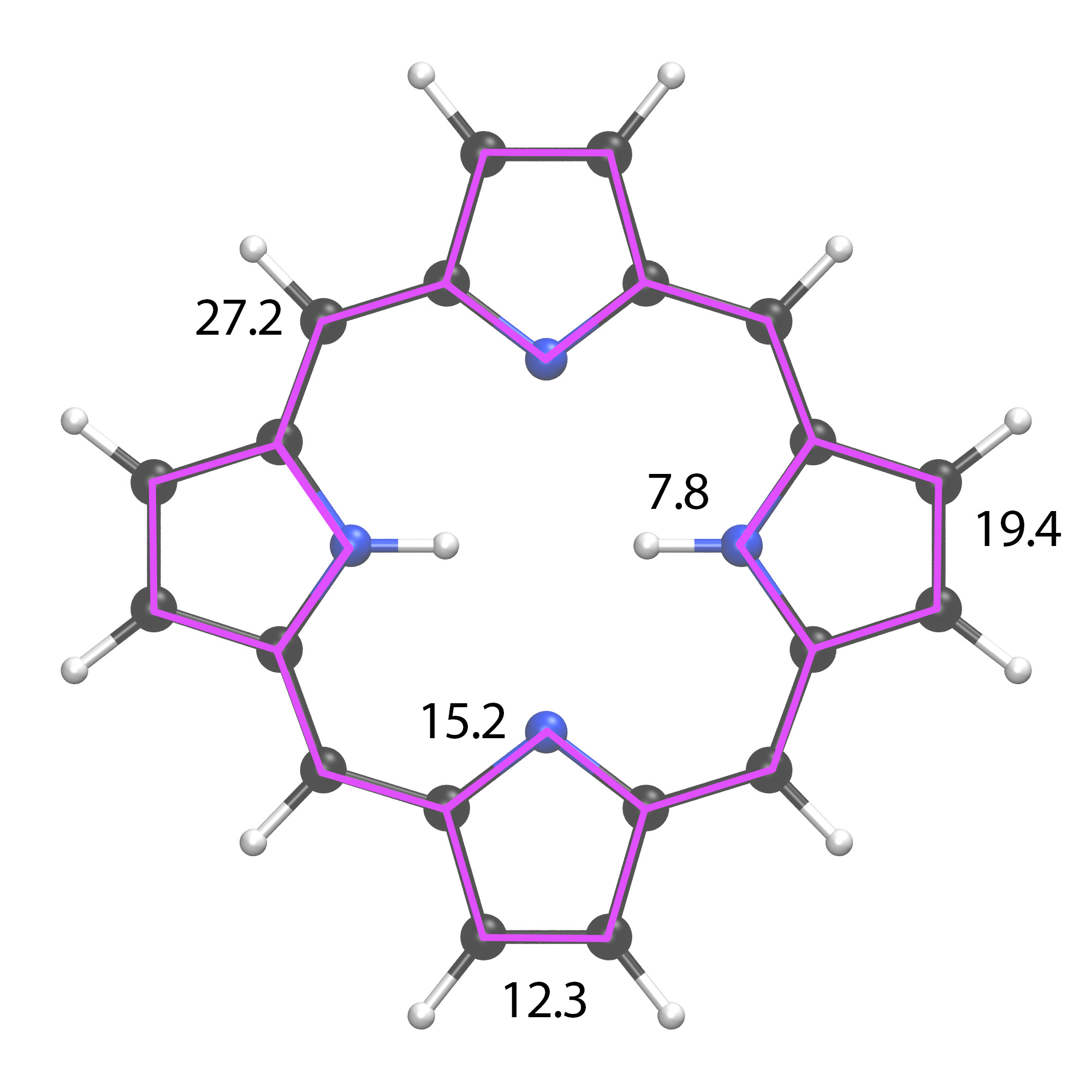}\label{fig:porphyrin}} 
    \subfigure[]{\includegraphics[width=0.40\linewidth]{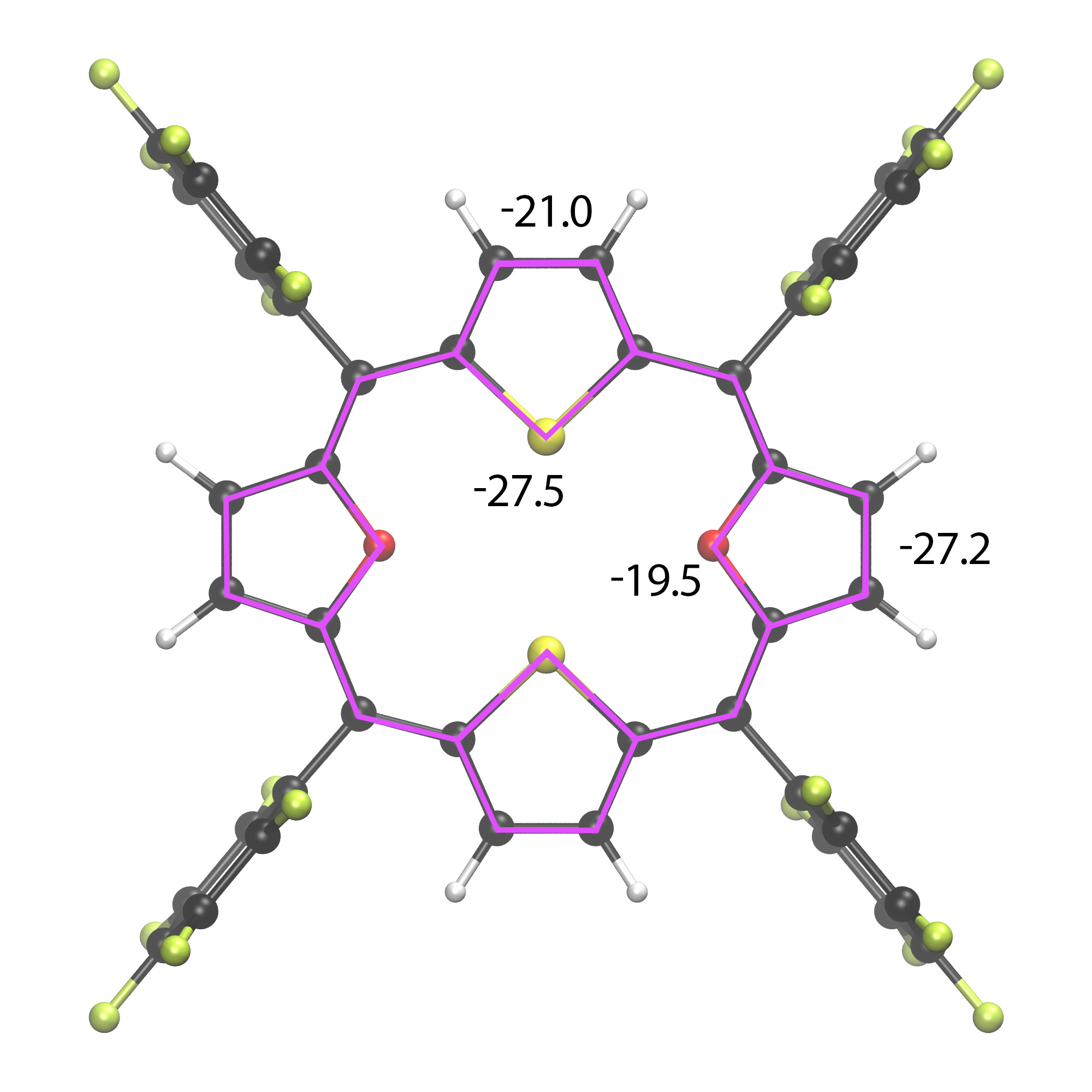}\label{fig:isophlorin}} 

\caption{The current-density pathways and ring-current strengths of (a)
    free-base porphyrin and (b) dioxa-dithiaisophlorin with pentafluorophenyl
    substituents in the \meso positions.  \label{fig:porphyrins} }

\end{figure}

The ring current of free-base porphyrin (\ce{H2P}) is divided into an inner and
an outer branch at the pyrrole rings as shown in Fig.\ \ref{fig:porphyrins}(a),
which means that all $\pi$ orbitals participate in its ring-current
pathways.\cite{Fliegl:12,Fliegl:18} Calculations of the orbital contributions to
the current density showed that the current density from four $\pi$ electrons
dominates the ring current of \ce{H2P}.  The remaining 22 $\pi$ electrons do
not contribute significantly to the ring current.\cite{Steiner:02} Integration
of the ring-current strengths along different routes shows that the
current-density pathway {\it via} the inner hydrogen atoms is slightly weaker
than along the outer route. The ring currents along the inner and outer branch
at the pyrrole rings without an inner hydrogen are of about the same
strength.\cite{Fliegl:12,Fliegl:18}

\noindent Individual current-density pathways can be identified by tracing the
streamlines and integrating the strength of the current density along different
routes. GIMIC calculations account for the electronic response of the molecule
to the external magnetic field and avoid assumptions of conjugation pathways,
which is necessary when employing the semi-empirical annulene
model.\cite{Aihara:12} 

\noindent Isophlorin is a free-base porphyrin with four inner hydrogens
fulfilling H{\"u}ckel's rule for antiaromaticity. Isophlorin was predicted 60
years ago and air-stable isophlorins with four oxygen atoms or with two
oxygen and two sulphur atoms replacing the inner \ce{NH} moieties were recently
synthesized.\cite{Woodward:60,Reddy:08,Reddy:17} Calculations of the current
density of the synthesized isophlorins showed that they are indeed strongly
antiaromatic, sustaining a strong paratropic ring current around the
porphyrinoid ring.\cite{Taubert:11b,Valiev:13} The integrated ring-current
strengths of dioxa-dithiaisophlorin in Fig.\ \ref{fig:porphyrins}(b) show that
the ring current is completely dominated by paratropic contributions. The ring
current  splits into inner and outer branches of almost equal strength at the
furan and the thiophene rings. Thus, all $\pi$ orbitals contribute to the
ring-current pathway. The current-density flux of the outer pathways is
slightly weaker than the inner ones. The current density is diatropic only at
the outermost edge of the isophlorin ring, whereas paratropic current-density
contributions completely dominate the inner part.\cite{Fliegl:18,Valiev:13}

\begin{figure}[H] 
    \centering
\subfigure[]{\includegraphics[width=0.90\linewidth]{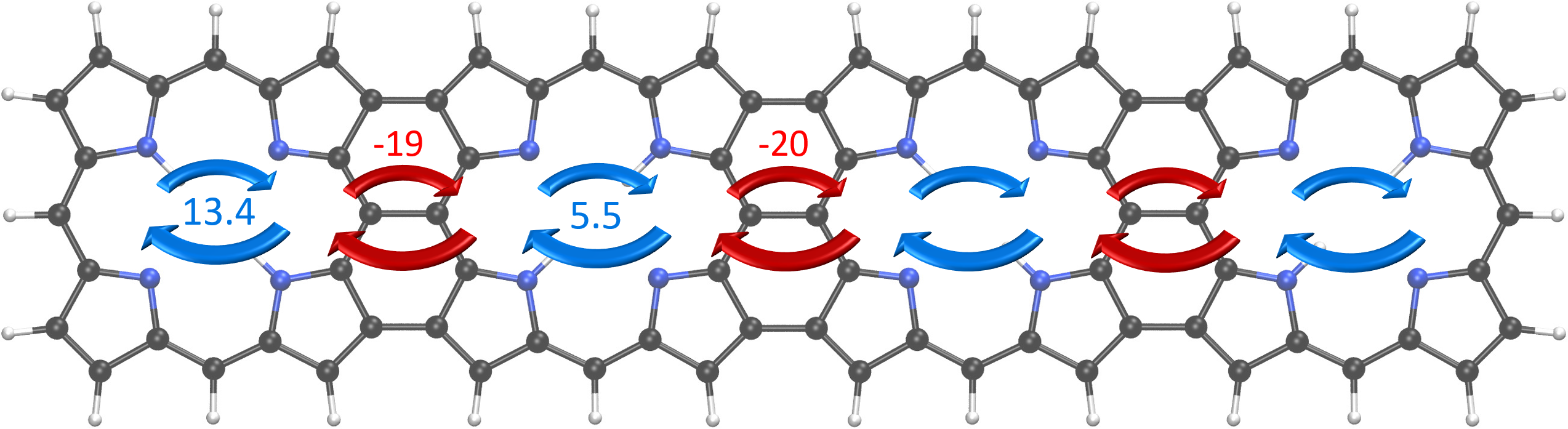}}\\ 
\subfigure[]{\includegraphics[width=0.90\linewidth]{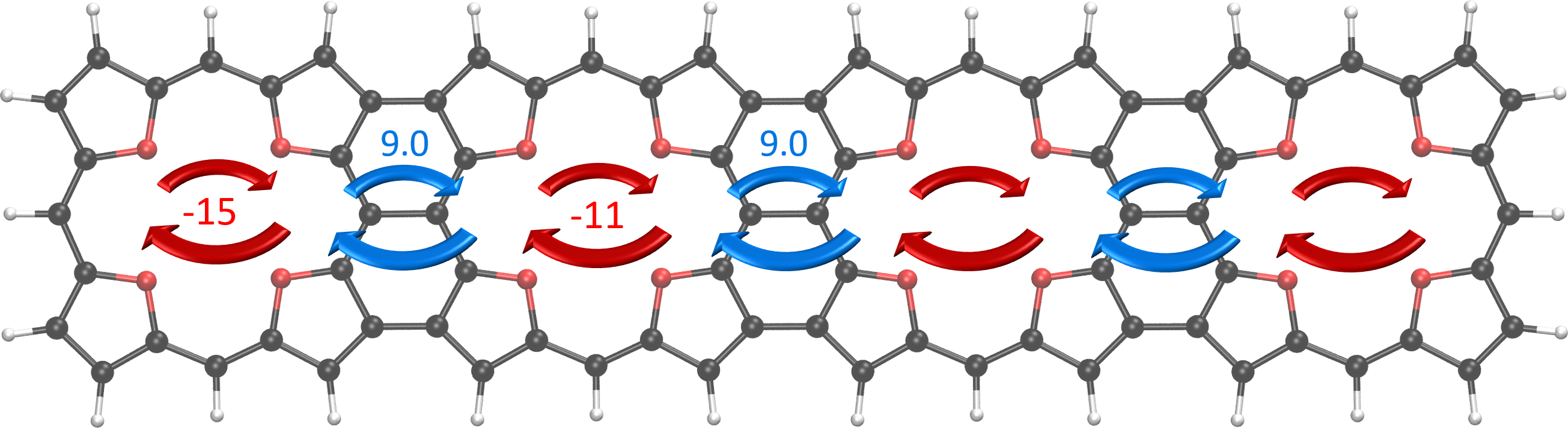}} 

\caption{Alternating diatropic and paratropic ring currents in annelated (a)
porphyrin and (b) isophlorin arrays. The blue arrows denote the local diatropic
ring currents and the local paratropic ones are marked with red arrows. The
strengths of the local ring currents in \nAT{} are also given.\cite{Valiev:20}
\label{fig:alternatingringcurrents} }

\end{figure}

\noindent Linear $\beta$-$\beta$, \meso-\meso, $\beta$-$\beta$-linked porphyrin
arrays have been synthesized.\cite{Tsuda:01,Ikeda:09} The linking of the
porphyrin rings leads to a molecular structure consisting of alternating
porphyrin rings and naphthalene moieties as shown in Fig.\
\ref{fig:alternatingringcurrents}(a). Current-density calculations showed that
the porphyrin rings sustain diatropic ring currents, whereas the
naphthalene-pyrrole moieties sustain paratropic ring currents.\cite{Valiev:20}
The analogous linear $\beta$-$\beta$, \meso-\meso, $\beta$-$\beta$-linked
isophlorin arrays shown in Fig.\ \ref{fig:alternatingringcurrents}(b) have not
been synthesized. Current-density analysis showed that the isophlorin arrays
also have alternating ring currents with paratropic ring currents in the
isophlorin rings and diatropic ring currents in the naphthalene
rings.\cite{Valiev:20} Even though isophlorin is paramagnetic, the isophlorin
arrays are diamagnetic, because the paratropic ring currents in the isophlorins
are weaker than \nAT{-20}, which is the lower threshold of the paratropic
ring-current strength of closed-shell paramagnetic porphyrin
rings.\cite{Valiev:20} The diatropic ring currents of the naphthalene moieties
also increase the diamagnetic contribution to the
magnetisability.\cite{Valiev:20} Large paramagnetic contributions would be
needed from all isophlorin rings to make the isophlorin arrays paramagnetic,
since the diamagnetic contribution to the magnetisability scales approximately
linearly with the size of the molecule.\cite{Pascal:10}

\subsection{Möbius-twisted molecules}
 
The topology of twisted molecular rings is characterised by the linking number,
$L_k$, which is equal to the sum of the twist $T_w$ and the writhe
$W_r$.\cite{Calugareanu:61,Pohl:68,White:69,Fuller:71} $T_w$ is a local
property, representing the one-dimensional twist of the molecular frame,
whereas $W_r$ is a measure of the global bending and deformation of the
molecular ring in 3D space.\cite{Rappaport:08,Wirz:18} $T_w$ and $W_r$ can
take any value for a given $L_k$ as long the relation $T_w + W_r= L_k$ is
fulfilled. 

\noindent A singly-twisted M{\"o}bius ring has an $L_k$ value of $1\pi$, where
$\pi$ is often omitted for simplicity. Molecules with linking numbers of
$L_k$ and $-L_k$ have the same topology but different
chirality.\cite{Rappaport:08} The H{\"u}ckel rule for even-twisted M{\"o}bius
molecules with $\modulus{L_k} =0,2,4,\ldots$ states that a ring is aromatic when
it has $(4n+2)~\pi$ electrons in the ring, and antiaromatic with $4n~\pi$
electrons. The opposite rule holds for rings with an odd $L_k$ value. 

\begin{figure}[H]
    \centering
\subfigure[]{\includegraphics[height=5cm]{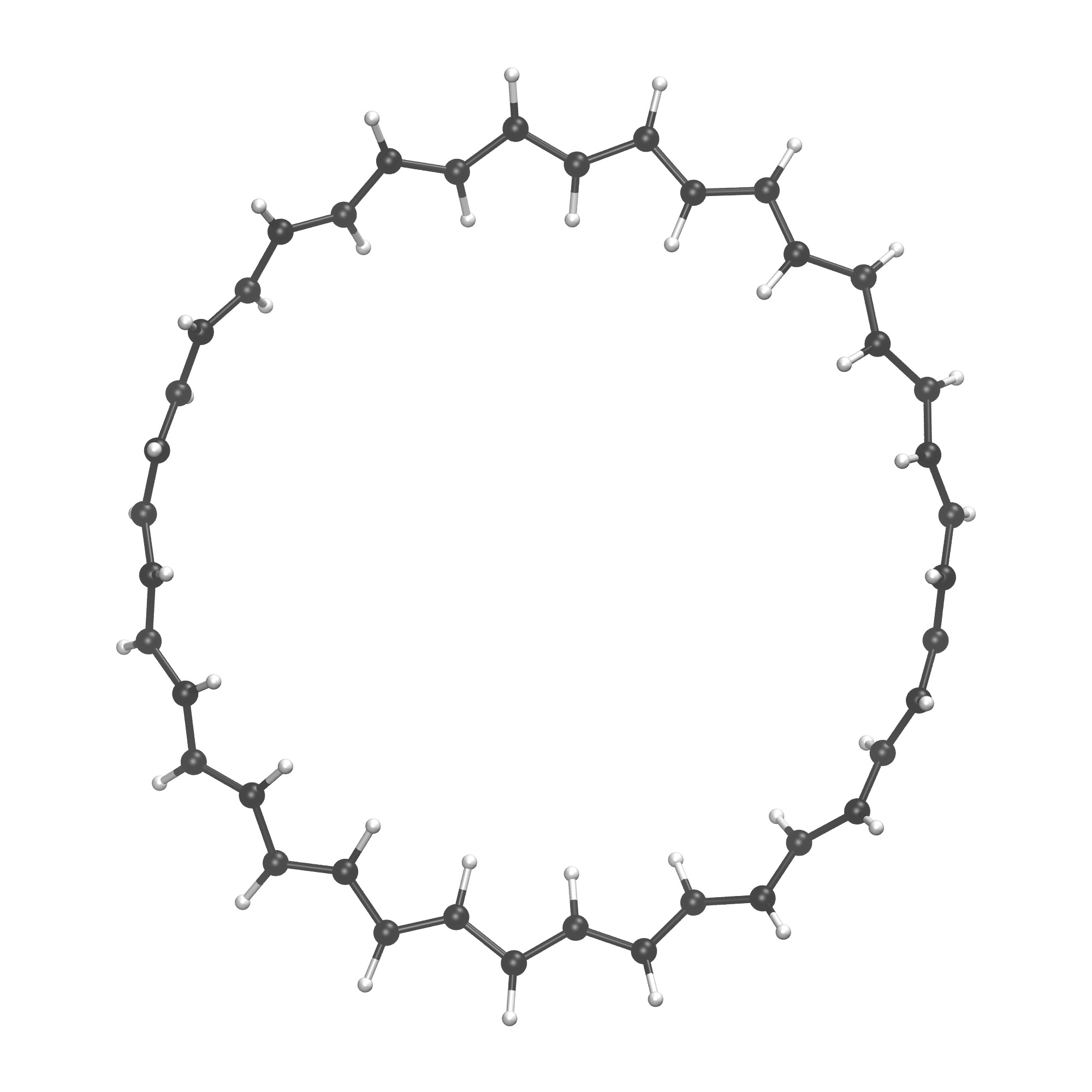}}
\hspace{11mm}
\subfigure[]{\includegraphics[height=5cm]{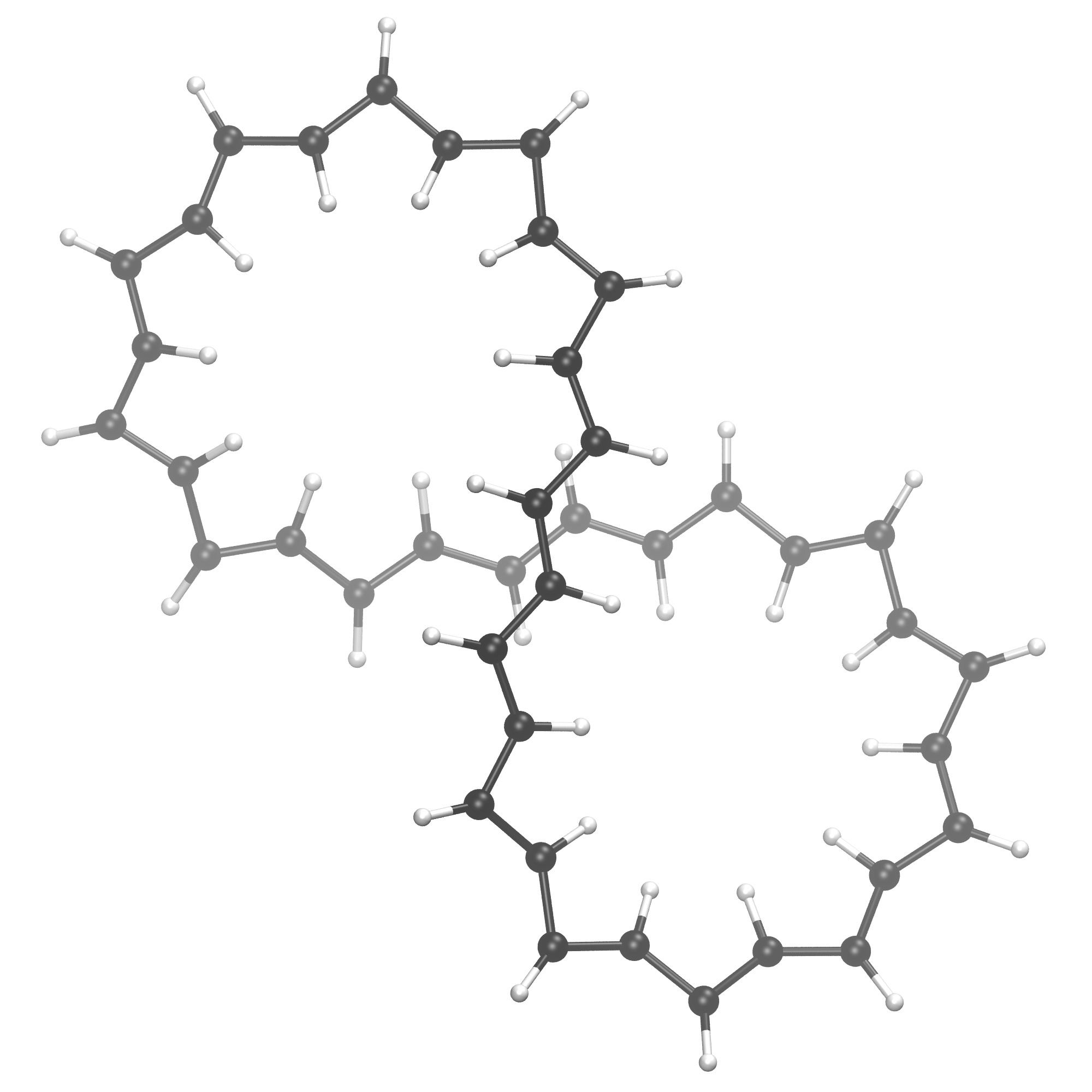}} 

\caption{(a) A planar double-twisted \trans-\ce{C40H40} ring with $L_k=2$,
    $T_w=2$ and $W_r=0$ and (b) a figure-eight-shaped \trans-\ce{C40H40} ring with
    $L_k=2$, $T_w=0.56$ and $W_r=1.44$. }

\label{fig:twistandwrite}
\end{figure}

\noindent In Fig.\ \ref{fig:twistandwrite}, two doubly M{\"o}bius-twisted
conformers of \trans-\ce{C40H40} are shown. The one with the shape of a circle
is doubly twisted without writhe, whereas the figure-eight structure is
less twisted but with a large writhe. M{\"o}bius-twisted molecules are generally
non-planar which means that there is no obvious direction for the external
magnetic field. The aromatic character can then be estimated by choosing the
direction of the magnetic field that leads to the largest projection area of
the ring, which is expected to yield the strongest ring-current strength.

\noindent Calculations on \trans-annulene rings and their dications with a
different topology showed that the strength of the ring current follows the
generalised H{\"u}ckel rule for M{\"o}bius-twisted molecules.\cite{Wirz:18}
However, the degree of aromaticity depends strongly on the partitioning of
$L_k$ between $T_w$ and $W_r$. The strongest ring currents were obtained for
the most twisted molecules with the smallest $W_r$, regardless of whether the
molecule is aromatic or antiaromatic. The most deformed rings with the largest
$W_r$ were found to be almost non-aromatic. The energetically lowest formally
aromatic and antiaromatic molecules were the most twisted structures that also
sustain the strongest diatropic and paratropic ring currents, respectively.

\noindent The first synthesized Möbius-twisted [16]annulene with $L_k=1$ was
suggested to be aromatic based on spectroscopic studies.\cite{Ajami:03} However,
calculations of the nucleus independent chemical shifts (NICS)\cite{Castro:05} and
ring-current strengths\cite{Taubert:09} showed that both the Möbius-twisted and
the corresponding untwisted (H{\"u}ckel) isomers are non-aromatic sustaining a
ring current of about \nAT{0.3}. Due to the twist of the M{\"o}bius isomer, the
carbon atoms at the opposite sides of the [16]annulene are close to each other,
enabling the current density to make a weak through-space
shortcut.\cite{Taubert:09}

\noindent Möbius-twisted expanded porphyrins with $L_k=1$ and $L_k=2$ have been
synthesized.\cite{Higashino:10,Shimizu:06} Spectroscopic studies of molecules
with a different number of $\pi$ electrons suggested that the aromatic
character of the synthesized hexaphyrins with $L_k=2$ follows the generalised
H{\"u}ckel rule for Möbius-twisted molecules.\cite{Shimizu:06,Rzepa:08}
Calculations of the current density of the doubly Möbius-twisted
\meso-trifluoromethyl-substituted hexaphyrins with formally 24, 26, 28 and
$30~\pi$ electrons showed that [26]hexaphyrin and [30]hexaphyrin are aromatic,
sustaining diatropic ring currents, whereas [28]hexaphyrin is antiaromatic and
[24]hexaphyrin is practically non-aromatic, sustaining a very weak paratropic
ring current.\cite{Fliegl:10} Singly Möbius-twisted monophosphorous complex of 
[28]hexaphyrin and bisphosphorous complex of [30]hexaphyrin were synthesized using
phosphorous oxide (\ce{PO}) moieties to force a single twist into the
hexaphyrin ring.\cite{Higashino:10} The second \ce{PO} introduces a strong
twist ($T_w=1.72$), which is compensated by deforming the ring with a writhe of
$-0.72$.  The second \ce{PO} unit also increases the number of $\pi$ electrons
in the conjugation pathway. The singly-twisted [28]hexaphyrin with $L_k=1$ is
expected to be aromatic according to the generalised H{\"u}ckel rule and
[30]hexaphyrin is antiaromatic. The aromatic character of the singly-twisted
hexaphyrins was confirmed by calculating the ring-current
strength.\cite{Fliegl:11c} The paratropic ring current of the bisphosphorous complex of 
[30]hexaphyrin prefers the inner pathways but it also takes routes across the
deformed molecular ring because the atoms on the opposite sides of the ring
come close to each other due to the deformation of the ring.

\subsection{Carbon nanostructures}

\subsubsection{Fullerene \ce{C60}}

\begin{figure}
    \centering
    \subfigure[Outside \ce{C60}]{\includegraphics[width=0.32\textwidth]{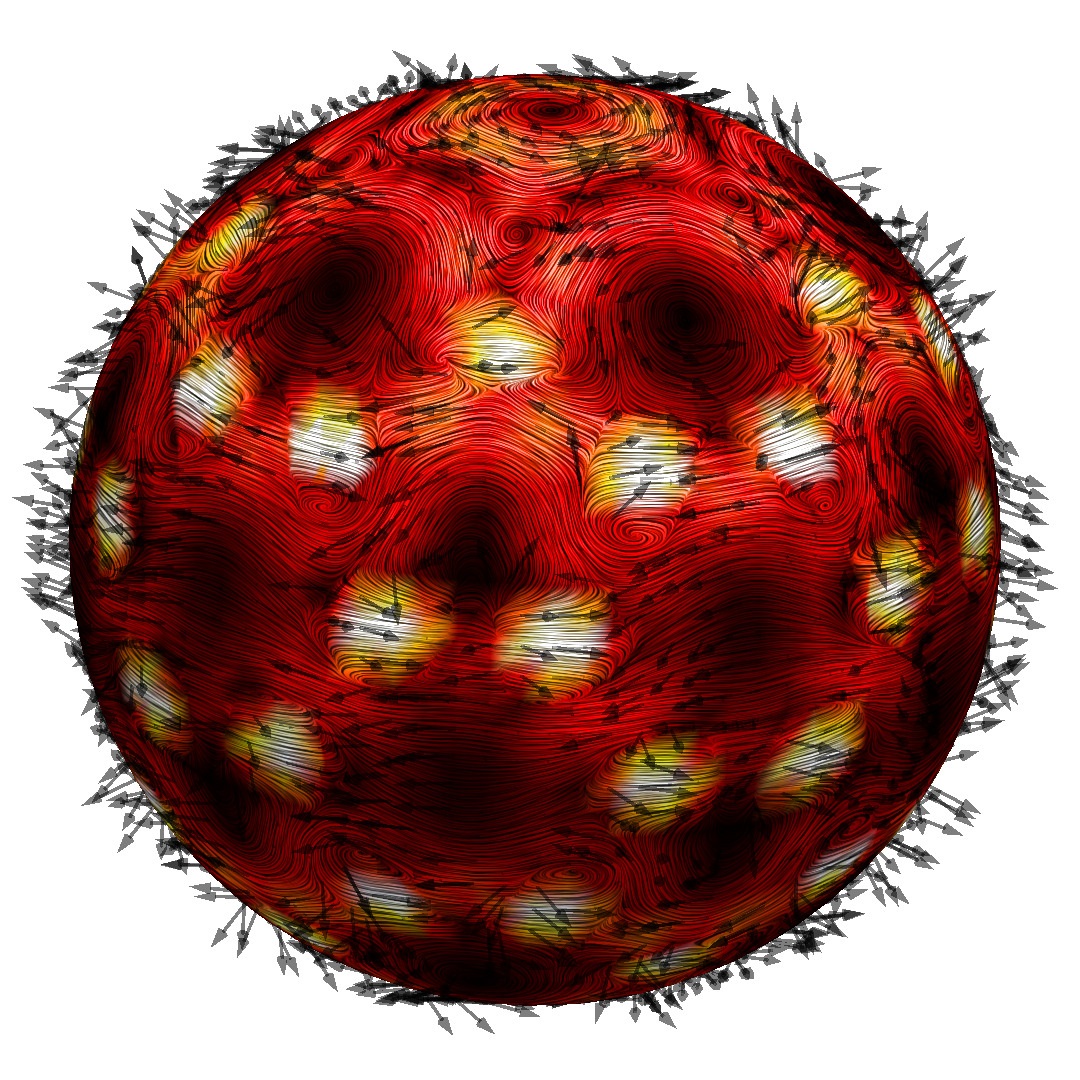}\label{fig:C60-0-outside}}
\hspace{0mm}
    \subfigure[Inside \ce{C60}]{\includegraphics[width=0.32\textwidth]{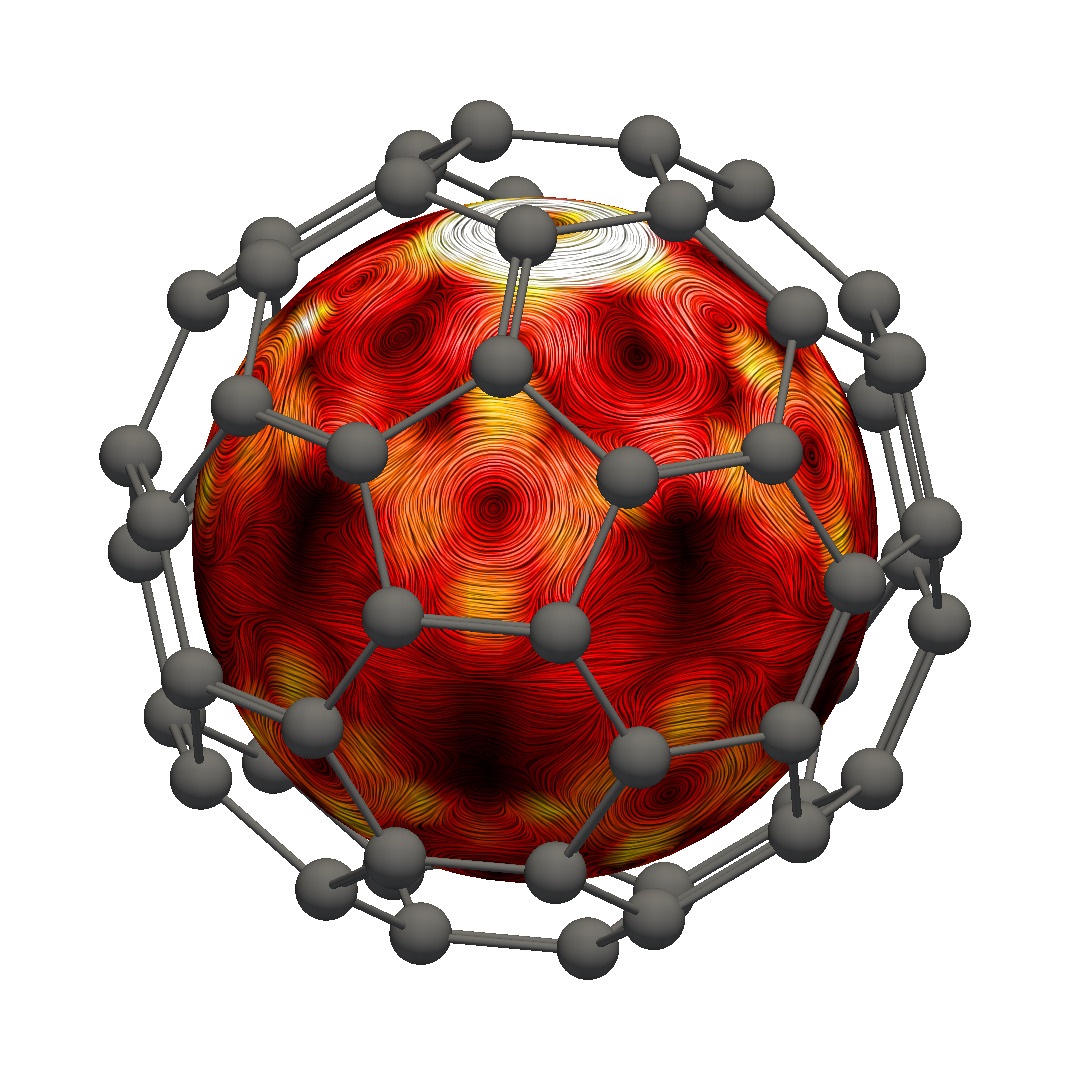}\label{fig:C60-0-inside}}
\hspace{0mm}
    \subfigure[Streamlines seen from above in \ce{C60}]{\includegraphics[width=0.3\textwidth]{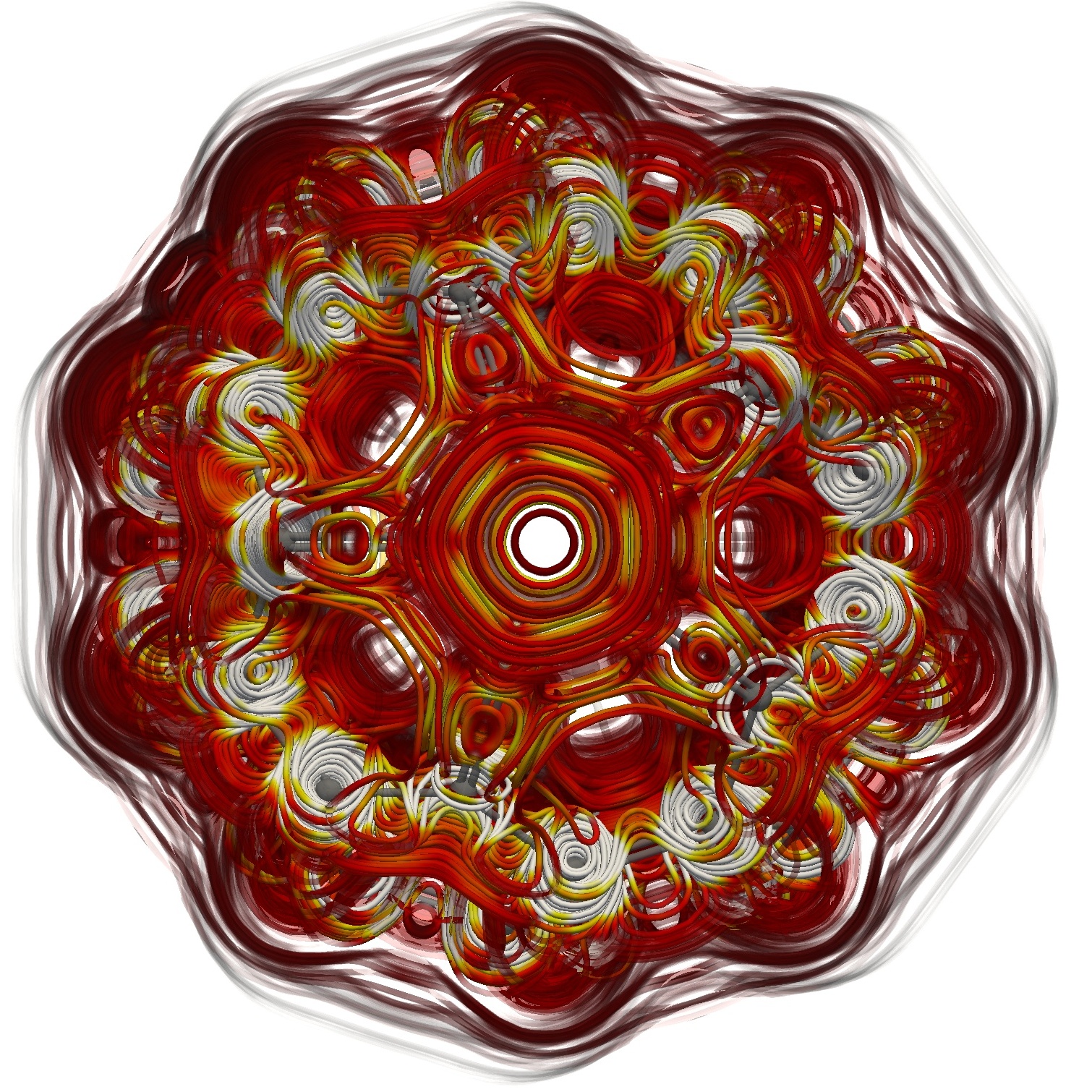}\label{fig:C60-0-spaghetti}}

\caption{The current density illustrated on a sphere with radius \bohr{1} larger (a) and small (b) than the surface of the \ce{C60} molecule, respectively. The
    colour scale indicates the strength of the current density in the range of
    $[3.57 \times 10^{-6}; 0.54 ]$ \nATA{}.  (c) The current density in the
    middle of \ce{C60}.  \label{fig:C60-currents}}

\end{figure}

In the seminal work by Kroto {\it et al.},\cite{Kroto:85} they stated that the
inner and outer surfaces of \ce{C60} are covered with a sea of $\pi$ electrons
and concluded that it appears to be aromatic.  The aromatic character of
fullerene (\ce{C60}) has since then been debated.  More recent studies of the
aromaticity of fullerene have reached the conclusion that \ce{C60} can
hardly be considered
aromatic.\cite{Elser:87,Zanasi:95b,Johansson:05,Chen:12,Munoz-Castro:15}
Fullerene sustains local current-density vortices that lead to a strong
diatropic current density on the outside and an equally strong paratropic
current density on the inside resulting in a weak net ring current 
around the whole molecule.\cite{Zanasi:95b,Johansson:05} Regardless of the
direction of the magnetic field, \ce{C60} has mainly local current-density
vortices, which appear as white spots in Fig.\ \ref{fig:C60-currents}. 

\noindent The current density of \ce{C60} is illustrated in Fig.\
\ref{fig:C60-currents} on a sphere with radius \bohr{1} larger and smaller than
the radius of the molecular surface. The paratropic ring current inside
\ce{C60} and the diatropic ring current outside it as well as the local
vortices in the middle of the molecule are seen in Fig.\
\ref{fig:C60-currents}(c).

\noindent The \ce{C60} fullerene fulfils the condition for spherical aromaticity when ten
electrons are removed. Approximately spherical molecules are spherical aromatic
when the number of $\pi$ electrons is $2(n+1)^2$,\cite{Hirsch:00,Buehl:01,Chen:05b}
which is analogous to the closed-shell configurations of pseudoatoms with a
spherical shell potential.\cite{Reiher:03} Current-density calculations showed
that \ce{C60^{10+}} is spherical aromatic sustaining strong diatropic
contributions to the current density around the entire molecule. 

\begin{figure}
    \centering
    \subfigure[$B \perp$ to five-membered ring]{\includegraphics[width=0.45\textwidth]{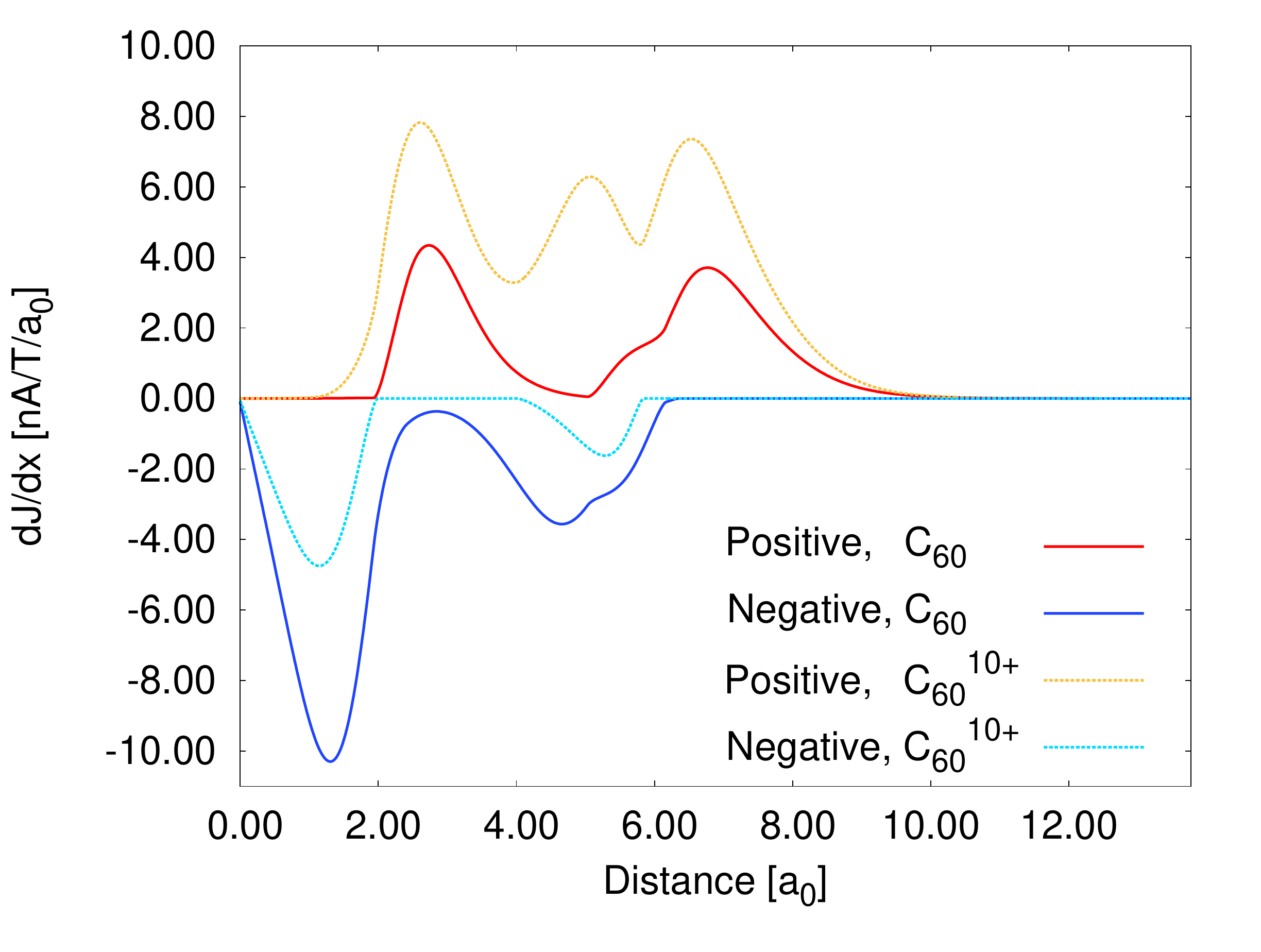}\label{fig:C60-profile-pentagon}}
    \subfigure[$B \perp$ to six-membered ring]{\includegraphics[width=0.45\textwidth]{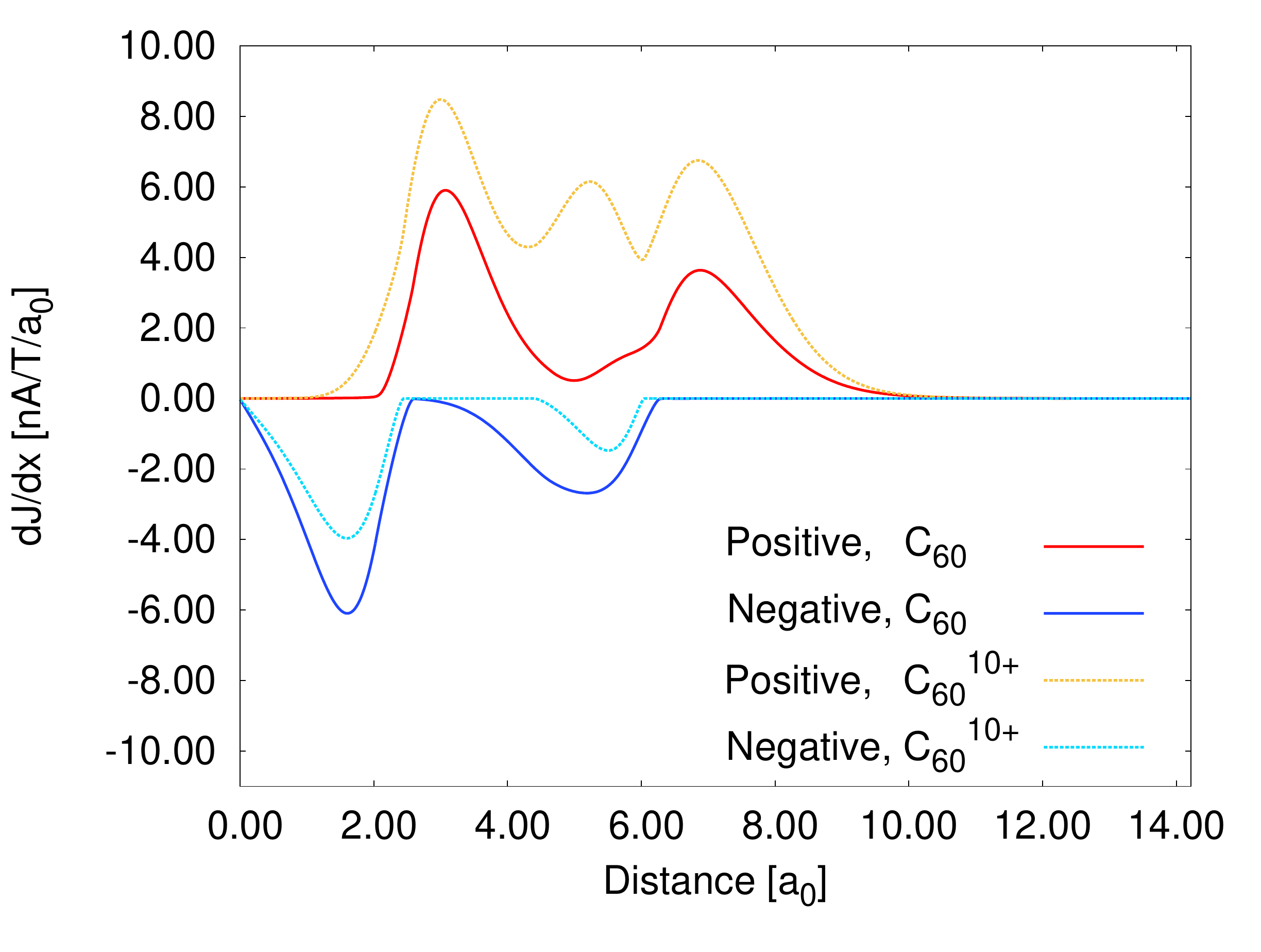}\label{fig:C60-profile-hexagon}}

\caption{The profile of the current density of \ce{C60} and \ce{C60^{10+}} when
    the magnetic field is perpendicular to (a) a five-membered ring (b) a
    six-membered ring. \label{fig:C60-profiles}}

\end{figure}

\noindent The profile of the current density passing through the upper half of
\ce{C60} and \ce{C60^{10+}} are shown in Fig.\ \ref{fig:C60-profiles} where one
sees that the current-density profiles are fairly independent of the direction
of the magnetic field.  The local ring current in the five-membered rings of
\ce{C60} is strongly paratropic when the magnetic field is perpendicular to
that ring.  There is a diatropic current-density flux in the six-membered ring
adjacent to the five-membered one.  When the magnetic field is perpendicular to
a six-membered ring, it sustains a local paratropic ring current inside it as
in benzene.  The local paratropic ring current of the five-membered ring is
much stronger than the one inside the six-membered one. Fullerene is
non-aromatic, since it sustains alternating diatropic and paratropic
contributions to the current density, whose strengths nearly cancel, whereas
\ce{C60^{10+}} is dominated by a strong global diatropic current-density flux.

\begin{figure}
    \centering
    \subfigure[Outside \ce{C60^{10+}}]{\includegraphics[width=0.32\textwidth]{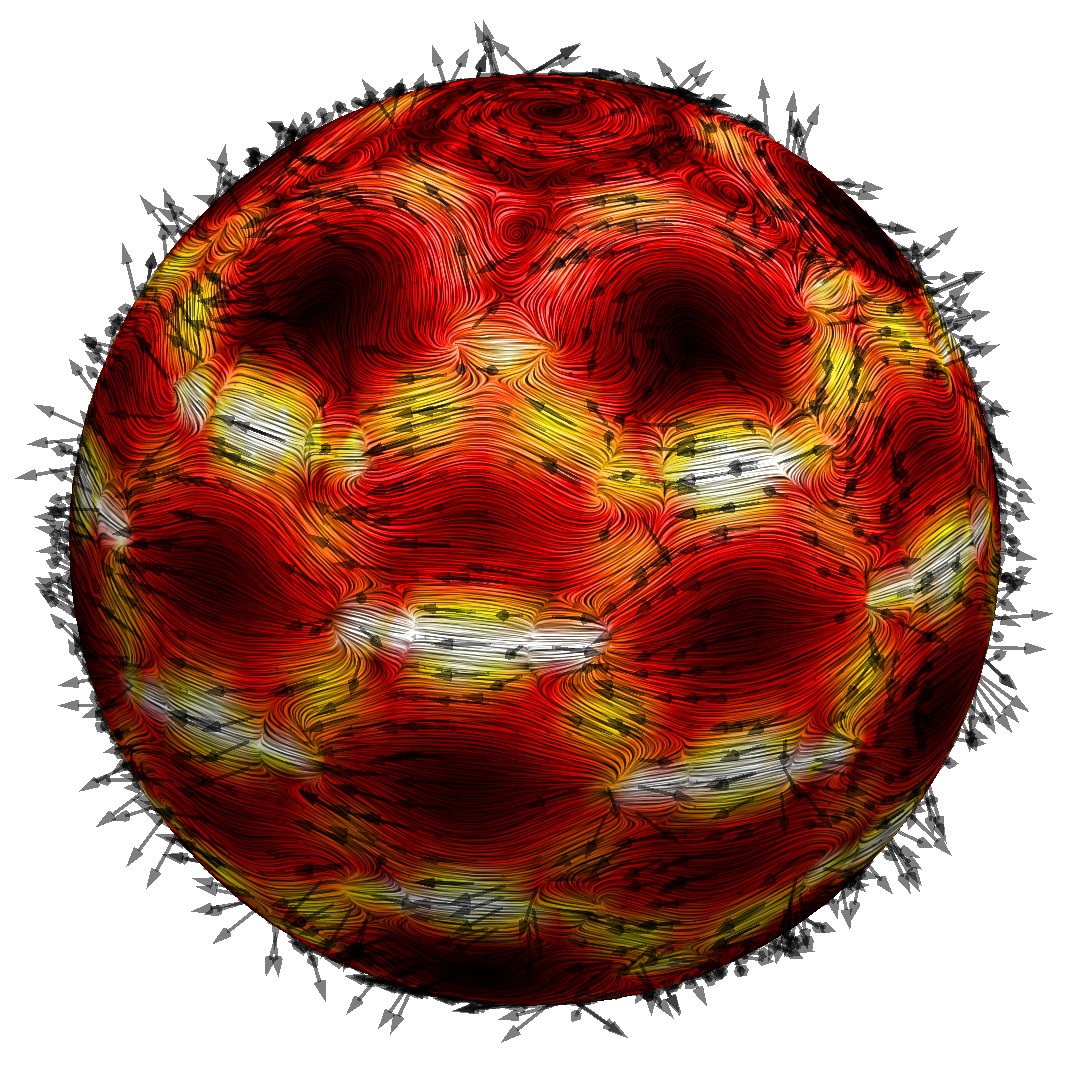}\label{fig:C60-10-outside}}
\hspace{0mm}
    \subfigure[Inside \ce{C60^{10+}}]{\includegraphics[width=0.32\textwidth]{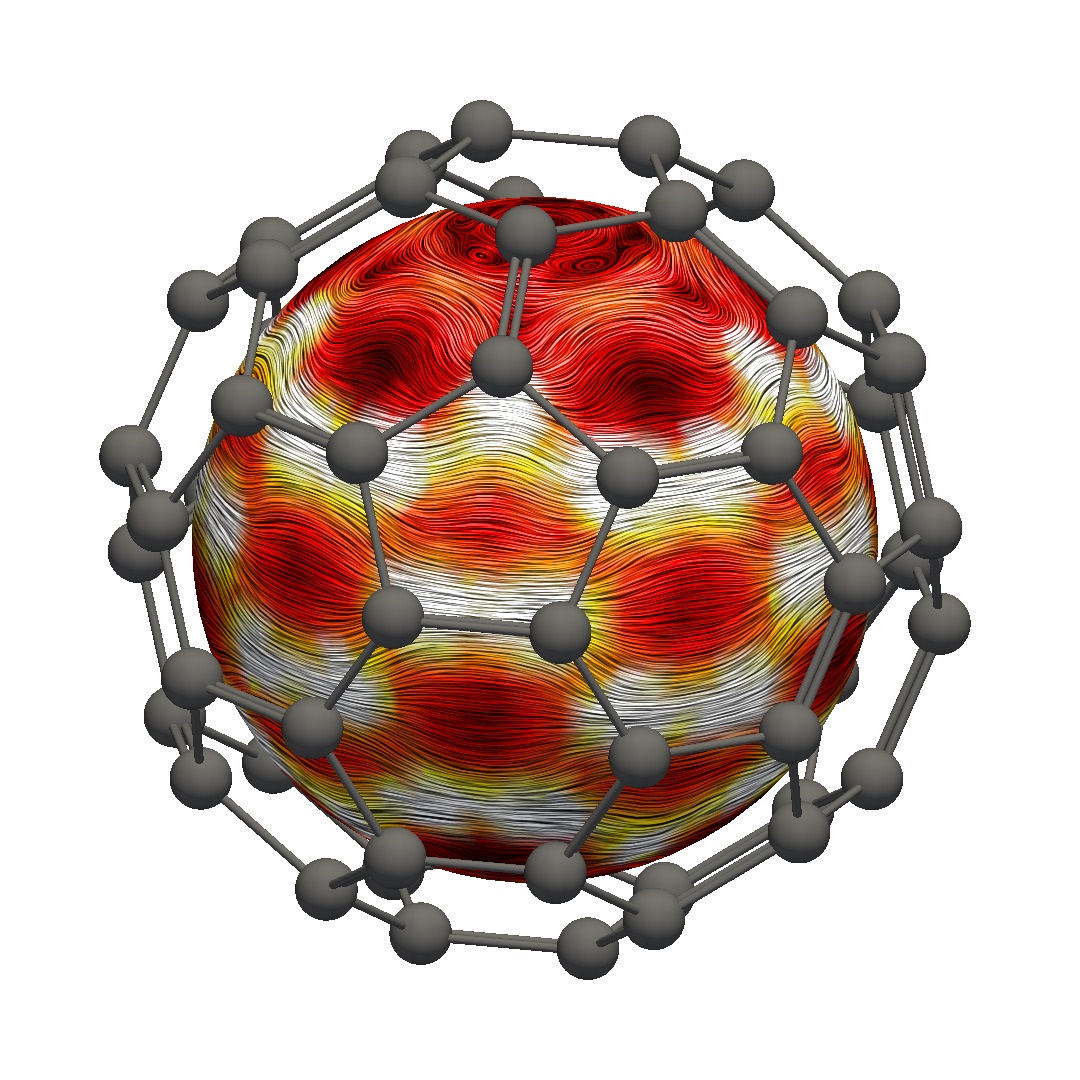}\label{fig:C60-10-inside}}
\hspace{0mm}
    \subfigure[Streamlines seen from above in \ce{C60^{10+}} ]{\includegraphics[width=0.3\textwidth]{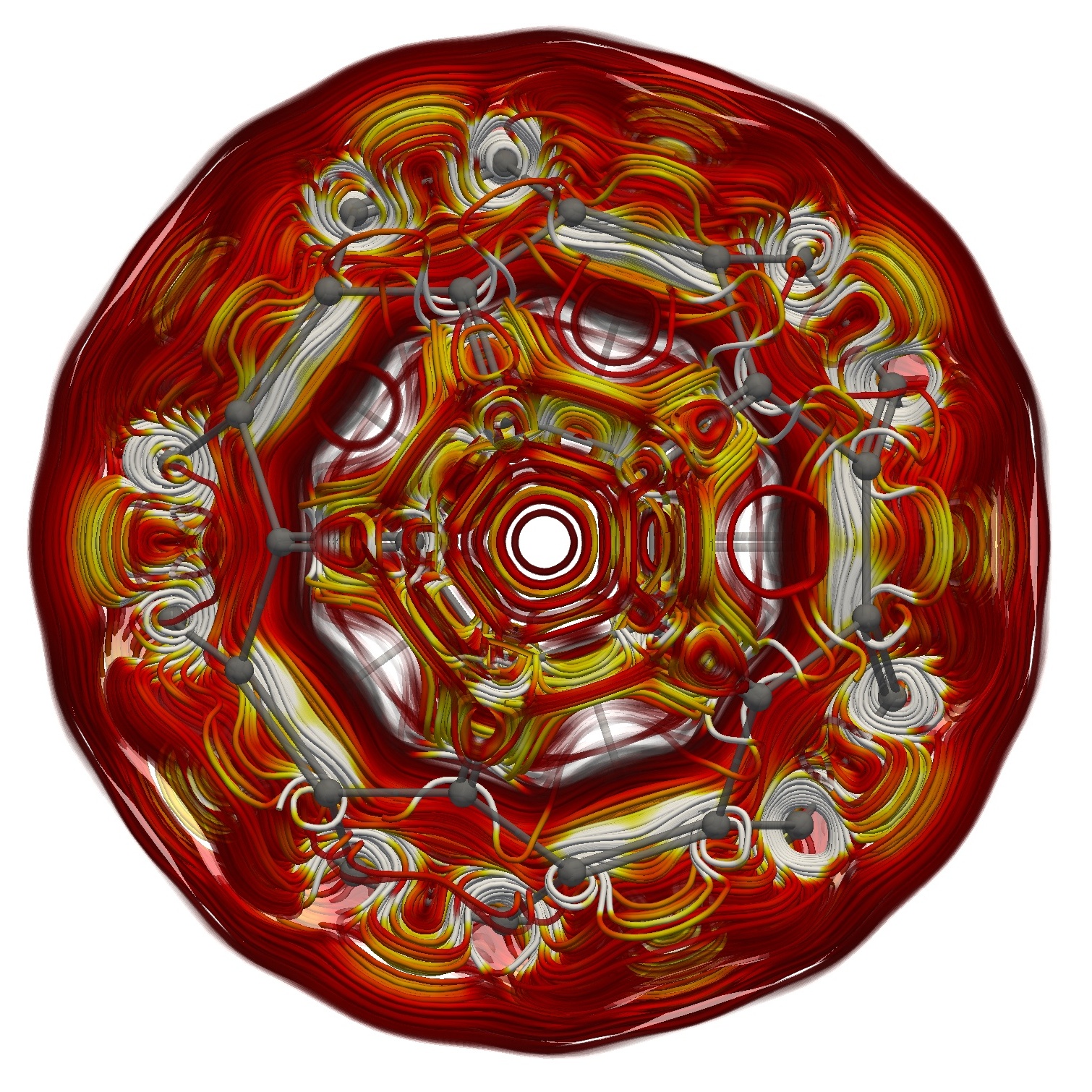}\label{fig:C60-10-spaghetti}}

\caption{The current density illustrated on a sphere with a radius \bohr{1} (a)
    larger and (b) smaller than the surface of the \ce{C60}$^{10+}$ molecule.
    The colour scale indicates the strength of the current density in the range
    of $[3.57 \times 10^{-6}; 0.54 ]$ \nATA{}.  (c) The current density in the
    middle of \ce{C60}$^{10+}$.  \label{fig:C60-10+-currents}}

\end{figure}

\noindent The current-density flux of \ce{C60^{10+}} is diatropic both inside
and outside the carbon framework as seen in Fig.\ \ref{fig:C60-10+-currents}.
The uniform diatropic current-density flux seen in Fig.\
\ref{fig:C60-10+-currents}(c) leads to a constant magnetic shielding inside the
molecule like a magnetic Faraday cage.\cite{Johansson:04}

\begin{table}[H]

\caption{Calculated diatropic and paratropic contributions to the net
ring-current strength susceptibility (in \nAT{}) of \ce{C60}, \ce{C60^{10+}} and
gaudiene (\ce{C72}). The calculations were performed at the B3LYP/def2-TZVP
level. \label{tab:currents-cages}} 

\begin{tabularx}{\textwidth}{LCCC}
\hline
\hline
Molecule &  Diatropic & Paratropic & Net current strength \\
\hline
\ce{C60}$~^a$ & 25.40 & -41.50& -16.00 \\
\ce{C60}$~^b$ & 31.74 & -26.46& ~~~5.26 \\
\ce{C60^{10+}}$^,~^a$ & 68.20 & -14.26 & 43.84 \\
\ce{C60^{10+}}$^,~^b$ & 71.26 & -13.22 & 58.04 \\
\ce{C72}$~^c$ & 56.82 & -10.98 & 45.84 \\
\hline
\hline
\end{tabularx}
$^a$ The magnetic field is perpendicular to a five-membered ring. \\ 
$^b$ The magnetic field is perpendicular to a six-membered ring. \\
$^c$ The magnetic field is perpendicular to a four-sided equilateral twelve-membered ring. 
\end{table}

\subsubsection{Gaudiene}

\noindent The cavernous gaudiene (\ce{C72}) molecule belonging to the
$O_\mathrm{h}$ point group consists of two different kinds of conjugated
twelve-membered rings.\cite{Sundholm:13,Sundholm:15,Rauhalahti:16} One of them
has alternating single, double and triple bonds and the other one is an
equilateral four-sided ring with single, triple and single bonds along the
sides. The molecular structure of gaudiene can formally be constructed from a
truncated octahedron by replacing eight of the twelve edges of the octahedron
with \ce{-C#C-} units.\cite{Sundholm:15} Gaudiene is aromatic, sustaining a
strong diatropic current-density flux of \nAT{45.84} around the cage structure.
The net current-density flux was integrated by placing a plane through the
whole molecule such that it begins at the centre of an equilateral
twelve-membered and continues outwards. 

\begin{figure}[H]
    \centering
\subfigure[Outside \ce{C72}]{\includegraphics[width=0.40\textwidth]{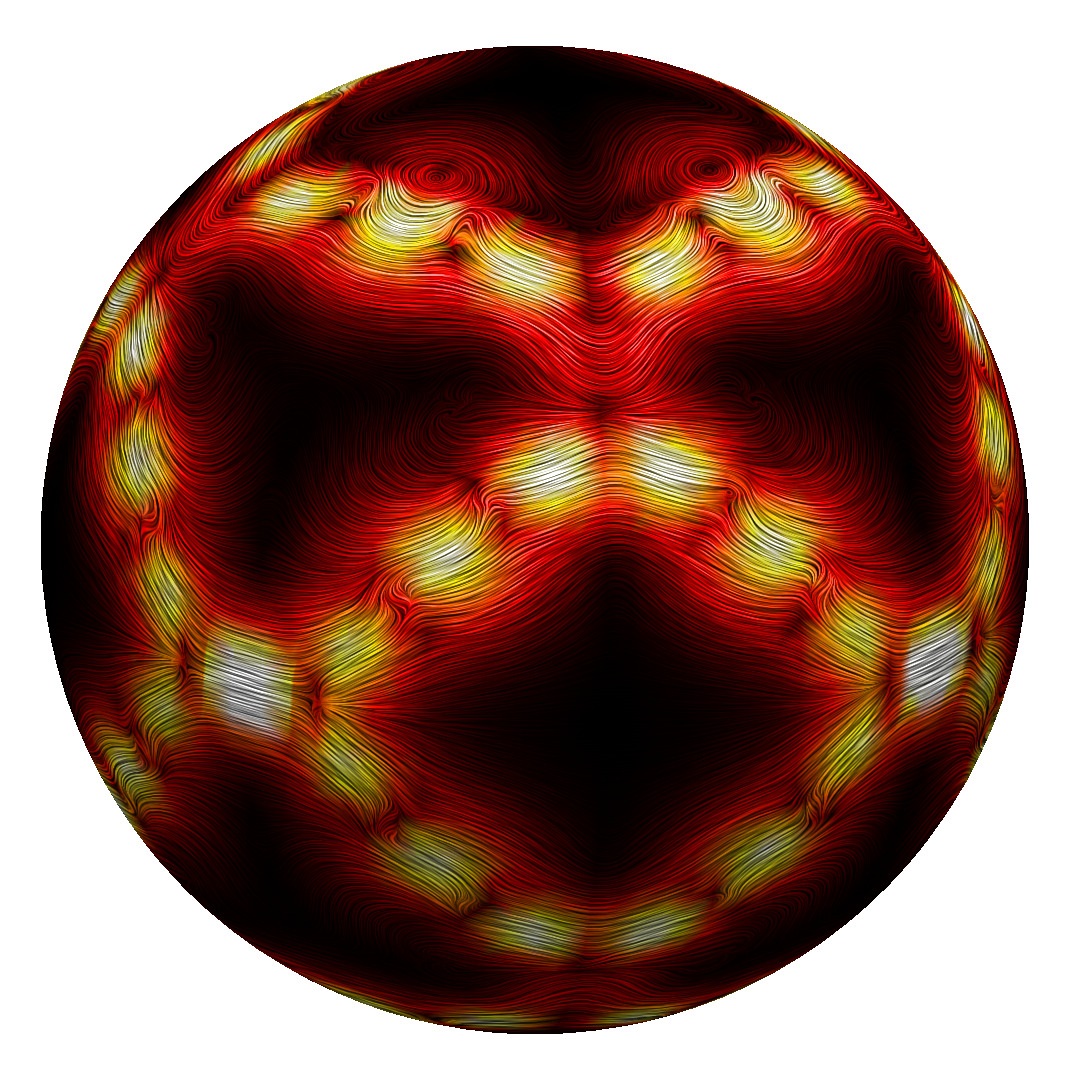}}
\hspace{11mm}
\subfigure[Inside \ce{C72}]{\includegraphics[width=0.40\textwidth]{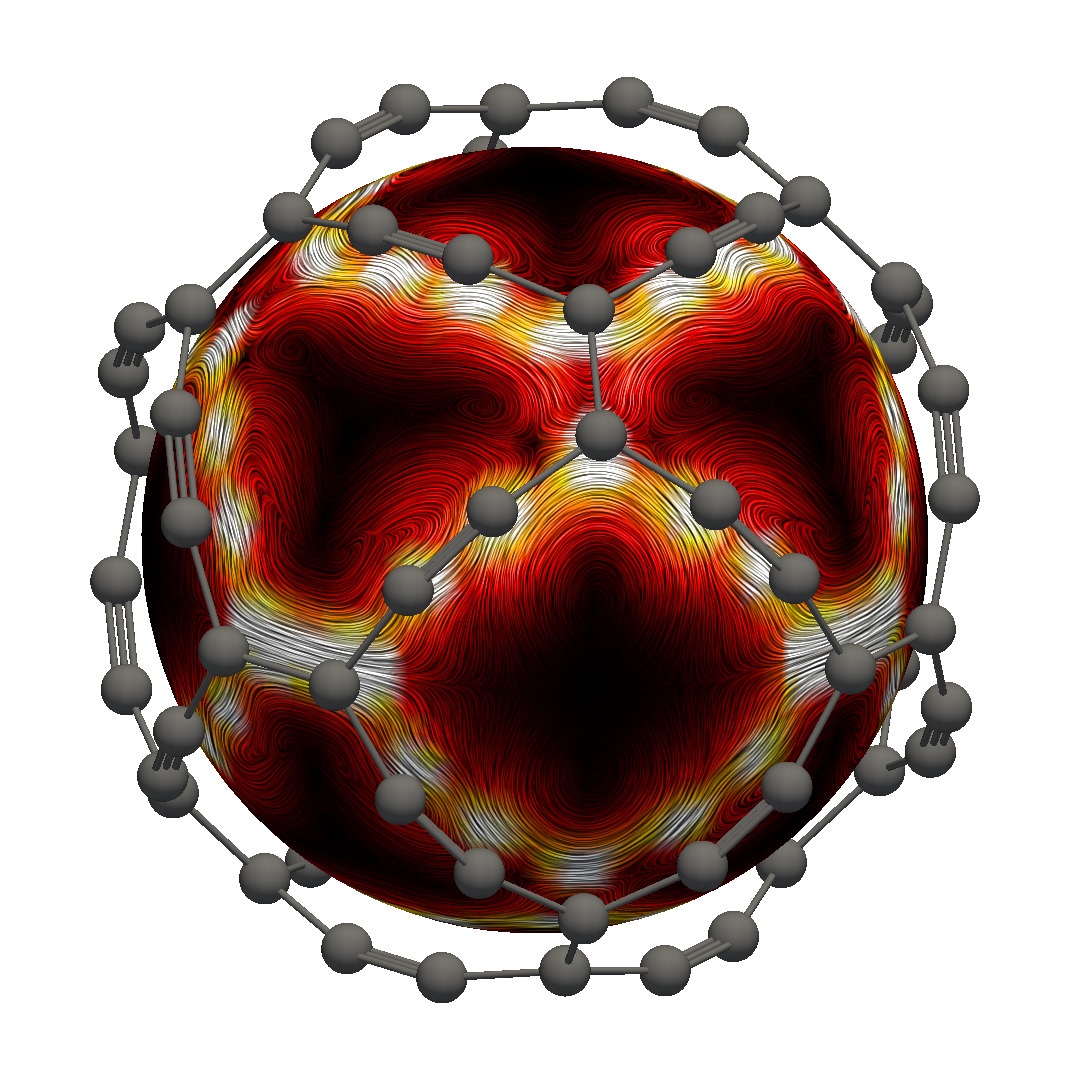}}

\caption{The current density of gaudiene illustrated on the surface of a sphere
at a distance of about \bohr{1} from the carbon framework (a) outside and (b)
inside of the \ce{C72} the cage. }

\label{fig:gaudiene}
\end{figure}

\noindent Even though gaudiene with 72 $\pi$ electrons fulfils the $2(n+1)^2$
rule of spherical aromaticity, it cannot be considered spherical aromatic
because the current density mainly follows the carbon framework as shown in
Fig.\ \ref{fig:gaudiene}, rather than forming a uniform current density on both
sides of the carbon framework as in the case of
\ce{C60^{10+}}.\cite{Sundholm:15,Rauhalahti:16} This is understandable, since
the cavities of \ce{C72} are significantly larger than in fullerene that is
constructed from five- and six-membered carbon rings.  \ce{C72} also lacks
dense horizontal pathways for the current density as in \ce{C60^{10+}}. 

\subsubsection{Toroidal carbon nanotubes}

\noindent Extensive computational and experimental studies of toroidal carbon
nanotubes (TCNT) were initiated after they were observed and unequivocally
confirmed.\cite{Sarkar:95,Endo:95,Terrones:96,Liu:97,Martel:99} The radii of
the observed TCNTs are about $300-400 \, \mathrm{nm}$, which is too large for
first-principle electronic structure calculations. TCNTs are expected to
exhibit unusual magnetic properties due to their topology. An external magnetic
field perpendicular to the major radius of the torus can induce a
current-density flux around the perimeter of the torus. In chiral tori, the
current-density pathways can also follow the nanotube forming the torus, giving
rise to an induced toroidal magnetic field perpendicular to the applied one inside the nanotube.
The induced magnetic field inside the volume of the torus along the axis of the
nanotube has no north and south poles. The current density around the nanotube
also leads to a magnetically induced anapole moment perpendicularly to the major
radius of the torus.\cite{Zeldovich:57,Ceulemans:98,Gray:10} Two enantiomers of
chiral TCNTs have an opposite sign of the anapole moment.\cite{Pelloni:11b}  

\noindent The magnetic properties of small TCNTs have been studied
computationally.\cite{Reiter:19,Ceulemans:98,Pelloni:11b,Papasimakis:09,Berger:12,Tellgren:13}
Optimisation of the molecular structure of TCNTs show that the cross-section
of the nanotube is elliptical, leading to reasonable \ce{C-C} distances on the
inside and the outside of the torus when the torus is large
enough.\cite{Reiter:19} Reasonable \ce{C-C} distances can be obtained for
smaller TCNTs by introducing seven-membered rings inside the torus and
five-membered rings on the outside of it.\cite{Reiter:19,Dunlap:92} 

\begin{figure}[H]
    \centering
    \includegraphics[width=0.75\linewidth]{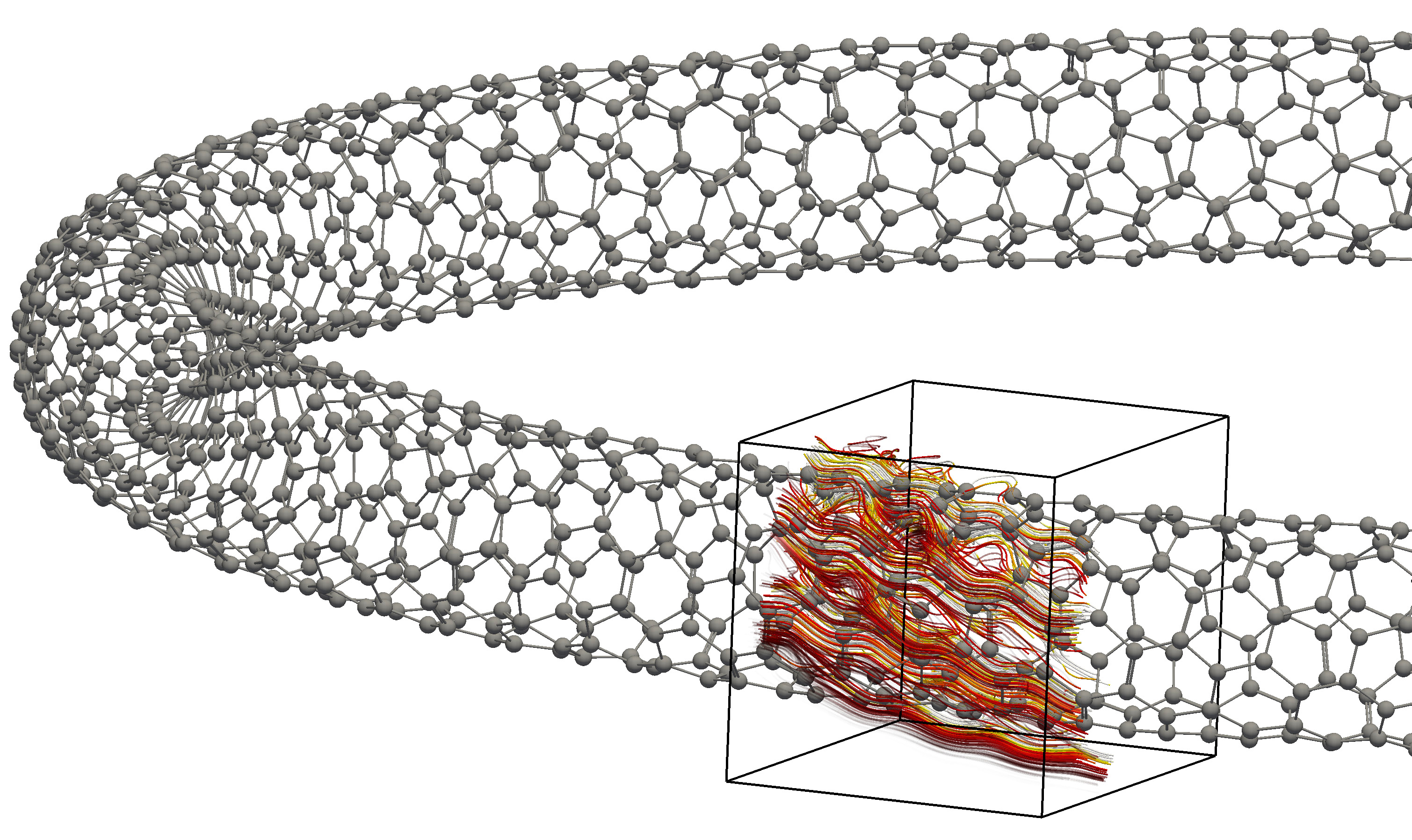}

\caption{The current density in the chiral polyhex $(6,3)$ toroidal carbon
nanotube \ce{C2016}.  \label{fig:torus} }

\end{figure}

\noindent Current-density calculations on the chiral polyhex $(6,3)$ \ce{C2016}
TCNT showed that it sustains a strong ring current of \nAT{292} along the
nanotube forming the torus and an even stronger ring current of \nAT{3307}
around it.\cite{Reiter:19} The current-density flux of \ce{C2016} is shown in
Fig.\ \ref{fig:torus}.  The chiral ring current around the nanotube results in
an induced magnetic field along the axis of the nanotube and a magnetically
induced anapole moment.  Calculations of the magnetically induced current
density of larger TCNTs are computationally expensive. The current density of
large TCNTs can be studied by using the computationally cheaper pseudo-$\pi$
model, which is a model system where the carbon atoms are replaced with
hydrogens.\cite{Fowler:02}

\subsection{Aluminium clusters}

Metal clusters may also sustain ring currents when they are exposed to an
external magnetic field. They can also be considered aromatic or antiaromatic
in accordance with the ring-current criterion, which is acceptable as long as
one is aware of the fact that aromaticity has a different meaning in organic
chemistry.\cite{Lazzeretti:04,Hoffmann:15} For example, the synthesized
\ce{Al4^{2-}} cluster is aromatic based on the tropicity and the strength of
the magnetically induced current
density.\cite{Li:01,Kuznetsov:03,Chen:03,Fowler:01a,Lin:05} The aluminium atoms
in \ce{Al4Li2} form a square-planar ring with a diatropic current-density
vortex in the centre of the ring. The vortex reaches out to the outer edge of
the \ce{Al4} ring where it becomes the global diatropic ring current.
Integration of the current density from the centre of the ring through a bond
and excluding the atomic current-density vortices of the lithium atoms yielded
a net current strength of \nAT{25.6}, which is twice the net ring-current strength
of benzene. No paratropic ring current was found. The \ce{Al4^{4-}} ring is
rectangular with alternating bond lengths, which suggests that it is
antiaromatic. Current-density calculations at the coupled-cluster level of
theory showed that the ring current has strong diatropic and paratropic
contributions.\cite{Chen:03,Lin:05} The paratropic contribution to the ring
current is slightly stronger leading to a net paratropic ring-current strength
of \nAT{-3.1} for \ce{Al4Li4}, indicating that it is non-aromatic or weakly
antiaromatic according to the ring-current criterion.  The \ce{Al4^{2-}} ring
is $\pi$- and $\sigma$-aromatic, whereas the \ce{Al4^{4-}} cluster sustains a
diatropic ring current in the $\sigma$ orbitals and a paratropic ring current
in the $\pi$ orbitals.\cite{Chen:03,Lin:05}

\subsection{Scope and limitations of the GIMIC method}

The GIMIC method is a powerful method for interpreting different aspects of
current-density susceptibilities and molecular magnetic properties. GIMIC is
under active development with new features being planned and implemented
continuously.  The current-density susceptibility is a tensor function with
nine components that are contracted to a vector function with three components
when specifying a given direction of the external magnetic field.  Vector
functions are complicated to visualise and to analyse. We have developed and
implemented novel computational and practical tools for interpreting calculated
current-density susceptibilities. Orbital contributions to the current density
are important, for example, when determining orbital contributions to nuclear
magnetic shieldings.  We plan to implement such features in future versions of
the GIMIC code.

\noindent The current-density susceptibility consists of diatropic and
paratropic contributions with opposite direction of rotation of the flux in the
current-density vortex. The tropicity can be assigned by following the
trajectories of the vector field, whereas in the present version of the GIMIC
method, we use the direction of the current-density flux across an integration
plane to assign the tropicity, which is an approximation that requires careful
considerations. 

\noindent Calculations with the GIMIC program can be performed at any level of
theory, as long as the unperturbed and the magnetically perturbed density
matrices are available. Current-density studies on large molecules are mainly
performed at the DFT level.  Calculations on aromatic organic molecules using
traditional functionals like B3LYP\cite{Becke:93,Lee:88} yield accurate
strengths of the current-density flux, whereas functionals with a larger
fraction of Hartree-Fock exchange should be used when studying magnetic
properties of strongly antiaromatic molecules, because  the B3LYP functional
significantly overestimates the strength of their paratropic ring
currents.\cite{Valiev:13,Valiev:18a} Similar conclusions have been reached in
recent comparisons of the degree of aromaticity calculated with different DFT
functionals using a variety of aromaticity
criteria.\cite{Szczepanik:17,Casademont-Reig:18,Casademont-Reig:20}

\bibliographystyle{achemso} 

\begin{footnotesize}

\providecommand{\latin}[1]{#1}
\makeatletter
\providecommand{\doi}
  {\begingroup\let\do\@makeother\dospecials
  \catcode`\{=1 \catcode`\}=2 \doi@aux}
\providecommand{\doi@aux}[1]{\endgroup\texttt{#1}}
\makeatother
\providecommand*\mcitethebibliography{\thebibliography}
\csname @ifundefined\endcsname{endmcitethebibliography}
  {\let\endmcitethebibliography\endthebibliography}{}

\end{footnotesize}
 
\end{document}